\documentclass{vldb}
\usepackage{graphicx}
\usepackage{balance}  

\AtBeginDocument{%
  \providecommand\BibTeX{{%
    \normalfont B\kern-0.5em{\scshape i\kern-0.25em b}\kern-0.8em\TeX}}}

\usepackage{times,amsmath,epsfig}

\long\def\comment#1{}

\usepackage{color}
\usepackage{graphicx}
\usepackage{xspace}
\usepackage{subfigure}
\usepackage{amsmath}
\usepackage{float}
\usepackage{amssymb}
\usepackage{latexsym}
\usepackage{graphicx}
\usepackage{url}
\usepackage{epsfig}
\usepackage{mathrsfs}
\usepackage{times}
\usepackage{multirow}
\usepackage{color}
\usepackage{epstopdf}
\usepackage{algorithmic}
\usepackage[ruled,vlined,linesnumbered]{algorithm2e}%

\newcounter{example}[section]
\renewcommand{\theexample}{\nthesection.\arabic{example}}
\newenvironment{example}{
     \refstepcounter{example}
     {\vspace{1ex} \noindent\bf  Example  \theexample:}}{
     \vspace{1ex}} 

\newcounter{definition}[section]
\renewcommand{\thedefinition}{\nthesection.\arabic{definition}}

\newcounter{theorem}[section]
\renewcommand{\thetheorem}{\nthesection.\arabic{theorem}}
\newenvironment{theorem}{\begin{em}
        \refstepcounter{theorem}
        {\vspace{1ex} \noindent\bf  Theorem  \thetheorem:}}{
        \end{em}\vspace{1ex}} 

\newcounter{lemma}[section]
\renewcommand{\thelemma}{\nthesection.\arabic{lemma}}
\newenvironment{lemma}{\begin{em}
        \refstepcounter{lemma}
        {\vspace{1ex}\noindent\bf Lemma \thelemma:}}{
        \end{em}\vspace{1ex}} 

\newcounter{remark}[section]
\renewcommand{\theremark}{\nthesection.\arabic{remark}}

\newcommand{\proofsketch}{\noindent{\bf Proof Sketch: }}

\newcommand{\nthesection}{\arabic{section}}

\newcommand{\eop}{\hspace*{\fill}\mbox{$\Box$}\vspace*{1ex}}

\newcommand{\stitle}[1]{\vspace{1ex} \noindent{\bf #1}}

\newcommand{\SWITCH}[1]{\STATE \textbf{switch} (#1)}
\newcommand{\ENDSWITCH}{\STATE \textbf{end switch}}
\newcommand{\CASE}[1]{\STATE \textbf{case} #1\textbf{:} \begin{ALC@g}}
\newcommand{\ENDCASE}{\end{ALC@g}}

\newcommand{\DEFAULT}{\STATE \textbf{default:} \begin{ALC@g}}
\newcommand{\ENDDEFAULT}{\end{ALC@g}}
\newcommand{\DEFAULTLINE}[1]{\STATE \textbf{default:} }

\newcommand{\CL}{{\sl DviCL}\xspace}
\newcommand{\cl}{{\sl cl}\xspace}
\newcommand{\DivideP}{{\sl DivideI}\xspace}
\newcommand{\DivideS}{{\sl DivideS}\xspace}
\newcommand{\CombineCL}{{\sl CombineCL}\xspace}
\newcommand{\CombineST}{{\sl CombineST}\xspace}
\newcommand{\Partition}{{\sl R}\xspace}
\newcommand{\CSG}{{\sl CSG}\xspace}
\newcommand{\Condense}{{\sl Condense}\xspace}
\newcommand{\Preprocessing}{{\sl Preprocessing}\xspace}
\newcommand{\CanonicalLabeling}{{\sl CanonicalLabeling}\xspace}
\newcommand{\Labeling}{{\sl Labeling}\xspace}
\newcommand{\SymmetryAndLabeling}{{\sl SymmetryAndLabeling}\xspace}
\newcommand{\DecTree}{{\sl DecTree}\xspace}
\newcommand{\PBBFS}{{\sl PBBFS}\xspace}

\newcommand{\SSM}{{\sl SSM-AT}\xspace}

\newcommand{\nauty}{{\sl nauty}\xspace}
\newcommand{\saucy}{{\sl saucy}\xspace}
\newcommand{\bliss}{{\sl bliss}\xspace}
\newcommand{\traces}{{\sl traces}\xspace}
 
\newcommand{\GreedyIM}{{\sl GreedyIM}\xspace}
\newcommand{\AutomorphismIM}{{\sl AutomorphismIM}\xspace}

\begin{document}

\title{Graph Iso/Auto-morphism: A Divide-\&-Conquer Approach}

\author{%
{Can Lu, Jeffrey Xu Yu, Zhiwei Zhang$^{\#}$, Hong Cheng}%
\vspace{1.6mm}\\
\fontsize{10}{10}\selectfont\itshape
The Chinese University of Hong Kong, Hong Kong, China;
$~^{\#}$ Hong Kong Baptist University, Hong Kong, China \\
\fontsize{9}{9}\selectfont\ttfamily\upshape
\{lucan,yu,hcheng\}@se.cuhk.edu.hk;  $~^{\#}$ cszwzhang@comp.hkbu.edu.hk
}

\def\thepage{\arabic{page}}
\pagestyle{empty}

\maketitle

\thispagestyle{empty}

\begin{abstract}
The graph isomorphism is to determine whether two graphs are
isomorphic.  A closely related problem is graph automorphism
(symmetry) detection, where an isomorphism between two graphs is a
bijection between their vertex sets that preserves adjacency, and an
automorphism is an isomorphism from a graph to itself.  Applications
of graph isomorphism and automorphism detection include database
indexing, network model, network measurement, network simplification,
and social network anonymization.
By graph automorphism, we deal with symmetric subgraph matching (SSM),
which is to find all subgraphs in a graph $G$ that are symmetric to a
given subgraph in $G$. An application of SSM is to identify multiple
seed sets that have the same influence power as a set of seeds
found by influence maximization in a social network.
%
%
%
To test two graphs for isomorphism, canonical labeling has been
studied to relabel a graph in such a way that isomorphic graphs are
identical after relabeling.
%
%
Efficient canonical labeling algorithms have been designed by
individualization-refinement. They enumerate all permutations of
vertices using a search tree, and select the minimum permutation as
the canonical labeling. The candidates are pruned by the minimum
permutation during enumeration.  Despite their high performance in
benchmark graphs, these algorithms face difficulties in handling
massive graphs, and the search trees used are for pruning purposes
which cannot answer symmetric subgraphs matching.

In this paper, we design a new efficient canonical labeling algorithm
\CL.
\CL designed is based on the observation that we can use the $k$-th
minimum permutation as the canonical labeling.
Different from previous algorithms, we take a divide-and-conquer
approach to partition a graph $G$. By partitioning $G$, an AutoTree is
constructed, which preserves symmetric structures as well as the
automorphism group of $G$. The canonical labeling for a tree node can
be obtained by the canonical labeling of its child nodes.  and the
canonical labeling for the root is the one for $G$.  Such AutoTree can
also be effectively used to answer the automorphism group, symmetric
subgraphs.
%
We conducted extensive performance studies using 22
large graphs, and confirmed that  \CL is much  more
efficient and robust than the state-of-the-art.
\end{abstract}

\section{Introduction}

Combinatorial objects and complex structures are often modeled as
graphs in many applications, including social networks and social
media \cite{papadopoulos2012community}, expert networks
\cite{lappas2009finding}, bioinformatics \cite{yeger2004network,
  zheng2011large}, and mathematical chemistry
\cite{bonchev1991chemical}.
%
%
Among many graph problems, graph isomorphism is a
problem to determine whether two graphs are isomorphic
%
%
\cite{read1977graph}. The graph isomorphism is an important issue in
practice since it has been used for deduplication and retrieval in
dealing with a collection of graphs, and is an important issue in
theory due to its relationship to the concept of NP-completeness.  A
closely related graph problem is automorphism (symmetry) detection, where an isomorphism between two graphs is a bijection between
their vertex sets that preserves adjacency, and an automorphism
(symmetry) is an isomorphism from a graph to itself.
%
%
Automorphism detection is also important in various graph problems.
On one hand, by automorphism, from a global viewpoint, two vertices
(or subgraphs) are equivalent in the sense that the entire graph
remains unchanged if one is replaced by the other. Therefore, with
automorphism, certain finding over a single vertex (or a subgraph) can
be applied to all other automorphic vertices (or subgraphs).  On the
other hand, as symmetries of combinatorial objects are known to
complicate algorithms, detecting and discarding symmetric subproblems can reduce
the scale of the original problems.

There are many applications of graph isomorphism and automorphism
detection, including database indexing \cite{randic1981computer}, network model \cite{macarthur2008symmetry,xiao2008emergence}, network
measurement \cite{xiao2008network}, network simplification \cite{xiao2008network}, and social network anonymization \cite{wu2010k}.
%
(a) Database Indexing: Given a large database of graphs (e.g.,
  chemical compounds), it assigns every graph with a certificate such
  that two graphs are isomorphic iff they share the same certificate
  \cite{randic1981computer}.
%
(b) Network Model: It studies the automorphism groups of a wide
  variety of real-world networks and finds that real graphs are richly
  symmetric~\cite{macarthur2008symmetry}. In \cite{xiao2008emergence},
  it claims that similar linkage patterns are the underlying
  ingredient responsible for the emergence of symmetry in complex
  networks.
%
(c) Network Measurement: In \cite{xiao2008symmetry}, it proposes a
  structure entropy based on automorphism partition to precisely
  quantify the structural heterogeneity of networks, and finds that
  structural heterogeneity is strongly negatively correlated to
  symmetry of real graphs.
%
(d) Network Simplification: In \cite{xiao2008network}, it utilizes
  inherent network symmetry to collapse all redundant information from
  a network, resulting in a coarse graining, known as ``quotient'',
  and claims that they preserve various key function properties such
  as complexity (heterogeneity and hub vertices) and communication
  (diameter and mean geodesic distance), although quotients can be
  substantially smaller than the original graphs.
%
(e) Social Network Anonymization: In \cite{wu2010k}, it proposes a
$k$-symmetry model to modify a naively-anonymized network such that
for any vertex in the network, there exist at least $k -$$1$
structurally equivalent counterparts, protecting against
re-identification under any potential structural knowledge about a target.
%
Below, we discuss how graph automorphism   is used for influence maximization
(IM)~\cite{arora2017debunking,chen2009efficient,kempe2003maximizing,ohsaka2014fast},
and discuss symmetric subgraph matching (SSM) by graph automorphism
and other SSM
applications~\cite{ferrante1987program,kuck1981dependence,liu2006gplag}.
%
%
%

%
%


\comment{
Below, we discuss how graph automorphisms discovered and the
AutoTree constructed by our approach \CL can
help deal with existing problems (e.g. influence maximization) and
novel problems (e.g. symmetric subgraph matching and $k$-symmetry),
respectively.
}

Influence maximization (IM) is widely studied in social networks and
social media to select a set $S$ of $k$ seeds s.t. the expected value
of the spread of the influence $\sigma(S)$ is maximized. In the
literature, almost all work in IM find a single $S$ with the maximum
influence. With graph automorphism, we can possibly find a set
${\mathcal S} = \{S_1, S_2, \cdots\}$ where each $S_i$ has the same
max influence as $S$ while contains some different vertices, and we are
able to select one $S_i$ in ${\mathcal S}$ that satisfies some
additional criteria (e.g., attributes on vertices in a seed set
and distribution of such seed vertices).
To show such possibilities, we compute IM by one of the best
performing algorithms, PMC \cite{ohsaka2014fast}, under the IC model
as reported by \cite{arora2017debunking}, over a large number of
datasets (Table~\ref{tbl:summarization}) using the parameters
following \cite{arora2017debunking}, where the probability to
influence one from another is treated as constant.  We conduct testing
to select a set of $k$ seeds, for $k = 10$ and $k = 100$.
We find that there are 8.82E+15 and 2.93E+15 candidate seed sets for
wikivote when $k=10$ and $k=100$, respectively, and the numbers for
Orkut are 4 and 2.9E+10, respectively.
To find ${\mathcal S}$ for $S$ found by IM can be processed as a
special case of symmetric subgraph matching (SSM), which we
discuss below.

Symmetric subgraph matching (SSM), we study in this paper, is closely
related to subgraph matching (or subgraph isomorphism). Given a query
graph $q$ and a data graph $G$, by subgraph matching, it finds all
subgraphs $g$ in $G$ that are isomorphic to $q$.  By SSM, $q$ is
required to be a subgraph that exists in $G$ and any $g$ returned
should be symmetric to $q$ in $G$, i.e., there is at least one
automorphism $\gamma$ of $G$ having $g=q^\gamma$. Note that all
subgraphs discussed here are induced.
The applications of SSM  include software plagiarism, program
maintenance and compiler optimizations
\cite{ferrante1987program,kuck1981dependence,liu2006gplag}, where an
intermediate program representation, called the {\sl program
  dependence graph (PDG)} is constructed for both control and data
dependencies for each operation in a program.  \comment{ In PDG,
  statements are represented by vertices and data and control
  dependencies between statements are represented by edges.  Software
  plagiarism attempts to detect copied programs from open source
  projects with modifications like statement reordering and code
  insertion, program maintenance finds all programs that share the
  same function or program logic with the program to update or to
  correct, and compiler optimization enhances the performance of
  programs by transformations like vectorization, pipeline and
  multiprocessing. These problems benefit from the invariance property
  of PDG, i.e., code changes regardless of dependencies are prone to
  errors and PDGs for programs sharing the same function and logic apt
  to be isomorphic and symmetric in the PDG of the whole software. In
  these problems, the query subgraph $q$ is required to be a subgraph
  of the given data graph $G$. We will discuss the algorithm in
  Section~\ref{sec:ssm}.  }

\comment{
The $k$-symmetry \cite{wu2010k} is a model proposed to modify a
naively-anonymized network such that for any vertex in the network,
there exist at least $k-1$ structurally equivalent counterparts,
protecting against re-identification under any potential structural
knowledge about a target. In other words, it requires each vertex has
at least $k-1$ automorphic counterparts in the reconstructed graph.
}


\comment{
\begin{algorithm}[t]
\small
\caption{SSM($(G,q,{\mathcal AT})$) }
\label{alg:SSM}
\begin{algorithmic}[1]
\STATE partition $q$ into vertex disjoint subgraphs $\{s_1, \ldots, s_{k1}\}$ according to leaves in AutoTree ${\mathcal AT}$, denote the leaf of ${\mathcal AT}$ containing $s_i$ by $l_i$;
\STATE construct the minimum subtree $T$ of $\mathcal AT$ containing leaves $\{l_1, \ldots, l_{k1}\}$;
\FOR {each $(s_i, l_i)$}
\STATE $R_i \leftarrow SM(s_i, l_i)$;
\STATE $L_i \leftarrow \{l_1^i, \ldots, l_{k2}^i\}$, where $l_j^i$ is a leaf of ${\mathcal AT}$ having the same signature as $l_i$;
\FOR {each $l_j^i \in L_i$}
\STATE mapping each subgraph in $R_i$ to a subgraph in $l_j^i$, resulting in $R_j^i$;
\ENDFOR
\ENDFOR
\STATE ${\mathcal R} \leftarrow \emptyset$;
\FOR {any $T' \in SM(T, \mathcal AT)$}
\STATE extracts the leaf nodes of $T'$, denoted as $(l_{i1}^1, l_{i2}^2, \ldots, l_{ik}^k)$;
\STATE ${\mathcal R} \leftarrow {\mathcal R} \cup R_{i1}^1 \times \ldots \times R_{ik}^k$;
\ENDFOR
\STATE ${\mathcal S} \leftarrow G[{\mathcal R}]$;
\RETURN $\mathcal S$;
\end{algorithmic}
\end{algorithm}

Before discussing Algorithm SSM, we revisit some properties of AutoTree $\mathcal AT$. Each node in $\mathcal AT$
corresponds to a subgraph in $G$, and associates with information including the automorphism group and canonical labeling of the subgraph as a signature. Two tree nodes with the same signature implies that the two corresponding subgraphs in $G$ are symmetric, and any two vertices with the same labeling are automorphic in $G$. These properties benefit SSM significantly.
Algorithm SSM, shown in Algorithm~\ref{alg:SSM}, follows divide-and-conquer paradigm. Query graph $q$ is divided into vertex disjoint subgraphs $\{s_1, \ldots, s_{k1}\}$ where all vertices in $s_i$ are contained in a leaf node $l_i$ in $\mathcal AT$ (Line~1).
SSM first finds mosaic subgraphs that are symmetric to each $s_i$ (Line~3-9) and reconstructs the resulting subgraphs symmetric to $q$ using these mosaics (Line~10-15).
For each $s_i$, its symmetric subgraphs can be found by first locating at $l_j^i$, a leaf node in $\mathcal AT$ that is symmetric to $l_i$ (Line~5) and then mapping subgraphs in $R_i=SM(l_i, s_i)$, symmetric to $s_i$ in $l_i$, to subgraphs in $l_j^i$ (Line~6-8). This mapping can be achieved with the canonical labeling of $l_i$ and $l_j^i$ easily.
To reconstruct subgraphs symmetric to $q$, any subtree matching of $T$, the minimum subtree of $\mathcal AT$ containing $\{s_1, \ldots, s_k\}$, helps to identify valid combinations.
In SSM, the most time consuming part is $SM(s_i, l_i)$ (Line~4) and $SM(T, {\mathcal AT})$ (Line11). For the former one, (1) both $l_i$ and $s_i$ are much smaller than $q$ and $G$, (2) orbit coloring of $l_i$ prunes significant unnecessary searching. For the latter one, since both $T$ and $\mathcal AT$ are trees, SM on trees are much easier than on general graphs.
}

\comment{
Below, we first discuss how graph automorphisms can help influence maximization (IM)
\cite{kempe2003maximizing,chen2009efficient,ohsaka2014fast,arora2017debunking},
which is an important issue in social networks and social media, as
%
%
%
%
it is to select a set $S$ of $k$ seeds s.t. the expected value of
the spread of the influence $\sigma(S)$ is maximized.
In the literature, almost all work in IM find a single $S$.
There are two questions to ask. The first question is how many sets
exist that have the same max influence, since such a set should not be
unique in practice. If we can find a set ${\mathcal S} = \{S_1, S_2,
\cdots\}$ where $S_i$ and $S_j$ are two different sets of $k$ seeds
that have the same max influence, it becomes possible to select one
set with additional criteria (e.g., connectivity among vertices in a
seed set and distribution of such seed vertices).
The second question is whether we can significantly speedup the up-to-date
IM algorithm, since IM is time-consuming. The fastest and best
performing algorithm under the IC model is PMC \cite{ohsaka2014fast},
whose time complexity is $O(kmnR)$ for a graph $G$ with $n$ vertices
and $m$ edges, if we run PMC in $R$ Monte-Carlo simulations to estimate the expected spread.
For the first question, we compute IM by PMC over a large number of
datasets (Table~\ref{tbl:summarization}) using the parameters
following \cite{arora2017debunking}, where the probability to
influence one from another is treated as constant.  We conducted
testing to select a set of $k$ nodes, for $k = 10$ and $k = 100$. We
show the number of different seed sets that have the same max
influence in Table~\ref{tbl:num_seedset}, which can be huge.
For the second question, recall PMC \cite{ohsaka2014fast} spends
$O(nmR)$ to find a single seed node. The time complexity for an IM
algorithm is high.
We design an algorithm, \AutomorphismIM,
that adds all vertices that are automorphic to the $v^*$ selected as
the best in every iteration. In other words, the existing only adds a
best vertex in an iteration, whereas \AutomorphismIM can add more than
one.
Fig.~\ref{fig:AutoIMEff} demonstrates the improvement in efficiency of
\AutomorphismIM over PMC. The y-axis shows the number of iterations
reduced.
%
%
As can be seen, incorporating graph automorphisms can improve the
efficiency significantly. On the other hand, Fig.~\ref{fig:AutoIMQua}
demonstrates its quality, in percentage,
%
%
i.e., $1-\sigma_A(S)/\sigma_G(S)$, where $\sigma_A(S)$ is the quality
by \AutomorphismIM and $\sigma_G(S)$ is the quality by PMC.  As shown
in Fig.~\ref{fig:AutoIMQua}, for most datasets, \AutomorphismIM
achieves comparative quality results.
%

}

\comment{
In the second attempt, we incorporate graph automorphisms into a IM
algorithm to improve its efficiency.  We focus on algorithms with
theoretical guarantees. In the following discussion, \GreedyIM refers
to an existing IM algorithm and \AutomorphismIM refers to the novel
algorithm incorporating graph automorphisms within
\GreedyIM. \AutomorphismIM shares the same schema with \GreedyIM,
while differs in choosing vertices to add into seed set $S$ in each
iteration. Let $v^*$ denote the vertex with maximum marginal gain each
iteration, i.e., $v^*=\arg max_{\forall v \in V} \{\sigma(S \cup
\{v\})-\sigma(S)\}$. \AutomorphismIM adds all vertices in $V$ that are
automorphic to $v^*$ while \GreedyIM only adds $v^*$ itself.

First, we show that with the automorphism group $Aut(G)$ of the given
graph $G$, a single seed set $S$ discovered by any ordinary IM
algorithm can be easily extended to a series of seed sets with the
same spread $\sigma(S)$. Table\ref{tbl:num_seedset} demonstrates the
results, where the 2nd column and the 3rd column show the numbers of
candidate seed sets when 10 and 100 seed nodes, respectively.  As can
been seen, in a majority of datasets, there are numerous candidates
that can replace the result of IM algorithms without decreasing
spread. Such attempt is worthy.  First, there are scenarios where
additional constrains, such as connectivity between seed vertices and
distribution of seed vertices in the graph, are required on the seed
set. Thus the single result returned by existing algorithms may not
meet such requirements. Second, due to the approximations and the
inefficiency of IM algorithms, designing an algorithm to enumerate all
seed sets with optimal spread is not practical. Whereas, with graph
automorphisms, any ordinary IM algorithm can achieve this goal with a
trivial and efficient postprocessing procedure.

We make two attempts to incorporate graph automorphisms within IM. The
first aims at  estimating the number of or enumerating seed sets
sharing the same spread with $S$,  given the automorphism group
$Aut(G)$. Here, $S$ is the resulting seed set  by  IM algorithms. The
second is to exploit graph automorphisms to improve the efficiency of
existing IM algorithms.

Specifically, all the edges of the graphs
are assigned with a constant probability $W(u,v)=0.1$, $R=10K$
Monte-Carlo simulations are performed to compute the expected spread,
and seeds number $k$ is set as 100.

Second, due to the approximations and the
inefficiency of IM algorithms, designing an algorithm to enumerate all
seed sets with optimal spread is not practical. Whereas, with graph
automorphisms, any ordinary IM algorithm can achieve this goal with a
trivial and efficient postprocessing procedure.

Worth noting that since IM is much
more time-consuming than automorphism detection and the time
complexity of PMC is $O(kmnR)$, we use the number of iterations to
approximate running times of IM algorithms.
}

\comment{
In IM, a {\sl social network} is modeled as an edge-weighted graph
$G(V, E, W)$, where $W(u,v)$ indicates the probability $u$ activates
$v$.  A {\sl seed node} acts as the source of information diffusion in
$G$, and the set of seed nodes is denoted as $S$. An {\sl active node}
$v$ is either (a) a seed node or (b) a node activated by a previous
active node. Once activated, $v$ is added to the set of activate nodes
$V_a$. The number of activated nodes, i.e., $|V_a|$, is called the
{\sl spread} of seed nodes $S$, denoted as $\Gamma(S)$. Therefore, the
{\sl influence maximization (IM)} refers the problem of selecting a
set $S$ of $k$ seeds s.t. the expected value of spread
$\sigma(S)=\mathbb E[\Gamma(S)]$ is maximized.
}

\comment{

\begin{table}[t]
{\footnotesize
    \begin{center}
   \begin{tabular}{|l|r|r|} \hline
       {\bf Graph}  & $k=10$  & $k=100$  \\    \hline\hline
        BerkStan &68 &7.63E+24 \\ \hline
        Epinions &2 &840  \\ \hline
        Google  &40 &2.86E+25  \\ \hline
        LiveJournal &30 &1.19E+37\\ \hline
        NotreDame &88 &63,4360  \\ \hline
        Pokec &1 &302,400  \\ \hline
        Slashdot0811 &1 &192    \\ \hline
        Slashdot0902 &2 &18,432 \\ \hline
        Stanford  &6 &4.92E+15   \\ \hline
        wikivote  &91,390  &3.21E+17  \\ \hline
        Orkut    &4 &2.91E+10  \\ \hline
        BuzzNet  &80 &3.12E+173 \\ \hline
        Delicious &19 &787,968 \\ \hline
        Foursquare  &6.63E+6 &4.44E+71   \\ \hline
    \end{tabular}
    \end{center}
\caption{Number of all candidate seed sets}
\vspace{-1cm}
\label{tbl:num_seedset}

}
\end{table}
}

\comment{
\begin{table}[h]
{
\caption{Number of candidate seed sets with spread $\sigma(S)$ given $Aut(G)$ and $S$}
    \begin{center}
   \begin{tabular}{|l|r|r||l|r|r|} \hline
       {\bf Graph}  & $k=10$  & $k=100$ &{\bf Graph}  & $k=10$  & $k=100$ \\    \hline\hline
        Amazon &1 &1   &WikiTalk  &1 &1  \\ \hline
        BerkStan &68 &7.63E24 & wikivote  &91,390  &3.21E17  \\ \hline
        Epinions &2 &840  & Youtube   &1 &1 \\ \hline
        Gnutella &1 &1  &Orkut    &4 &2.91E10  \\ \hline
        Google  &40 &2.86E25 &BuzzNet  &80 &3.12E173 \\ \hline
        LiveJournal &30 &1.19E37 & Delicious &19 &787,968 \\ \hline
        NotreDame &88 &63,4360  & Digg  &1 &1 \\ \hline
        Pokec &1 &302,400  & Flixster &1 &1  \\ \hline
        Slashdot0811 &1 &192   & Foursquare  &6.63E6 &4.44E71   \\ \hline
        Slashdot0902 &2 &18,432 & Friendster &1 &1   \\ \hline
        Stanford  &6 &4.92E15  & Lastfm  &1 &1 \\ \hline
    \end{tabular}
    \end{center}
    \vspace*{-0.4cm}

\label{tbl:num_seedset}
}
\end{table}
}

\comment{

\begin{figure}[t]
\begin{center}
  \includegraphics[scale=0.3]{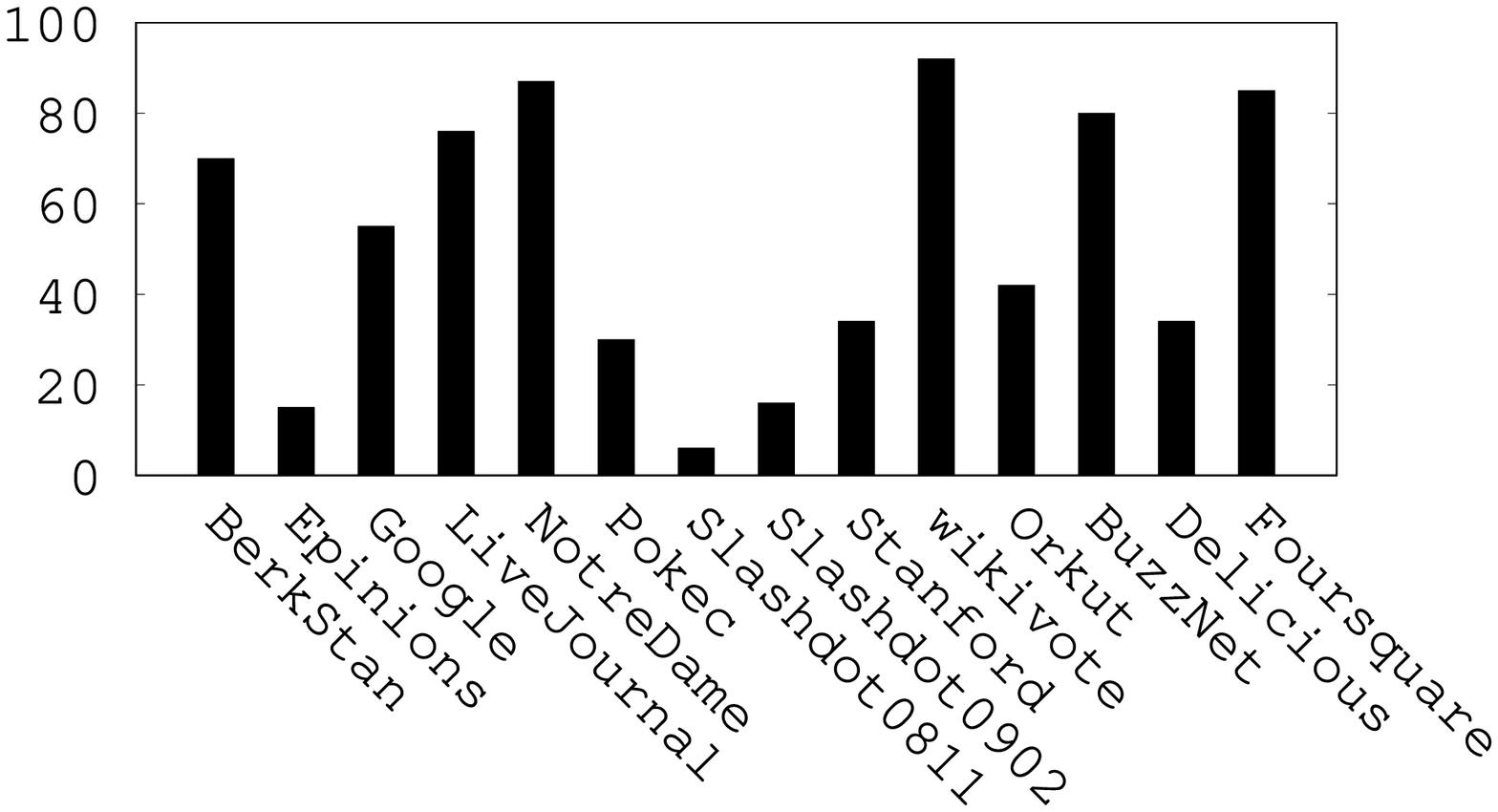}
\end{center}
\vspace{-0.4cm}
\caption{The efficiency of AutomorphismIM} 
\label{fig:AutoIMEff}
\vspace{-0.4cm}
\end{figure}

\begin{figure}[t]
\begin{center}
  \includegraphics[scale=0.3]{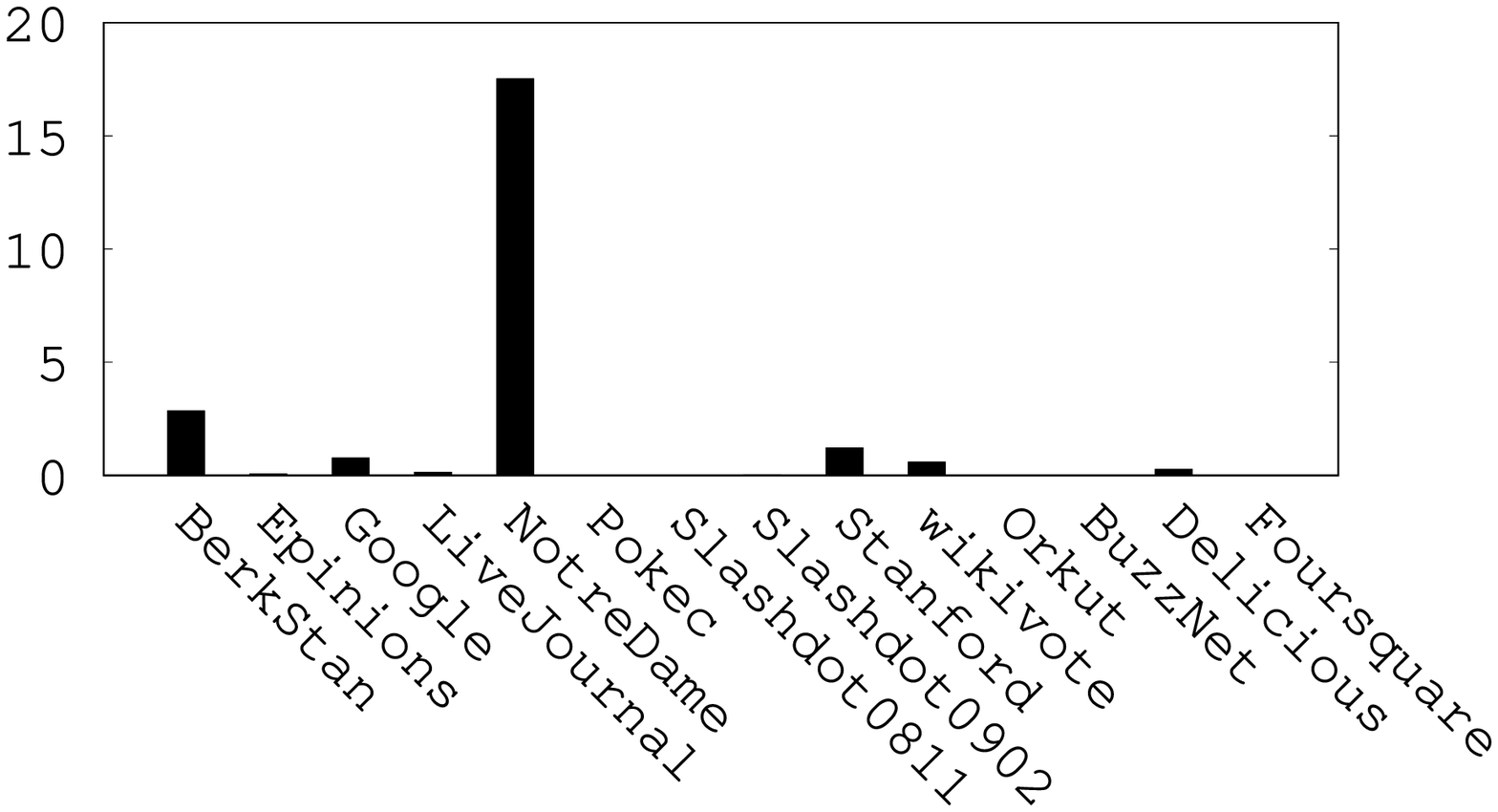}
\end{center}
\vspace{-0.4cm}
\caption{The quality of AutomorphismIM} 
\label{fig:AutoIMQua}
\vspace{-0.4cm}
\end{figure}
}

In the literature, to check if two graphs are isomorphic, the most
practical approach is {\it canonical labeling}, by which a graph is
relabeled in such a way that two graphs are isomorphic iff their
canonical labeling are the same.
%
%
%
Since the seminal work \cite{mckay1978computing,mckay1981practical} by
McKay in 1981, \nauty has become a standard for canonical
labeling and has been incorporated into several mathematical software
tools such as GAP \cite{gap2007gap} and MAGMA \cite{bosma1997magma}.
Other canonical labeling approaches, such as \bliss
\cite{junttila2007engineering, junttila2011conflict} and \traces
\cite{piperno2008search}, address possible shortcomings of \nauty
closely following \nauty's ideas.  Despite their high performance, these   approaches face
difficulties in handling today's massive graphs. As shown in our
experimental studies (Table~\ref{tbl:performance}), \nauty fails in
all but one datasets, \traces fails in nearly half datasets, and
\bliss is inefficient in most datasets.
Due to the
lack of efficient canonical labeling algorithms for massive graphs, to
the best of our knowledge, merely have any algorithms incorporated graph isomorphism or
graph automorphism.

In this work, we propose a novel efficient canonical labeling
algorithm for massive graphs. We observe that the state-of-the-art
algorithms (e.g., \nauty \cite{mckay1978computing,mckay1981practical},
\traces \cite{piperno2008search} and \bliss
\cite{junttila2007engineering, junttila2011conflict}) discover the
canonical labeling following ``individualization-refinement"
schema. These algorithms enumerate all possible permutations and
select the minimum $G^\gamma$ as the canonical labeling.
Here, graph $G$ as well as the permutated $G^\gamma$ can be represented as elements from a totally ordered set, for instance, $G$ ($G^\gamma$) can be represented by its sorted edge list.
At first
glance, choosing the minimum $G^\gamma$ as the target is probably the
most efficient for branch-and-bound algorithms. However, the minimum
$G^\gamma$ is not always the best choice for any graph $G$. For
instance, if all vertices in $G$ can be easily distinguished,  the permutation $\gamma$ based
on sorting is a better choice.  Our main idea is to divide the given
graph into a set of subgraphs satisfying that (1) two isomorphic
graphs $G$ and $G'$ will be divided into two sorted subgraph sets
$\{g_1, \ldots, g_k\}$ and $\{g'_1, \ldots, g'_k\}$, such that $g_i$ is isomorphic to $g'_i$ for $1 \leq i \leq
k$; (2)
%
the canonical labeling of the original graph $G$ can be easily
obtained by canonical labeling of the subgraphs. Note that canonical
labeling of each subgraph $g_i$ can be defined arbitrarily, not
limited to the minimum $g_i^\gamma$.  As a consequence, our approach
returns the $k$-th minimum $G^\gamma$ as the canonical labeling. Note
that $k$ is not fixed for all graphs, and we do not need to know what
the $k$ value is when computing the canonical labeling.  Applying such
idea to each subgraph $g_i$, our approach \CL follows
divide-and-conquer paradigm and constructs a tree index, called
AutoTree ${\mathcal AT}$. Here, a tree node in ${\mathcal AT}$ corresponds to a
subgraph $g_i$ of $G$, and contains its canonical labeling as well as
automorphism group. The root   corresponds to
$G$.

By the AutoTree, we can easily detect symmetric vertices and
subgraphs in $G$.
\comment{
To deal with SSM, we  convert SSM in a general graph into subtree
matching in AutoTree ${\mathcal AT}$.
The conversion is possible based on the
following property of AutoTree. If two vertices, $u$ and $u'$, are
automorphic in $G$, there are two subgraphs, $g$ containing $u$ and
$g'$ containing $u'$, represented as two leaf nodes $t$ and $t'$ in
${\mathcal AT}$, that are isomorphic and symmetric in $G$.
%
%
}
Take the maximum clique as an example.
Given a graph $G$, for a maximum clique $q$ found \cite{lu2017finding},
%
%
we can efficiently identify 4 candidate maximum cliques in Google and
16 candidate maximum cliques in LiveJournal
(Table~\ref{tbl:summarization}) using the AutoTree constructed,
respectively. Algorithm \SSM for symmetric subgraph matching is given in Section~\ref{sec:ssm}.
%
%
%
%
%
\comment{
To deal with IM, for a seed set $S$ found, we can identity all
automorphic vertices for every single vertex in $S$.
}
For $k$-symmetry \cite{wu2010k}, with AutoTree, each subtree of root
can be duplicated to have at least $k-1$ symmetric
siblings. As a consequence, each vertex has at least $k-1$ automorphic
counterparts in the reconstructed graph.

\comment{
In this work, we study the graph isomorphism problem for massive
graphs, and focus on designing a novel efficient canonical labeling
algorithm. Different from the state-of-the-art algorithms (e.g. \nauty
\cite{mckay1978computing,mckay1981practical}, \traces
\cite{piperno2008search} and \bliss \cite{junttila2007engineering,
  junttila2011conflict}) which discover the automorphism group and
canonical labeling directly on the original graph, we design a new
algorithm \CL by divide-and-conquer. With \CL, an index, called
AutoTree, is constructed for a given graph $G$. By the AutoTree,
symmetric vertices and subgraphs in $G$ can be easily detected.
}

\comment{
Specifically, given a graph $G$,
\CL recursively divides $G$ into smaller subgraphs such that canonical
labeling is only necessary for some non-singleton subgraphs. Additionally,
the divide-and-conquer tree rearranges components of $G$ in a
tree-shape structure such that symmetric subgraphs in $G$ can be
easily detected.
}


The main contributions of our work are summarized below.
First, we propose a novel canonical labeling algorithm \CL following
the divide-and-conquer paradigm. \CL can efficiently discover the
canonical labeling and the automorphism group for   massive
graphs.
%
%
%
Second, we construct an AutoTree for a graph $G$ which provides an
explicit view of the symmetric structure in $G$ in addition to the
automorphism group and canonical labeling.
Such AutoTree can also be used to  solve  symmetric subgraph matching   and
social network anonymization.  Third, we conduct
extensive experimental studies to show the efficiency and robustness of \CL.


The preliminaries and the problem statement are given in
Section~\ref{sec:problemdefinition}. We discuss related works in Section~\ref{sec:relatedworks}, and review the previous algorithms
in Section~\ref{sec:previous}.  We give an overview in
Section~\ref{sec:overview}, and discuss the algorithms in
Section~\ref{sec:cl}. We conduct comprehensive experimental studies
and report our findings in Section~\ref{sec:exp}. We conclude this
paper in Section~\ref{sec:conclusion}.

\comment{
\section{Graph Symmetry and Information Maximization}
\label{sec:IM}

Influence maximization (IM) \cite{kempe2003maximizing,chen2009efficient,ohsaka2014fast,arora2017debunking} on social networks and social media is one of the most active areas of research in computer science.
In influence maximization, it is assumed that a user $u$ can {\sl directly} influence user $v$ if there is an edge from $u$ to $v$. For example, $u$ positing a positive review on a movie may result in $v$ actually watching the movie. This event may in turn result in $v$ influencing his/her own friends. The IM problem is to identify a set of {\sl seed nodes} so that the total number of users influenced is maximized. To be self-contained, we formally give the definitions of some concepts of vital importance as follows.

\stitle{Social Network:} A social networks can be modeled as an edge-weighted graph $G(V,E,W)$, where $V$ is the set of nodes, $E$ is the set of directed relationships, and $W$ is the set of edge-weights corresponding to each edge in $E$. $In(v)$ and $Out(v)$ denote the set of incoming and outgoing neighbors of vertex $v$, respectively.

\stitle{Seed node:} A node $v \in V$ that acts as the source of information diffusion in the graph $G(V, E, W)$ is called a seed node, and the set of seed nodes is denoted as $S$.

\stitle{Active node:} A node $v \in V$ is deemed active if either (a) it is a seed node ($v \in S$) or (b) it receives information from a previously active node $u \in V_a$. Once activated, the node $v$ is added to the set of active nodes $V_a$.

\stitle{Independent Cascade Model (IC):} Under the IC model, time unfolds in discrete steps. At any time-step $i$, each newly active node $u \in V_a$ gets one independent attempt to active each of its outgoing neighbors $v \in Out(u)$ with a probability $p(u,v)=W(u,v)$.

\stitle{Linear Threshold Model (LT):} Under the LT model, every node $v$ contains an activation threshold $\theta_v$, which is chosen uniformly at random from $[0,1]$. Further, LT dictates that the summation of all incoming edge weights is at most 1, i.e., $\sum_{\forall u \in In(v)} W(u,v) \leq 1$. $v$ gets activated if  $\sum_{\forall u \in In(v) \cap V_a} W(u,v) \leq \theta_v$.

\stitle{Spread:} Given an information diffusion model $\mathcal I$, the spread $\Gamma(S)$ of a set of seed nodes $S$ is defined as the total number of nodes that are active, including both the newly activated nodes and the initially active set $S$, at the end of the information diffusion process. Mathematically, $\Gamma(S)=|V_a|$.

\stitle{Influence Maximization (IM):} Given an integer $k$ and a social network $G$, select a set $S$ of $k$ seeds, i.e., $S \subset V$ and $|S|=k$, such that the expected value of spread $\sigma(S)= \mathbb E[\Gamma(S)]$ is maximized.

\begin{algorithm}[t]
\caption{GreedyIM($G(V,E,W), k, {\mathcal I}$) }
\label{alg:GreedyIM}
\begin{algorithmic}[1]
\STATE  $S \leftarrow \emptyset$;
\WHILE {$|S| <k$}
\STATE $v^* \arg max_{\forall v \in V} \{\sigma(S \cup \{v\})-\sigma(S)\}$ under $\mathcal I$;
\STATE $S \leftarrow S \cup \{ v^*\}$;
\ENDWHILE
\RETURN $S$;
\end{algorithmic}
\end{algorithm}

\begin{algorithm}[t]
\caption{AutomorphismIM($G(V,E,W), k, {\mathcal I}$) }
\label{alg:AutomorphismIM}
\begin{algorithmic}[1]
\STATE  $S \leftarrow \emptyset$;
\WHILE {$|S| <k$}
\STATE $v^* \arg max_{\forall v \in V} \{\sigma(S \cup \{v\})-\sigma(S)\}$ under $\mathcal I$;
\FOR {any $u$ that is automorphic to $v^*$}
\STATE $S \leftarrow S \cup \{ u\}$;
\ENDFOR
\ENDWHILE
\RETURN $S$;
\end{algorithmic}
\end{algorithm}

It is proved by \cite{kempe2003maximizing} that the IM problem is NP-hard under both IC and LT models. Since the spread function $\Gamma(\cdot )$ and its expectation $\sigma(\cdot)=\mathbb E[\Gamma(\cdot)]$ are  monotone and submodular, greedy algorithm achieves the optimal solution within a factor of $(1-\frac{1}{e})$ \cite{nemhauser1978analysis}. Ignoring implementation details and optimizations, Algorithm~\ref{alg:GreedyIM} shows the outline of most state-of-the-art approaches with theoretical guarantees.

To the best of our knowledge, merely has any existing work exploited graph structure information when studying IM. In this section, we attempt to utilize automorphisms discovered to answer questions related to IM, and start with two elementary questions, namely, whether automorphisms can improve the efficiency of existing algorithms and whether automorphisms can improve the quality of the results obtained by existing algorithms.

The answer to the first question is positive. We give an algorithm that incorporates graph automorphisms within  Algorithm~\ref{alg:GreedyIM}, shown in Algorithm~\ref{alg:AutomorphismIM}.
Algorithm AutomorphismIM differs from Algorithm GreedyIM in choosing  vertices to add into seed set $S$ in each iteration. Let $v^*$ denote the element with maximal marginal gain in each iteration. AutomorphismIM adds all vertices in $G$ that are automorphic to $v^*$ while GreedyIM only adds $v^*$ itself. The intuition is as follows.
Let $v_i$ denote the vertex selected in the $i$th iteration, for $1 \leq i \leq k$. Similar, let $S_i$ denote the seed set at the beginning of the $i$th iteration.
Consider two continuous iterations in GreedyIM, the $j$th iteration with $v^*=v_j$ and the $(j+1)$th iteration with $v^*=v_{j+1}$. Suppose vertex $v_{j'}$ is automorphic to $v_j$, and $v_{j+1}$ is not automorphic to $v_j$ without loss of generality. According to submodularity property,
$\sigma(S_j \cup \{v_j\}) =\sigma(S_j \cup \{v_{j'}\})> \sigma(S_j \cup \{v_{j+1}\})$,

We conduct experimental studies to demonstrate the efficiency and the quality of the results obtained by Algorithm AutomorphismIM. The experiments are conducted under the IC model, and similar results can be obtained under the LT model. We apply PMC \cite{ohsaka2014fast}, which is claimed to be the fastest and best performing algorithm under the IC model by \cite{arora2017debunking}, as the baseline algorithm. Detailed description of the datasets is given in Table~\ref{tbl:summarization}. We set parameters following \cite{arora2017debunking}. Specifically, all the edges of the graphs are assigned with a constant probability $W(u,v)=0.1$, $R=10K$ Monte-Carlo simulations are performed to compute the expected spread, and seeds number $k$ is set as 100. Worth noting that since IM is much more time-consuming than automorphism detection and the time complexity of PMC is $O(kmnR)$, we use the number of iterations to approximate running times of IM algorithms, then the iterations saved by AutomorphismIM can be regarded as improvement in efficiency of AutomorphismIM over GreedyIM.

Regard the second question, namely, whether automorphisms detected can improve the quality of the results obtained by existing algorithms, the answer is negative under the ordinary IM problem setting. However, graph automorphisms enable us dealing with novel IM problems. For instance, almost all current IM algorithms discover only one seed set $S$. With

}

\section{Problem Definition}
\label{sec:problemdefinition}

In this paper, we discuss our approach on an undirected graph $G=(V,E)$
without self-loops or multiple edges, where $V$ and $E$
denote the sets of vertices and edges of $G$, respectively.
We use $n$ and $m$ to denote the numbers of vertices and edges
of $G$, respectively, i.e., $n=|V|$ and $m=|E|$.
For a vertex $u \in V$, the neighbor set of $u$ is denoted as
 $N(u)=\{v~{}|~{}(u,v) \in E\}$,
and the degree of $u$ is denoted as $d(u)=|N(u)|$. In the following, we discuss some concepts and notations  using an example graph $G$ shown in Fig.~\ref{fig:expgraph}.

\comment{
\begin{figure}[t]
\begin{center}
  \includegraphics[width=0.45\columnwidth,height=2.2cm]{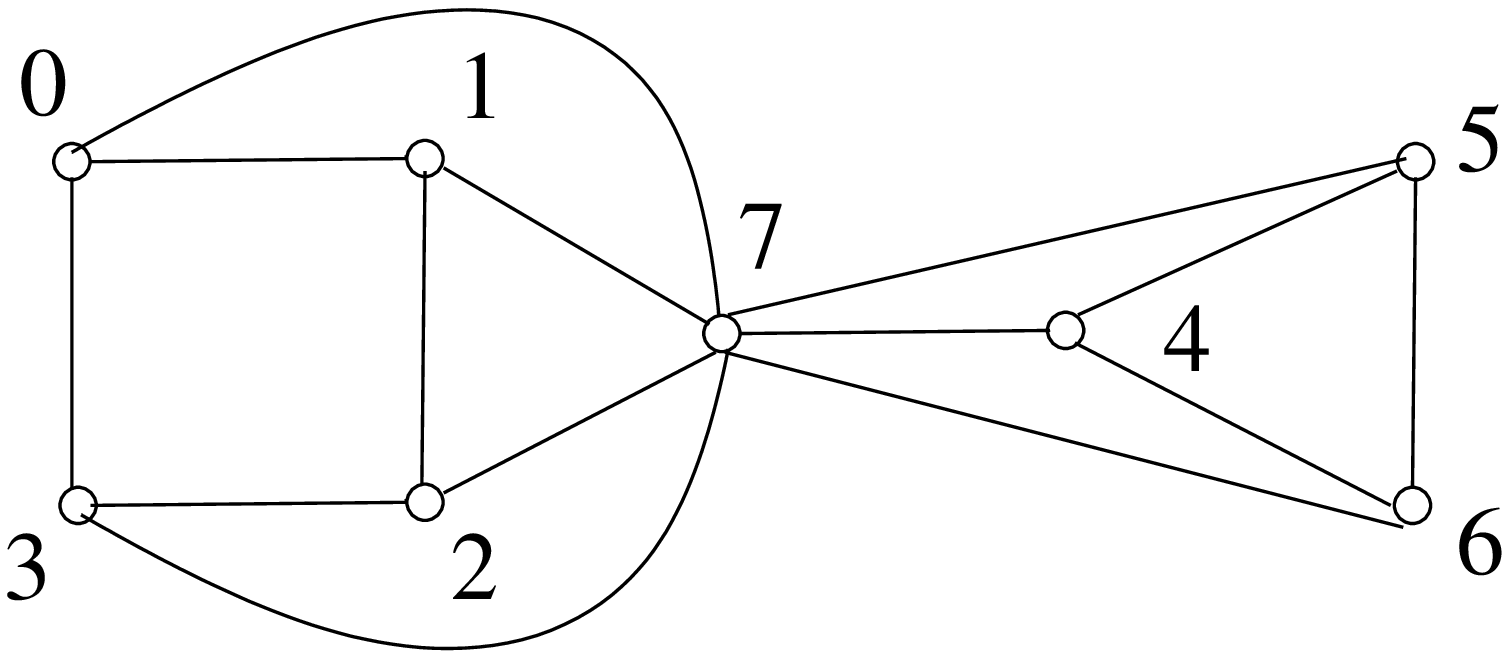}
\end{center}
\vspace{-0.4cm}
\caption{An example graph}
\label{fig:expgraph}
\vspace{-0.4cm}
\end{figure}
}

\begin{figure*}[t]
\begin{center}
\begin{tabular}[t]{c}
\hspace{-0.8cm}
   \subfigure[An example graph]{
     \includegraphics[width=0.6\columnwidth,height=2.5cm]{figure/exp.eps}
     \label{fig:expgraph}
    }
    \hspace{0.5cm}
    \subfigure[Backtrack search tree ${\mathcal T}(G, \pi)$ by \bliss for the graph in Fig.~\ref{fig:expgraph}]{
      \includegraphics[scale=0.33]{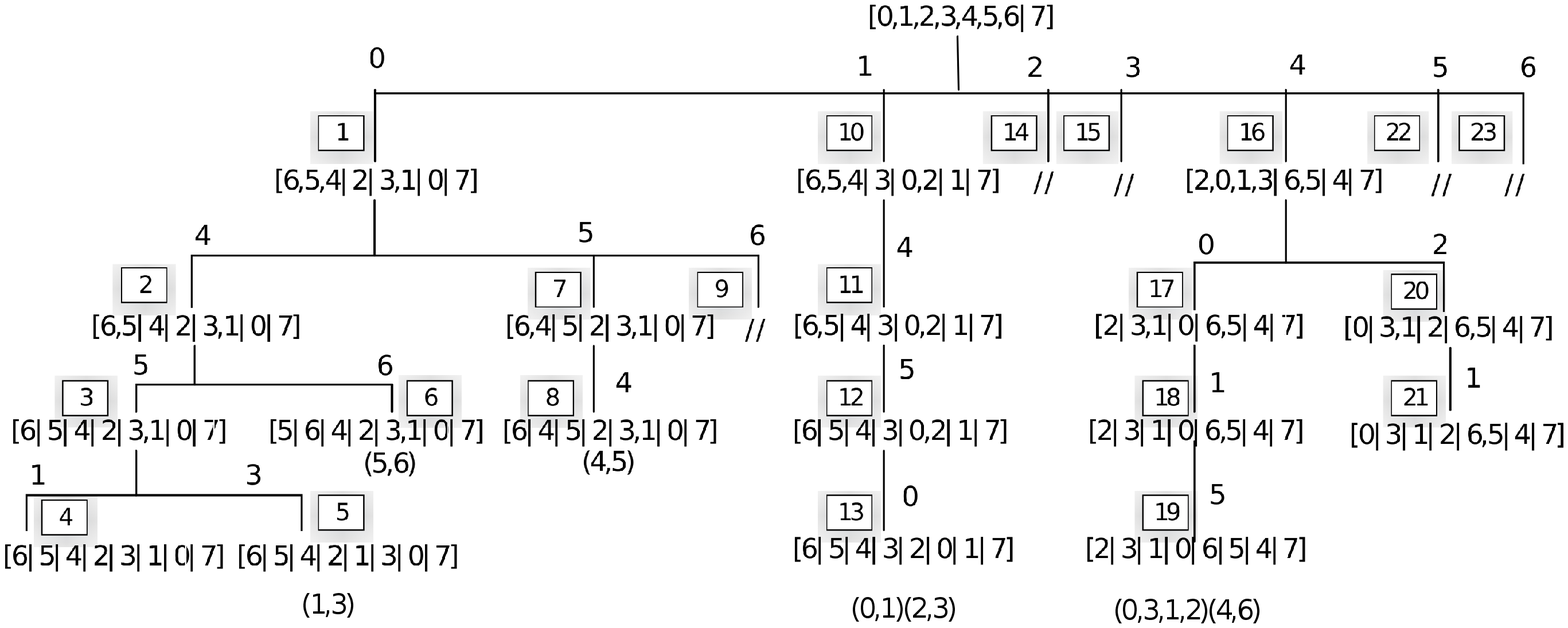}
      \label{fig:st_bliss}
    }
\end{tabular}
\end{center}
\vspace*{-0.4cm}
\caption{An example graph and a backtrack search tree by \bliss}
\vspace*{-0.4cm}
\end{figure*}

\stitle{Permutation}: A permutation of $V$, denoted as $\gamma$, is a bijection function from $V$ to itself.
We use $v^\gamma$ to denote the image of $v \in V$ under a permutation $\gamma$.
By a permutation $\gamma$ to a graph $G$, it permutes vertices in $V$ of  $G$ and produces a graph $G^\gamma=(V^\gamma, E^\gamma)$, here $V^\gamma=V$ and $E^{\gamma}=\{(u^\gamma, v^\gamma)| (u,v)\in E\}$.
Following the convention used in the literature, we use the cycle notation to represent permutations.
In a permutation $\gamma$, $(v_1, v_2, \ldots, v_k)$ means $v_i^{\gamma} =v_{i+1}$ for $1\leq i \leq k-1$ and $v_k^\gamma=v_1$.
For simplicity, we may only show permutation for a subset of vertices using the cycle notation, with the assumption that other vertices will be permuted to themselves. Consider the graph $G$ in Fig.~\ref{fig:expgraph}, the permutation  $\gamma_1=(4,5,6)$ is to relabel 4 as 5, 5 as 6, and 6 as 4, where all the other vertices are permutated to themselves. It produces a graph $G^{\gamma_1}=(V, E^{\gamma_1})$, where $E^{\gamma_1}=E$. For the same graph $G$, the permutation $\gamma_2=(0,1)$ relabels 0 as 1 and 1 as 0, and produces $G^{\gamma_2}=(V, E^{\gamma_2})$, where $E^{\gamma_2}= E \cup \{(0,2), (1,3)\} \setminus \{(0,3), (1,2)\}$.
All permutations of $V$ ($n!$ for $n = |V|$) consist of a {\sl
  symmetry group} with the permutation composition as the group
operation, denoted as $S_n$.

\stitle{Automorphism}: An automorphism
of a graph $G = (V, E)$ is
a permutation $\gamma~(\in S_n)$ that preserves $G$'s edge relation,
i.e., $G^{\gamma}=G$, or equivalently, $E^{\gamma}=E$.  In graph $G$
(Fig.~\ref{fig:expgraph}), $\gamma_1=(4,5,6)$ is an automorphism of
$G$ whereas $\gamma_2=(0,1)$ is not.
Similarly, all automorphisms of $G$, consist of an {\sl automorphism
  group} with permutation composition as the group operation, denoted
as $Aut(G)~(\subseteq S_n) $.  Each graph $G$ has a trivial
automorphism, called {\sl identity}, denoted as $\iota$, that maps
every vertex to itself. For two distinct vertices $u$ and $v$ in $G$,
if there is an automorphism $\gamma$ mapping $u$ to $v$, i.e.,
$u^{\gamma}=v$, we say vertices $u$ and $v$ are {\sl automorphic
  equivalent}, denoted as $u \sim v$. For instance, automorphism
$\gamma_1=(4,5,6)$ indicates that vertices 4, 5 and 6 are automorphic
equivalent.

\stitle{Structural equivalent}: In a graph $G$, two distinct vertices
$u$ and $v$ are structural equivalent if they have the same neighbor
set, i.e., $N(u)=N(v)$. Obviously, if two vertices are structural
equivalent, they must be automorphic equivalent, while the converse
does not always hold.  For $G$ in Fig.~\ref{fig:expgraph},
vertices 0 and 2 are structural equivalent since they have the same
neighbor set. Similarly, vertices 1 and 3 are also
structural equivalent.  Vertices 4 and 5 are not structural
equivalent, although they are automorphic equivalent.

\stitle{Isomorphism}: Two graphs $G_1$ and $G_2$ are isomorphic iff
there exists a permutation $\gamma$ s.t., $G_1^{\gamma} = G_2$, and we
use $G_1 \cong G_2$ to denote $G_1$ and $G_2$ are isomorphic.

To check whether two graphs are isomorphic, canonical labeling (also
known as canonical representative or canonical form) is used. A
canonical labeling is a function, $C$, to relabel all vertices of a
graph $G$, such that $C(G) \cong G$, and two graphs, $G$ and $G'$, are
isomorphic iff $C(G) = C(G')$. A common technique used in the
literature to determine a canonical labeling is by coloring. Below, we
introduce coloring, colored graph, and canonical labeling by
coloring. In brief, to get a canonical labeling for a graph $G$, we
first get a colored graph $(G, \pi)$ for given color $\pi$, and we get
the canonical labeling for $(G, \pi)$ using coloring to prune
unnecessary candidates.  The canonical labeling obtained for $(G,
\pi)$ is the canonical labeling for the original graph $G$.

\stitle{Coloring}: A coloring $\pi=[V_1 | V_2 | \ldots | V_k]$ is a
disjoint partition of $V$ in which the order of subsets matters. We
use $\Pi(V)$, or simply $\Pi$, to denote the set of all colorings of
$V$.  Here, a subset $V_i$ is called a {\sl cell} of the coloring, and
all vertices in $V_i$ have the same color. In other words, $\pi$ is to
associate each $v \in V$ with the color $\pi(v)$, where $\pi(v)
\leftarrow \sum_{0<j<i} |V_j|$ if $v \in V_i$.
%
%
A cell is called a {\sl singleton } cell if it contains only one
vertex, otherwise it is called a {\sl non-singleton } cell. $\pi$ is
called an {\sl unit} coloring if there is only one partition of $V$
($k=1$) and $\pi$ is called a {\sl discrete} coloring if there are $n$
partitions for a graph with $n$ vertices ($k=n$).  $\pi$ is {\sl
  equitable} with respect to graph $G$, if for every $v_1, v_2 \in
V_i$ ($1 \leq i \leq k$),
$v_1$ and $v_2$ have
the same number of neighbors in $V_j$ ($1 \leq j \leq k$), for any $i$
and $j$. Consider $G$ in Fig.~\ref{fig:expgraph}. The coloring
$\pi_1=[0,1,2,3,4,5,6 | 7]$ has two cells, $V_1$ and $V_2$, where
$V_2=\{7\}$ is a singleton cell. For every two vertices in
$V_1=\{0,1,2,3,4,5,6\}$, they have 2 neighbors in $V_1$, and 1
neighbor in $V_2$. In a similar way, $\pi_2=[0,1,2,3|4,5,6|7]$ is also
equitable. However, $\pi_3=[0,1,2,3|4,5,6,7]$ is not equitable, since
not every two vertices in the cell of $V_2=\{4,5,6,7\}$ have the same
number of neighbors in $V_1$ even though they have the same number of
neighbors in $V_2$. For example, 7 in $V_2$ has 4 neighbors in cell
$V_1$ but the other vertices in $V_2$ have no neighbors in $V_1$.

A coloring $\pi=[V_1 | V_2 | \ldots | V_k]$ represents $\Pi_{1\leq
  i\leq k}|V_i|!$ permutations. A discrete coloring corresponds to a
single permutation $\overline \pi: x \rightarrow \pi(x)$, where every
vertex has a unique color.  For instance, the discrete coloring
$[0|3|2|1|4|6|5|7]$ corresponds to the permutation $(1,3)(5,6)$. The
concept of equitable coloring is proposed to reduce the search space
for discovering automorphism group and canonical labeling.  A partial
order is defined over colorings.  Given two colorings $\pi$ and
$\pi'$, $\pi'$ is {\sl finer than or equal to} $\pi$, denoted as $\pi'
\preceq \pi$, if each cell of $\pi'$ is a subset of a cell of $\pi$.
If $\pi' \preceq \pi$ and $\pi' \neq \pi$, then $\pi'$ is {\sl finer
  than} $\pi$, denoted as $\pi' \prec \pi$.  For instance,
$\pi_2=[0,1,2,3|4,5,6|7]$ is finer than $\pi_1=[0,1,2,3,4,5,6|7]$. A
permutation $\gamma$ can be applied to a coloring $\pi$, denoted as
$\pi^\gamma$, which results in a coloring such that
$\pi^\gamma(v)=\pi(v^\gamma)$ for any $v\in V$. Suppose $\pi_3=[0,1,2
  |3,4,5,6 | 7]$ and $\gamma_3=(1,3)(5,7)$,
$\pi_3^{\gamma_3}=[0,2,3|1,4,6,7|5]$.

An orbit coloring is a coloring where each cell contains all vertices
that are automorphic.

\stitle{Colored graph}: A colored graph is a pair $(G, \pi)$, where
$\pi$ is a coloring of $G$. Note that coloring $\pi$ be used to represent labels/attributes on vertices s.t. two vertices are in the same cell iff they share the same labels/attributes.
A graph $G$ itself is a colored graph $(G,
\pi)$ where the coloring $\pi$ is unit, or in other words, all
vertices have the same color.
Similarly, two colored graphs $(G_1,
\pi_1)$ and $(G_2, \pi_2)$ are isomorphic if there exists a
permutation $\gamma$ s.t. $(G_1, \pi_1)=(G_2, \pi_2)^\gamma$, denoted
as $(G_1, \pi_1) \cong (G_2, \pi_2)$. Note that $(G,
\pi)^\gamma=(G^\gamma, \pi^\gamma)$.


\stitle{Canonical labeling (by coloring)}:
Let ${\mathcal G}$ and $\Pi$ denote the set of graphs and colorings, a canonical representative
(or a canonical form) is a function $C: {\mathcal G} \times \Pi
\rightarrow {\mathcal G} \times \Pi $, such that for any colored graph $(G, \pi) \in
            {\mathcal G} \times \Pi$ and permutation $\gamma$ of
            $V$, the following two properties are satisfied. First,
            the canonical representative of a colored graph is
            isomorphic to the colored graph, i.e., $C(G, \pi) \cong
            (G, \pi)$. Second, the canonical representative of a
            colored graph by $\gamma$, $(G^\gamma,
            \pi^\gamma)=(G, \pi)^\gamma$, is the same as the canonical
            representative of graph $(G, \pi)$, i.e., $C(G^\gamma,
            \pi^\gamma) = C(G, \pi)$, meaning that the canonical
            representatives of two isomorphic graphs are the
            same. There are many candidates for canonical
            representative. A canonical labeling of $(G, \pi)$ is a
            permutation $\gamma^*$ satisfying $ (G^{\gamma^*},
            \pi^{\gamma^*})=C(G,\pi)$. For simplicity, we use
            canonical labeling to represent canonical
            representative. For $G$ in Fig.~\ref{fig:expgraph}, if
            $\pi$ is a unit coloring and canonical labeling
            function $C$ is defined as $C(G, \pi)= (G^{\gamma^*},
            \pi^{\gamma^*})$ where $\gamma^* =arg min_{\gamma \in S_n}
            E^\gamma $, then $\gamma^*=(0,7)(1,5)(2,4)(3,6)$ is a
            candidate.

\comment{
\stitle{Canonical labeling}: A {\sl canonical representative}, or a {\sl canonical form}, is a function
\begin{displaymath}
C:  {\mathcal G} \times \Pi \rightarrow   {\mathcal G} \times \Pi
\end{displaymath}
such that for any graph $G \in {\mathcal G} $, coloring $\pi \in \Pi$ and permutation $\gamma$ of $V$,
\begin{itemize}
\item   $C(G, \pi) \cong (G, \pi)$,
\item   $C(G^{\gamma}, \pi^{\gamma})= C(G, \pi)$.
\end{itemize}
Any permutation $\gamma^*$ satisfying $(G^{\gamma^*}, \pi^{\gamma^*})=C(G, \pi)$ is called a canonical labeling of $(G,\pi)$. For simplicity, we reuse canonical labeling to represent canonical representative.

Consider the example graph in Fig.~\ref{fig:expgraph}, if $\pi$ is unit and the canonical labeling function $C$ is defined as $C(G, \pi)= (G^{\gamma^*}, \pi^{\gamma^*})$ where $\gamma^* =arg max_{\gamma \in S_n} E^\gamma $, then  $\gamma^*=(0,7)(1,5)(2,4)(3,6)$ is one candidate canonical labeling.
}

We summarize the discussions on coloring, permutation, automorphism,
and canonical labeling. A coloring represents a set of permutations
and a discrete coloring corresponds to a single permutation. A
permutation $\gamma$ is a relabeling of the vertices such that
$\gamma$ on a colored graph $(G,\pi)$ results in a relabeled graph
$(G,\pi)^\gamma$ that is isomorphic to $(G, \pi)$. All permutations in
$S_n$ are classified into several subgroups s.t. all permutations in
each subgroup generate the same relabeled colored graph.
A signature of the corresponding subgroup is the relabeled colored
graph, $(G, \pi)^\gamma$. The subgroup with the signature $(G,
\pi)^\gamma= (G,\pi)$ forms the automorphism group, in which each
permutation is an automorphism. By defining a total order among such
signatures, a permutation  with the minimum signature
is the canonical labeling.


\section{Related Works}
\label{sec:relatedworks}


Graph isomorphism is an equivalence relation on graphs by which all
graphs are grouped into equivalence classes. By graph isomorphism, it
allows us to distinguish graph properties inherent to the structures
of graphs   from properties associated with graph
representations: graph drawings, graph labeling, data structures, etc.

%

From the theoretical viewpoint, the graph isomorphism problem is one
of few standard problems in computational complexity theory belonging
to NP, but unknown if it belongs to either of P or NP-complete. It is
one of only two, out of 12 total, problems listed in
\cite{garey2002computers} whose complexity remains unresolved.
%
%
NP-completeness is considered unlikely since it would imply collapse
of the polynomial-time hierarchy \cite{goldreich1991proofs}. The best
currently accepted theoretical algorithm is due to
\cite{babai1983computational, babai1983canonical}, whose time
complexity is $e^{O(\sqrt{nlogn})}$.
%
%
Although the graph isomorphism problem is not generally known to be in
P or NP-complete, they can be solved in polynomial time for special
classes of graphs, for instance, graphs of bounded degree
\cite{luks1982isomorphism}, bounded genus
\cite{filotti1980polynomial,miller1980isomorphism}, bounded
tree-width \cite{bodlaender1990polynomial}, and with high probability
for random graphs \cite{babai1979canonical}. However, most of these
algorithms are unlikely to be useful in practice.

In practice,
\comment{
 to test two graphs if they are isomorphism, most
approaches relabel vertices in each graph to obtain a canonical
labeling. Among all such approaches, the most successful ones adopt an
``individualization-refinement'' schema, i.e., fixing vertices
together with refinement of partitions of the vertex set.
}
the first
practical algorithm to canonically labeling graphs with hundreds of
vertices and graphs with large automorphism groups was \nauty
\cite{mckay1978computing,mckay1981practical}, developed by
McKay. Observing that the set of symmetries of a graph forms a group
under functional composition, \nauty integrates group-theoretical
techniques and utilizes automorphisms discovered to prune the search
tree. Motivated by
\nauty, a number of algorithms, such as \bliss
\cite{junttila2007engineering, junttila2011conflict} and \traces
\cite{piperno2008search} are proposed to address possible shortcomings
of \nauty's search tree, which we will discuss in Section~\ref{sec:previous}.
Another algorithm worth noting is \saucy
\cite{darga2004exploiting}. The data structures and algorithms in
\saucy take advantage of both the sparsity of input graphs and the
sparsity of their symmetries to attain scalability.  Different from
\nauty-based canonical labeling algorithms, \saucy only finds graph
symmetries, precisely, a generating set of the automorphism
group.
All algorithms mentioned above are difficult to deal with real-world
massive graphs, and the search tree used are for pruning purposes
not for answering SSM queries.

\comment{
\stitle{Relationships between coloring, permutation,
    automorphism and canonical labeling}: A coloring represents a set
  of permutations and a discrete coloring corresponds to a single
  permutation; A permutation $\gamma$ is a relabeling of the
  vertices. $\gamma$ on a colored graph $(G, \pi)$ results in a
  relabeled colored graph $(G^\gamma, \pi^\gamma)$ that is isomorphic
  to $(G, \pi)$.  All permutations in $S_n$ are classified into
  several subgroups s.t. all permutations in each subgroup generate
  the same relabeled colored graph. Take the relabeled colored graph
  as a signature of the corresponding subgroup. The subgroup with
  signature $(G,\pi)$ is the automorphism group, in which each
  permutation is an automorphism.  Define a total order among these
  signatures, then the maximum signature and any permutation in the
  corresponding subgroup is chosen as the canonical labeling.

\stitle{Canonical Labeling for Massive Graphs}: In this work, we study
the problem of Canonical Labeling for Massive Graphs, i.e., give a
colored graph ($G, \pi$), where $\pi$ is predefined or is unit if only
$G$ is given, find a canonical labeling $C(G,\pi)$.  }

\comment{
\stitle{Graph Isomorphism Problem and Canonical Labeling Problem}: In this paper, we study the Graph Isomorphism Problem, i.e., given two  graphs $G_1$ and $G_2$, determine whether these two  graphs are isomorphic.
To achieve this goal, we study the Canonical Labeling Problem, i.e.,
given a graph $G$, or equivalently, a colored graph $(G,\pi)$ where $\pi$ is unit, find a canonical representative $C(G,\pi)$,
such that every graph that is isomorphic to $(G,\pi)$ has the same canonical representative. Therefore, the graph isomorphism problem is equivalent to whether $C(G_1, \pi)$ equals $C(G_2, \pi)$.
}

\comment{
\stitle{Graph Canonization Problem}: In this paper, we study Graph Canonization Problem, i.e., given a graph $G$, find a canonical form (a canonical labeling) $C(G)$ that is isomorphic to $G$, such that every graph that is isomorphic to $G$ has the same canonical form as $G$.
}

\section{The Previous Algorithms}
\label{sec:previous}

In this section, we outline the main ideas of the three
state-of-the-art algorithms, namely, \nauty, \bliss and \traces, that
enumerate all permutations in the symmetry group $S_n$, add all
permutations $\gamma$ satisfying $(G^\gamma, \pi^\gamma)=(G, \pi)$
into the automorphism group $Aut(G, \pi)$ and choose the colored graph
$(G^\gamma, \pi^\gamma)$ with the minimum value under some specific
function as the canonical labeling.  Such enumeration of permutations
in $S_n$ is done by a search tree.
In the search tree, each node corresponds to a coloring, and each edge
is established by individualizing a vertex in a non-singleton cell in
the coloring of the parent node.  Here,  individualizing a vertex
means to assign this vertex a unique color. For instance,
individualizing vertex $4$ in $\pi=[0,1,2,3|4,5,6|7]$ results in
$\pi'=[0,1,2,3|4|5,6|7]$. The coloring of the child node is definitely
finer than the coloring of the parent node, and each leaf node
corresponds to a discrete coloring, which is equivalent to a
permutation in $S_n$.  By the search tree, each permutation is
enumerated once and only once, which implies that the whole search
tree contains as many as $n!$ leaf nodes.

\comment{
Following, we describe main
components of \nauty, \traces and \bliss, which act as pruning
techniques to remove fruitless subtrees of the search tree.

Afterwards, we describe the main components of the algorithms by means
of an associated backtrack search tree, which is at the heart of
essentially all algorithms relying on the individualization-refinement
paradigm.
}

\comment{
\begin{figure*}[t]
\begin{center}
  \includegraphics[scale=0.44]{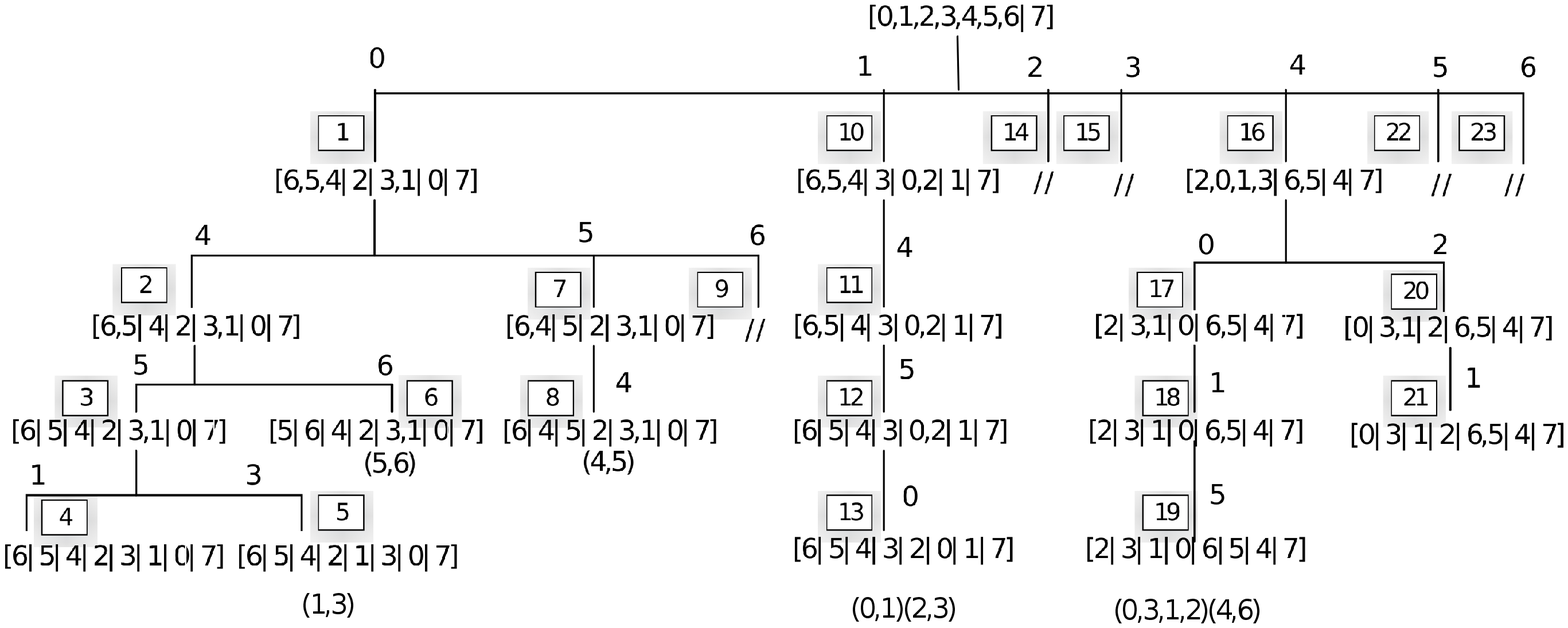}
\end{center}
\vspace{-0.4cm}
\caption{Backtrack search tree ${\mathcal T}(G, \pi)$ constructed by \bliss for the graph in Fig.~\ref{fig:expgraph}}
\label{fig:st_bliss}
\vspace{-0.4cm}
\end{figure*}
}


\comment{
We first review some concepts related to sequence, which is important
in discussing the backtrack search tree. Let $\Sigma$ be an ordered
set. A sequence $\nu=(a_1, a_2, \ldots, a_k)$ is a tuple consisting of
elements in $\Sigma$, and the set of finite sequences is denoted as
$\Sigma^*$. For $\nu \in \Sigma^*$, $|\nu|$ denotes the length, i.e.,
the number of components of $\nu$. For a sequence $\nu=(a_1, \ldots,
a_k) \in \Sigma^*$ and an element $b \in \Sigma$, $\nu || b$ denotes
$(a_1, \ldots, a_k, b)$ and $[\nu]_s=(a_1, \ldots, a_s)$ for $1 \leq s
\leq k$. The ordering $\leq$ on sequences is the lexicographic order.
}

\comment{
  As an overview, the backtrack search tree constructed by
  previous algorithms, denoted as ${\mathcal T}(G, \pi)$, is a rooted
  tree with labels on both nodes and edges. Fig.~\ref{fig:st_bliss}
  shows the search tree constructed by \bliss on the example graph in
  Fig.~\ref{fig:expgraph}, and the nodes are labeled in the order
  they are traversed.  In ${\mathcal T}(G,\pi)$, each node is
  associated with a sequence of vertices that are individualized from
  the root to this node and each edge is labeled with a vertex that
  are individualized from the parent node to the child node. For
  instance, node 3 is associated with sequence 045, consisting of the
  edge labels from the root to node 3 and tree edge (2,3) is labeled
  with 5. For clarity, we do not explicitly show sequences associated
  nodes in Fig.~\ref{fig:st_bliss}. Therefore, the root node
  corresponds to the empty sequence and the sequences becomes longer
  as we move down the tree.  For each node in ${\mathcal T}(G, \pi)$,
}

We give the details on the search tree. The search tree, denoted as
%
%
${\mathcal T}(G, \pi)$,   is a rooted label tree with labels on
both nodes and edges.
Here, a node-label is a coloring by individualizing from the root to
the node concerned, and an edge-label is a vertex in $G$ that is
individualized from the node-label of the parent node to the
node-label of the child node in ${\mathcal T}(G, \pi)$.
%
%
%
Fig.~\ref{fig:st_bliss} shows the search tree ${\mathcal T}(G, \pi)$
constructed by \bliss for the graph $G$ (Fig.~\ref{fig:expgraph}),
in which a node in ${\mathcal T}(G, \pi)$ is shown as $\fbox{x}$ where
$x$ is a node identifier. The node identifiers indicate the order they
are traversed.  In Fig.~\ref{fig:st_bliss}, the root node is labeled
by an equitable coloring $[0,1,2,3,4,5,6|7]$, which has 7 child nodes
by individualizing one of the vertices in the non-singleton cell.
The node $\fbox{1}$ is a child node of the root by individualizing
vertex $0$ in $G$. Here, the individualization of 0 is represented as
the edge-label of the edge from the root to the node $\fbox{1}$. The
node-label of $\fbox{1}$ represents a finer equitable coloring
$[6,5,4|2|1,3|0|7]$ comparing the coloring of $[0,1,2,3,4,5,6|7]$ at
the root.
In ${\mathcal T}(G, \pi)$, the edge-label sequence (or simply
sequence) from the root to a node shows the order of
individualization. In Fig.~\ref{fig:st_bliss}, node $\fbox{3}$ is
associated with a sequence 045 and has a node-label coloring
$[6|5|4|2|3,1|0|7]$.
In the following, we also use $ (G, \pi, \nu)$ to identify
a node in the search tree by the sequence $\nu$ from the root to the
node.
The node $\fbox{4}$ is the leftmost leaf node in the search tree with
a discrete coloring $\pi_0=[6|5|4|2|3|1|0|7]$ whose corresponding
permutation is $\gamma_0=(0,6)(1,5)(2,3,4)$.
In ${\mathcal T}(G, \pi)$, the leftmost leaf node (corresponds to a colored
graph $(G^{\gamma_0}, \pi^{\gamma_0})$ with some specific permutation
$\gamma_0$) is taken as a reference node. Any automorphism,
$\gamma'\gamma_0^{-1}$, will be discovered when traversing a leaf node with
permutation $\gamma'$ having $(G^{\gamma_0}, \pi^{\gamma_0})=
(G^{\gamma'}, \pi^{\gamma'})$. Here, $\gamma_0^{-1}$ denotes the inverse element of $\gamma_0$.
Reconsider Fig.~\ref{fig:st_bliss}, by taking the leftmost leaf node
$\fbox{4}$ as a reference node, an automorphism $(1,3)$ is discovered
when traversing the node $\fbox{5}$.

The three state-of-the-art algorithms, \nauty, \bliss and \traces
exploit three main techniques, namely, {\sl refinement function} $R$,
{\sl target cell selector} $T$ and {\sl node invariant} $\phi$ to
construct the search tree ${\mathcal T}(G, \pi)$ and prune fruitless
subtrees in ${\mathcal T}(G, \pi)$.
In brief, the refinement function $R$ aims at pruning subtrees whose
leaf nodes cannot result in any automorphisms with the reference node,
the target cell selector $T$ selects a non-singleton cell from a
coloring at a node for its children in the search tree, and
the node invariant $\phi$ is designed to prune subtrees where no new
automorphisms can be found or the canonical labeling cannot
exist.

\stitle{The refinement function $R$}: For every tree node with a
sequence $\nu$ (the edge-label sequence from the root to the node),
the refinement function, $R: {\mathcal G} \times \Pi \times V^*
\rightarrow \Pi$, specifies an equitable coloring corresponding to
$\nu$ and $\pi$. In specific, the refinement is done by giving the vertices in
the sequence unique colors and then inferring a coloring of the other
vertices s.t., the resulting coloring is equitable.  Mathematically, a
refinement function is a function, $R: {\mathcal G} \times \Pi \times
V^* \rightarrow \Pi$, such that for any $G \in \mathcal G$, $\pi \in
\Pi$ and $\nu \in V^*$, we have the following.  (i) $R(G, \pi, \nu)
\preceq \pi$.  (ii) If $v \in \nu$, then $\{v\}$ is a cell of $R(G,
\pi, \nu)$. (iii) For any $\gamma \in S_n$, $R(G^\gamma, \pi^\gamma,
\nu^\gamma) = R(G, \pi, \nu)^\gamma$.

\comment{
**We show **
Let ${\mathcal T}(G, \pi, \nu)$ denote the subtree rooted at the node with sequence $\nu$.
Consider two leaf nodes rooted at ${\mathcal T}(G, \pi, \nu)$, one with discrete coloring $\pi_1$ and permutation $\gamma_1$, and the other with $\pi_2$ and $\gamma_2$. If these two nodes derives an automorphism, i.e., $(G^{\gamma_1}, \pi^{\gamma_1}) = (G^{\gamma_2}, \pi^{\gamma_2})$, then $\gamma_2 \gamma_1^{-1} \in Aut(G, \pi)$. Let $u$ and $v$ be two vertices share the same color $c$ in $\pi_1$ and $\pi_2$, respectively. Recall that a discrete coloring $\pi$ derives a permutation $\gamma=\overline \pi: x \rightarrow \pi(x)$. Then $\gamma_1(u)=c$, $\gamma_1^{-1}(c)=u$ and $\gamma_2(v)=c$,  implying that $\gamma_2 \gamma_1^{-1}(u)=v$
********
}

Revisit the search tree ${\mathcal T}(G, \pi)$ in
Fig.~\ref{fig:st_bliss}. Refinement function $R$ refines the empty
sequence and the unit coloring of the root node by differentiating
vertex 7 from the others in $G$ (Fig.~\ref{fig:expgraph}). The node
$\fbox{1}$ can be identified by a sequence 0 from the root. $R(G, \pi,
0)$ individualizes vertex $0$ from the coloring associated with
root node, i.e., $[0,1,2,3,4,5,6|7]$, resulting in
$[1,2,3,4,5,6|$ $0|7]$, which is further refined to an equitable coloring
$[6,5,4|2|1,3|$ $0|7]$.

\stitle{Target cell selector $T$}: For a tree node $(G, \pi, \nu)$
that is identified by a sequence $\nu$, the target cell selector $T:
{\mathcal G} \times \Pi \times V^* \rightarrow 2^V$ selects a
non-singleton cell from the coloring by $R(G, \pi, \nu)$ to specify
its children, where each child node is generated by individualizing a
vertex in the non-singleton cell selected, if the coloring $R(G, \pi,
\nu)$ is not discrete. Mathematically, a target cell selector is a
function, $T: {\mathcal G} \times \Pi \times V^* \rightarrow 2^V$,
such that for any $G \in {\mathcal G}$, $\pi \in \Pi$ and $ \nu \in
V^*$, the following three holds.
(i) If $R(G, \pi, \nu)$ is discrete, then $T(G, \pi, \nu) =\emptyset$.
(ii) If $R(G, \pi, \nu)$ is not discrete, then $T(G, \pi, \nu)$ is a
non-singleton cell of $R(G, \pi, \nu)$.
(iii) For any $\gamma \in S_n$, $T(G^\gamma, \pi^\gamma, \nu^\gamma) =
T(G, \pi, \nu)^\gamma$.

The choice of a target cell has a significantly effect on the shape of
the search tree. Some
\cite{mckay1981practical} uses the first
smallest non-singleton cell, while some others
\cite{kocay1996writing} use the first non-singleton cell
regardless of the size.
%
%
In Fig.~\ref{fig:st_bliss}, we follow the suggestion of
\cite{kocay1996writing}. For instance, target cell selector $T$ on the
node $\fbox{1}$ chooses the first non-singleton cell $\{6,5,4\}$, and
generates three child nodes ($\fbox{2}$, $\fbox{7}$, and $\fbox{9}$) by
individualizing vertices 4,5, and 6, respectively.

\stitle{Node invariant $\phi$}: It assigns each node in the search
tree with an element from a totally ordered set, and $\phi$ is
designed with the following properties: (a) $\phi$ is
isomorphic-invariant on tree nodes, i.e., $\phi(G^\gamma, \pi^\gamma,
\nu^\gamma)=\phi(G, \pi, \nu)$ for any $\gamma \in S_n$; (b) $\phi$
acts as a {\sl certificate} on leaf nodes, i.e., two leaf nodes share
the same value under $\phi$ iff they are isomorphic; (c) $\phi$
retains the partial ordering between two subtrees rooted at the same
level.  Mathematically, let $\Omega$ be some totally ordered set. A
node invariant is a function,
$\phi: {\mathcal G} \times \Pi \times V^* \rightarrow \Omega$,
such that for any $\pi \in \Pi$, $G \in {\mathcal G}$, and distinct
$\nu, \nu' \in {\mathcal T}(G, \pi_0)$, we have the following.
(i) If $|\nu|=|\nu'|$, and $\phi(G, \pi, \nu) < \phi(G, \pi, \nu')$,
then for every leaf $\nu_1 \in {\mathcal T}(G, \pi, \nu)$ and leaf
$\nu_1' \in {\mathcal T}(G, \pi, \nu')$, we have $\phi(G, \pi, \nu_1)
< \phi(G, \pi, \nu_1')$;
(ii) If $\pi=R(G, \pi, \nu)$ and $\pi'=R(G, \pi, \nu')$ are discrete,
then $\phi(G, \pi, \nu) = \phi(G, \pi, \nu') \Leftrightarrow G^\pi =
G^{\pi'}$;
(iii) For any $\gamma \in S_n$, we have $\phi(G^\gamma, \pi^\gamma,
\nu^\gamma)=\phi(G, \pi, \nu)$.
By the node invariant $\phi$, three types of pruning operations are
possible.
(1) $P_A(\nu, \nu')$ removes subtree ${\mathcal T}(G, \pi, \nu')$ that
contains no automorphisms with the reference node, when $\phi(G, \pi,
\nu')$ on some node $\nu'$ does not equal to $\phi(G, \pi,
\nu)$. Here, $\nu$ is the node on the leftmost path having
$|\nu|=|\nu'|$.
(2) $P_B(\nu, \nu')$ removes subtree  ${\mathcal T}(G, \pi, \nu')$
that does not contain the canonical labeling, when $\phi(G, \pi, \nu')
<\phi(G, \pi, \nu)$. Here $\nu$ is the node on the path whose leaf
node is chosen as the current canonical labeling, and $|\nu|=|\nu'|$.
(3) $P_C(\nu, \nu')$ removes subtree ${\mathcal T}(G, \pi, \nu')$ that
contains no new automorphisms, when $\nu'=\nu^\gamma$ where $\gamma$
is an automorphism discovered or can be composed by discovered
automorphisms.

\comment{

\section{The Previous Algorithms}
\label{sec:previous}
In this section, we review the main ideas of \nauty-based approaches and \saucy. For \nauty-based approaches, we choose \bliss as a reference, and the discussion can be extended to other \nauty-based approaches.
Due to the limited space and the complexity of the algorithms, instead of discussing each component of these algorithms in detail, we show an example graph (Fig.~\ref{fig:expgraph}) as well as its search trees by \bliss (Fig.~\ref{fig:st_bliss}) and  \saucy (Fig.~\ref{fig:st_saucy}), and use these two search trees to better explain the main ideas and main components in each algorithm.
Note that the nodes of the search trees in Fig.~\ref{fig:st_bliss} and Fig.~\ref{fig:st_saucy} are labeled in the order they are constructed.

We first introduce some concepts of vital importance in explaining the algorithms.

\stitle{group}: Both of the set of permutations of $V$, denoted as $Per(V)$, and the set of automorphisms of $G$, denoted as $Aut(G)$, form groups under functional composition. A {\sl generating set}, or a {\sl set of generators}, of $Aut(G)$ is a subset of $Aut(G)$ whose combination under functional composition generate $Aut(G)$.

\stitle{subgroup}: A subgroup of $Aut(G)$ is a subset of $Aut(G)$ that forms a group under functional composition. For instance, the {\sl stabilizer subgroup} of $v\in V$, denoted as $Aut_v(G)$, is a subgroup of $Aut(G)$ that fixes vertex $v$, i.e., $Aut_v(G)= \{\gamma \in Aut(G) | v^\gamma =v\}$.

\stitle{coset}: Elements of $Aut(G)$, when composed with elements of $Aut_v(G)$, partition $Aut(G)$ into equally-sized cosets, i.e., non-empty pair-wise disjoint subsets whose union is $Aut(G)$. Choosing one element from each coset yields a set of {\sl coset representatives}. Each coset representative composed with $Aut_v(G)$ can generate the entire coset.

\stitle{orbit partition}: Two vertices $u$ and $v$ share the same {\sl orbit} if $u \sim v$, i.e., $u$ and $v$ are automorphic. The $\sim$ operation, i.e., automorphism relation, partitions $V$ into a orbit partition.

\stitle{Ordered partition pair (OPP)}: An ordered partition pair $\Pi$ is specified as
\begin{equation}
\Pi=
\begin{bmatrix}
\pi_T \\
\pi_B
\end{bmatrix}
=
\begin{bmatrix}
T_1 &|T_2 &|\ldots &|T_m \\
B_1 &|B_2 &|\ldots &|B_k
\end{bmatrix}
\end{equation}
with $\pi_T$ and $\pi_B$ referred to, respectively, as the top and bottom ordered partitions of $\Pi$. $\Pi$ is {\sl isomorphic} if $m=k$ and $|T_i|=|B_i|$ for $i \in [1, m]$, otherwise $\Pi$ is {\sl non-isomorphic}. An isomorphic OPP is {\sl matching } if its corresponding non-singleton cells are identical, and an OPP is discrete (resp. unit) if its top and bottom partitions are discrete (resp. unit).

\subsection{Two trivial algorithms}

\comment{
We first introduce a trivial algorithm for the graph isomorphism problem, and then discuss how to modify this algorithm to generate the automorphism group
as well as discover a canonical labeling
for a given graph $G$. Let $Sym(V)$ denote the group of all permutations of $V$ and $Aut(G)$ denote the group of all automorphisms of a graph $G$. Given two graphs $G_1$ and $G_2$, according to the definition of isomorphism,
a trivial algorithm is developed by enumerating permutations $\lambda \in Sym(V)$, either find a permutation $\lambda$ such that $G_1^\lambda = G_2$ or verify that $G_1^\lambda \neq G_2$ for all $\lambda \in Sym(V)$.
To generate the automorphism group for a given graph $G$, we set $G_1 = G_2 =G$ in the trivial graph isomorphism algorithm and $Aut(G)$ contains all permutations $\lambda \in Sym(V)$ such that $G^\lambda = G$. On the other hand, in order to discover a canonical labeling for graph $G$,

First, for \nauty and \nauty-based approaches, each node in the search tree is an ordered equitable partition, representing a partial labeling.
}

To better understand previous algorithms, We first introduce two trivial algorithms, one generates the automorphism group and the other discovers a canonical representative for a given graph $G$. Recall that $Per(V)$ denotes the group of all permutations of $V$ and $Aut(G)$ denotes the group of all automorphisms of a graph $G$.
For the automorphism group generation algorithm, according to the definition of automorphism, it enumerates all permutations $\gamma \in Per(V)$ and adds $\gamma$ into $Aut(G)$ if $G^\gamma = G$.
For the canonical representative discovery algorithm, we first define a certificate $c(G,\gamma)$ on a specific permutation $\gamma$ for graph $G$. One popular certificate is the  lexicographically sorted edge list of $G^\gamma$. The canonical representative discovery algorithm enumerates all permutations $\gamma \in Per(V)$ and returns the certificate with the maximum value. We prove its correctness. Assume $G_1 \cong G_2$ and $G_1^\lambda =G_2$, then
$\{ c(G_2, \gamma) | \gamma \in Per(V) \} =\{ c(G_1^\lambda, \gamma)  | \gamma \in Per(V) \} = \{ c(G_1, \lambda\gamma) | \gamma \in Per(V) \}=\{ c(G_1, \gamma)  | \gamma \in Per(V)\} $, implying that isomorphic graphs have the same certificate value set.
Therefore, this canonical representative discovery algorithm returns the same result for isomorphic graphs.
Easy to see, these two trivial algorithms are too expensive for large graphs since the permutation group $Per(V)$ contains $|V|!$ elements, increasing exponentially with the graph size. To deal with the huge search space, previous algorithms exploit group-theoretical pruning mechanisms to avoid enumerating all permutations in $Per(V)$.

\subsection{\nauty-based approaches}

We first review \bliss, one \nauty-based approach. \bliss discovers a canonical representative for a given graph $G$, meanwhile generates the automorphism group $Aut(G)$ as a byproduct. Different from the trivial canonical representative discovery algorithm, which enumerates permutations in $Per(V)$ directly, \bliss constructs a search tree whose leaf nodes correspond to permutations in $Per(V)$,
and uses the automorphisms discovered to prune fruitless subtrees.

Fig.~\ref{fig:st_bliss} demonstrates the search tree, or the canonical labeling tree,  constructed by \bliss for the graph in Fig.~\ref{fig:expgraph}. The nodes of the search tree in Fig.~\ref{fig:st_bliss} are ordered partitions, each representing a (partial) labeling. Here, a labeling is achieved by renaming each vertex with  its position in the ordered partition. Specifically, each leaf node of the search tree is a discrete partition, and the corresponding labeling is a permutation in $Per(V)$. For instance, at node 4, vertices 0,1,2,3,4,5,6,7 are at indices 6,5,3,4,2,1,0,7, respectively, and hence, node 4 represents the labeling obtained by the permutation (6 0)(5 1)(3 4 2). Similar to the trivial canonical representative discovery algorithm, each node in the search tree is also associate with a certificate, which can be used to discover automorphisms and prune search space. For instance, node 4 and 5 have the same certificate, i.e., $\{(0,1), (0,2), (0,7), (1,2), (1,7), (2,7), (3,4), \\(3,5), (3,7), (4,6), (4,7), (5,6), (5,7), (6,7) \}$, discovering an automorphism (1,3).

The search of \bliss starts by initializing a unit ordered partition as the root of the search tree, and refining it using partition refinement.
In Fig.~\ref{fig:expgraph}, vertex 7 has larger degree than the other vertices, thus partition refinement separates vertex 7 as a single cell in the root of the search tree in Fig.~\ref{fig:st_bliss}.
At each non-leaf node, which is a non-discrete partition, \bliss selects a non-singleton cell and individualizes all vertices in this cell. For instance, vertices in the first non-singleton cell of node 1, i.e., vertices 4,5 and 6, are individualized, resulting in tree nodes 2, 7 and 9. Each individualization is followed by partition refinement to discard impossible permutations and invalid labeling. This individualization-refinement continues until the partition becomes discrete, which is a leaf node. The first leaf node (node 4) is used as a reference node to discover automorphisms and initializes the canonical representative. An automorphism is discovered if another leaf node has the same certificate as the reference node and the canonical representative is updated whenever finding a better certificate. For instance, node 5 has the same certificate as the reference node (node 4), generating an automorphism $(1,3)$, and  node 19 updates the canonical representative. \bliss returns the best certificate as the canonical representative of the give graph $G$.

\bliss prunes search space using two group-theoretical pruning mechanisms, namely {\sl coset pruning} and {\sl orbit pruning}. Coset pruning results from the concept of coset representatives: one generator per coset is sufficient to generate all symmetries in the coset.
To enable coset pruning, the leftmost path of \bliss search tree corresponds to a sequence of subgroup stabilizers, and whenever find an automorphism, the corresponding subtree can be pruned.
For example, node 6, 8 and 13 are coset representatives of their corresponding subtrees rooted at node 6, 7 and 10, respectively. Therefore, these subtrees can be pruned by coset pruning. Orbit pruning relies on automorphisms discovered to eliminate redundant generators. Here, orbit partition is exploited to efficiently store and represent automorphisms. For example, node 9 is pruned by orbit pruning because vertices 4 and 6 share the same orbit.


\subsection{\saucy}
We discuss \saucy. Different from \nauty-based approaches, \saucy only finds automorphisms, or more precisely, a generating set of the automorphism group. Note that \saucy can be exploited as a preprocessing step for \nauty-based approaches.

Fig.~\ref{fig:st_saucy} illustrates the search tree, or the permutation tree, constructed by \saucy for the graph in Fig.~\ref{fig:expgraph}. Different from the search tree of \bliss (Fig.~\ref{fig:st_bliss}), the nodes of the search tree in Fig.~\ref{fig:st_saucy} are ordered partition pairs (OPPs), each encoding a set of permutations. In general, a non-isomorphic OPP contains no permutations and an isomorphic OPP
\begin{equation}
\Pi=
\begin{bmatrix}
T_1 &|T_2 &|\ldots &|T_m \\
B_1 &|B_2 &|\ldots &|B_m
\end{bmatrix}
\end{equation}
contains $\Pi_{1 \leq i \leq m}|T_i|!$ permutations. Specifically, a discrete OPP contains only one permutation and an unit OPP contains $n!$ permutations. For example, the root of the search tree in Fig.~\ref{fig:st_saucy} contains 5,040 permutations and the OPP at node 2 contains 4 permutations $\{ \iota, (4,6), (0,2), (0,2)(4,6) \}$, recall $\iota$ represents the identity permutation.
While \bliss finds automorphisms between two partitions with the same certificate, \saucy discovers each automorphism at a single  discrete OPP or a single matching OPP. For instance, node 5 and  node 6 represent  automorphisms $(0,2)$ and  $(4,6)$, respectively.

Similarly, the search of \saucy starts by initializing a unit OPP as the root of the search tree, and refining it using partition refinement. At each tree node, which is not a discrete OPP, \saucy chooses a vertex from a non-singleton cell of the top partition and maps it to all the vertices in the corresponding cell of the bottom partition. For instance, consider node 1,  vertex 5 in the first non-singleton cell in the top partition is mapped to vertices 4,5,6 in the bottom partition, generating nodes 2,7,9. This mapping procedure continues until the OPP becomes discrete (node 5), matching (node 6) or non-isomorphic (node 10). The search terminates when all possible mappings are explored.

Similar to \bliss, \saucy exploits coset pruning and orbit pruning to prune search space.  For instance, the automorphisms found at node 6, 8, 14 are coset representatives of their corresponding subtrees rooted at node 6, 7, 13, respectively. And the subtree rooted at node 9 is pruned by orbit pruning since vertices 4 and 6 share the same orbit. On the other hand, the OPP structure provides \saucy with additional pruning mechanisms, including {\sl non-isomorphic OPP pruning} and {\sl matching OPP pruning}. Non-isomorphic OPP pruning indicates that a non-isomorphic OPP contains no permutations. For instance, the OPP at node 10 is non-isomorphic, implying that mapping of 3 to 4 is conflict, the subtree rooted at node 10 contains none automorphisms.  Whereas, matching OPP pruning represents an early automorphism constructed by mapping the vertices of non-singleton cells identically. Combined with coset pruning, this automorphism can be returned as the coset representative of the subtree. For example, the OPPs at nodes 6, 8 and 14 are matching, and are returned as the coset representatives of the subtree rooted at nodes 6,7 and 13, respectively.

\comment{
\begin{figure*}[t]
\begin{center}
  \includegraphics[width=1.8\columnwidth,height=6.4cm]{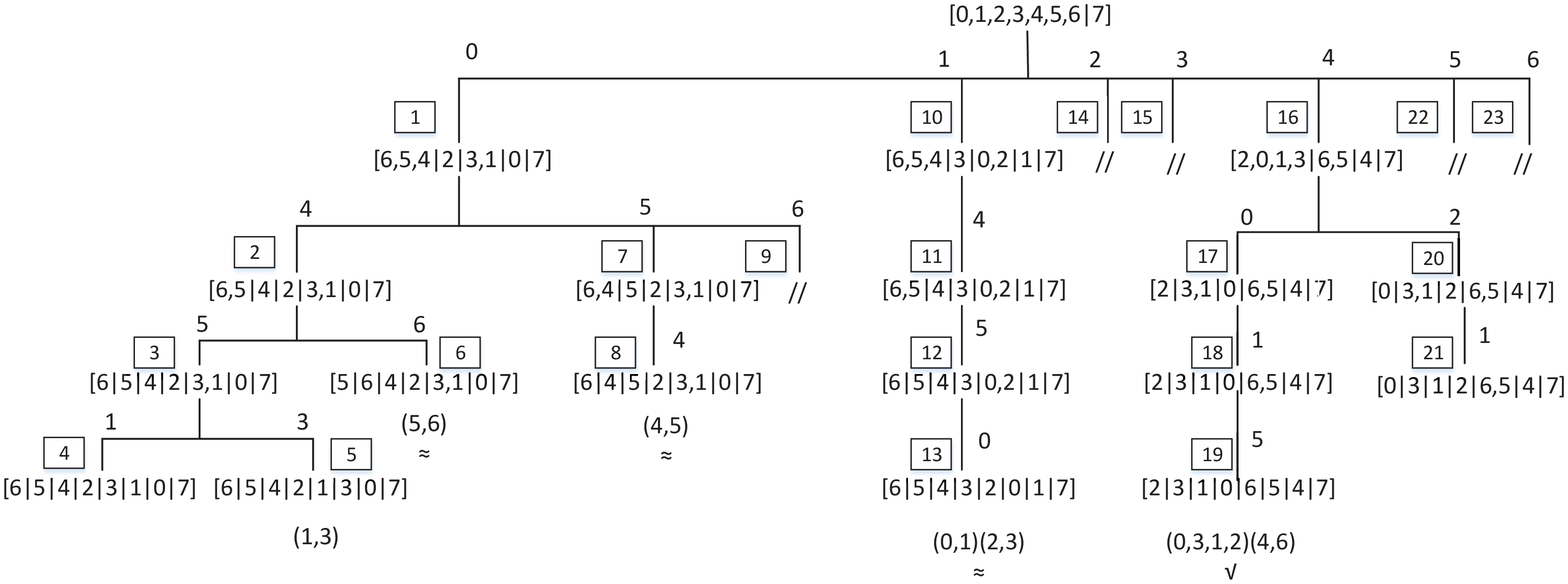}
\end{center}
\vspace{-0.6cm}
\caption{Canonical labeling tree constructed by \bliss for the graph in Fig.~\ref{fig:expgraph} }
\label{fig:st_bliss}
\vspace{-0.4cm}
\end{figure*}
}

\begin{figure*}[t]
\begin{center}
  \includegraphics[scale=0.48]{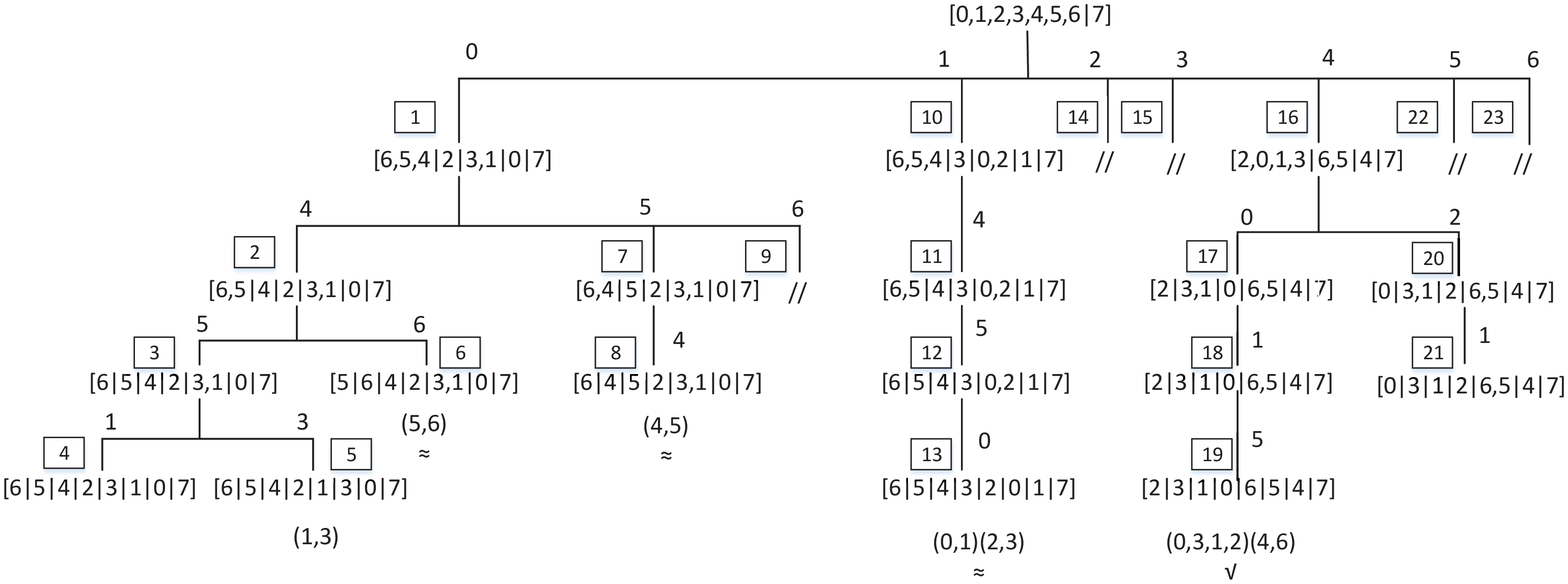}
\end{center}
\vspace{-0.6cm}
\caption{Canonical labeling tree constructed by \bliss for the graph in Fig.~\ref{fig:expgraph}. In the search tree, $\approx$ means coset pruning, $//$ means orbit pruning and $\surd$ means better certificate }
\label{fig:st_bliss}
\vspace{-0.4cm}
\end{figure*}

\comment{
\begin{figure*}[t]
\begin{center}
  \includegraphics[width=1.8\columnwidth,height=6.4cm]{figure/st_nauty.png}
\end{center}
\vspace{-0.6cm}
\caption{Canonical labeling tree constructed by \bliss for the graph in Fig.~\ref{fig:expgraph} }
\label{fig:st_bliss}
\vspace{-0.4cm}
\end{figure*}
}

\begin{figure*}[t]
\begin{center}
  \includegraphics[width=1.8\columnwidth,height=6.8cm]{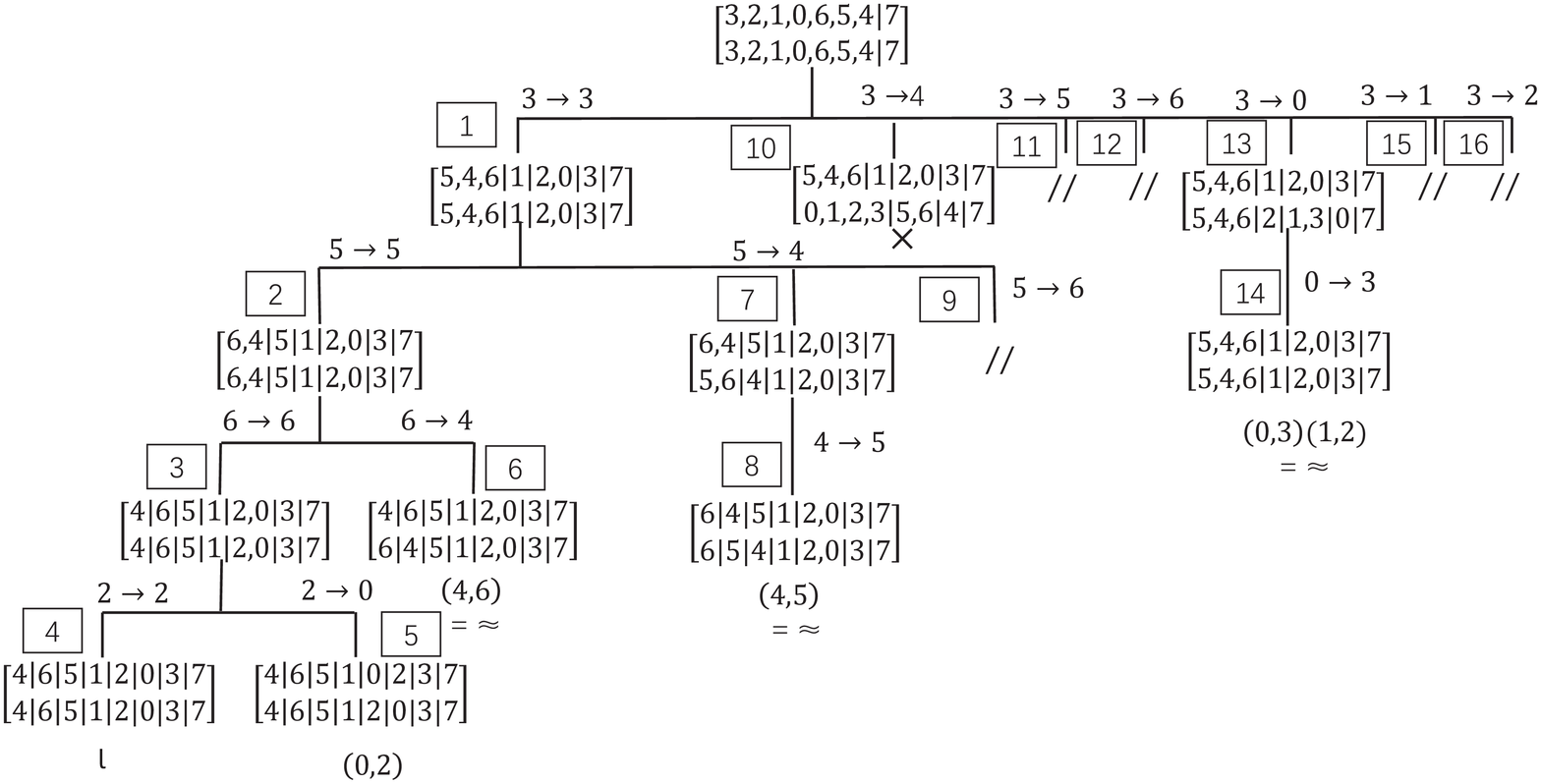}
\end{center}
\vspace{-0.6cm}
\caption{Permutation tree constructed by \saucy for the Graph in Fig.~\ref{fig:expgraph}. In the search tree, $\approx$ means coset pruning, $//$ means orbit pruning, $=$ means matching OPP pruning and $\times$ means non-isomorphic OPP pruning }
\label{fig:st_saucy}
\vspace{-0.3cm}
\end{figure*}
}

\section{An Overview of Our Approach}
\label{sec:overview}

Previous algorithms enumerate all permutations and select the
minimum $(G,\pi)^\gamma$ as the canonical labeling $C(G,\pi)$. There
are two things. The first is that the algorithms use the minimum
$(G,\pi)^\gamma$ as the target to prune candidates during the
enumeration, and the second is that the minimum $(G,\pi)^\gamma$ is
used for any graph.  From a different angle, we consider if we can use
the $k$-th minimum $(G,\pi)^\gamma$ as the canonical labeling
$C(G,\pi)$, where the minimum $(G,\pi)^\gamma$ is a special case when
$k = 1$. Recall that $(G,\pi)^\gamma$ is represented as the sorted edge list,
in other words, all possible $(G,\pi)^\gamma$ form  a totally ordered set.
We observe that there is no need to fix a certain $k$ for
any graph or even to know what the $k$ value is when computing the
canonical labeling. We only need to ensure that there is such a $k$
value based on which two graphs are isomorphic iff their corresponding
$k$-th minimum $(G,\pi)^\gamma$ are the same. Different from previous
algorithms which are designed to prune candidates, we take a
divide-and-conquer approach to partition a graph. We
discuss an axis by which a graph is divided, the AutoTree ${\mathcal
  AT}(G,\pi)$ and its construction.


\stitle{Axis}:
We partition $(G,\pi)$ into a set of vertex disjoint subgraphs,
$\{(g_1,$ $\pi_1),$ $(g_2, \pi_2), ..., (g_k, \pi_k)\}$. We ensure
that by the partition, all automorphisms in $(G,\pi)$ can be composed
by the automorphisms in every $(g_i,\pi_i)$, and the isomorphisms
between subgraphs $(g_i,\pi_i)$ and $(g_j,\pi_j)$.  In other words,
the automorphisms in $(g_i, \pi_i)$ and the isomorphisms between
$(g_i,$ $\pi_i)$ and $(g_j, \pi_j)$ form a generating set for the
automorphism  group of $(G, \pi)$.
We then compute canonical labeling $C(G,\pi)$ by  $C(g_i, \pi_i)$ for every $(g_i, \pi_i)$.
%

We discuss how to partition $(G,\pi)$ into subgraphs by symmetry
according to an axis, which satisfies the requirements mentioned
above.  Note that two subgraphs, $(g_i, \pi_i)$ , $(g_j, \pi_j)$,
are symmetric in $(G, \pi)$, if there is an automorphism $\gamma$ that
maps $(g_i, \pi_i)$ to $(g_j, \pi_j)$. The {\bf axis} by $\gamma$ includes
all vertices $v$ having $v^\gamma=v$, since they are invariant under
$\gamma$. We partition $(G,\pi)$ by such an axis. By removing vertices
in the axis and their adjacent edges from $(G, \pi)$, $(g_i, \pi_i)$
and $(g_j, \pi_j)$ are connected components,
%
%
and all symmetries by $\gamma$ in $(G, \pi)$ are preserved due to the
fact that $(g_i, \pi_i)$ and $(g_j, \pi_j)$ are isomorphic.
We preserve all symmetries by any such automorphism $\gamma$ with an
equitable coloring. Recall that, in an equitable coloring, vertices in
singleton cells cannot be automorphic to any other vertices, and thus
such vertices, as the common part of all axes, preserve the symmetries of
$Aut(G,\pi)$.

\begin{figure}[t]
\begin{center}
   \includegraphics[width=0.9\columnwidth,height=3cm]{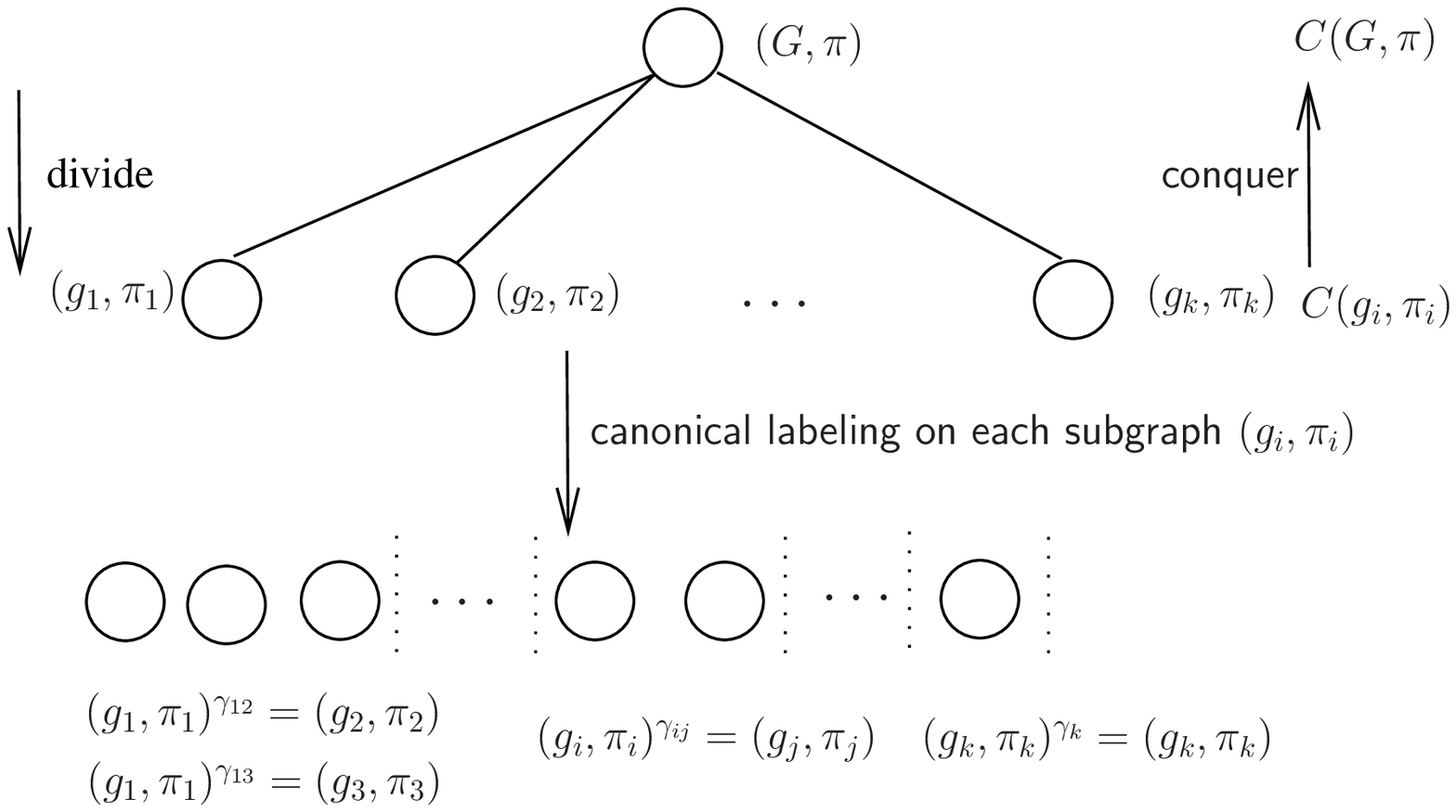}
\end{center}
\vspace{-0.4cm}
\caption{An Overview of Our Approach}
\label{fig:overview}
\vspace{-0.4cm}
\end{figure}

\stitle{The AutoTree ${\mathcal AT}(G,\pi)$}:
We illustrate the main idea of our approach in Fig.~\ref{fig:overview}.  First, graph $(G,\pi)$ is divided into a set of vertex disjoint colored subgraphs  $\{(g_1, \pi_1), \ldots, (g_k,\pi_k)\}$. Such  partition can be achieved by common symmetries given in an equitable coloring
obtained by a refinement function $R$ on $(G,\pi)$.
Giving canonical labeling $C(g_i,\pi_i)$ for every subgraph
$(g_i,\pi_i)$, all subgraphs can be sorted and divided into subsets,
where subgraphs having the same canonical labeling are grouped
in a subset (divided by vertical dash lines in
Fig.~\ref{fig:overview}). The subgraphs in the same subset are
symmetric in $(G,\pi)$ since they are partitioned by symmetry.
For instance, in Fig.~\ref{fig:overview}, suppose two subgraphs
$(g_1,\pi_1)$ and $(g_2,\pi_2)$ are with the same canonical labeling,
then they are in the same subset. They are isomorphic, and there is
a permutation $\gamma_{12}$ such that $(g_1, \pi_1)^{\gamma_{12}} =
(g_2, \pi_2)$ by definition.
%
%
In general, for $(g_i, \pi_i)^{\gamma_{ij}} = (g_j, \pi_j)$, such
$\gamma_{ij}$ will derive an automorphism in $Aut(G, \pi)$, and in
addition, every automorphism in a single subgraph $(g_i, \pi_i)$ is also an automorphism in $Aut(G, \pi)$.  In such sense, we have a
generating set of the  automorphism group $Aut(G, \pi)$, i.e., $Aut(G,
\pi)$ is completely preserved.
%
%
%
%
Note that, two graphs, $(G,\pi)$ and $(G',\pi')$, are isomorphic,
iff they generate the same sorted subgraph sets, resulting in the
same canonical labeling. As a consequence, \CL
discovers the $k$-th minimum $(G,\pi)^\gamma$ as the canonical
labeling.

\begin{figure}[t]
\begin{center}
  \includegraphics[width=1\columnwidth,height= 5cm]{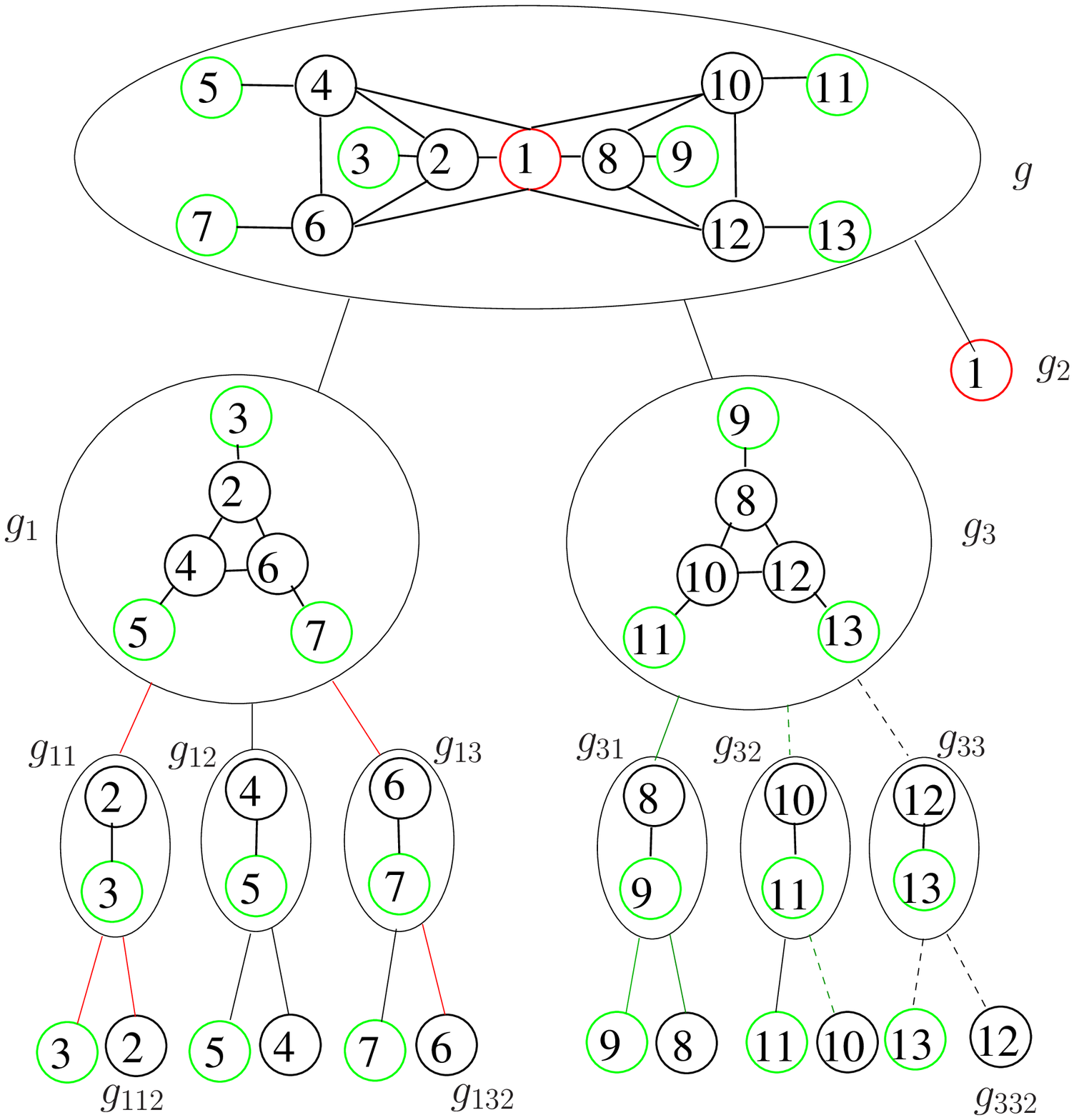}
\end{center}
\vspace{-0.4cm}
\caption{An AutoTree Example
}
\label{fig:axis}
\vspace{-0.4cm}
\end{figure}

\comment{
and automorphisms of each subgraphs are permutations
in symmetry group $S_n$ and the composition refers to permutation
composition, i.e., group operation of $S_n$.

such that the automorphisms of the
graph $G$ can be composed by the automorphisms in each subgraph,
$g_i$, and isomorphisms between two subgraphs, $g_i$ and $g_j$.
}

\comment{
As discussed, previous algorithms enumerate all permutations  and select the minimum $(G,\pi)^\gamma$ value as the canonical labeling $C(G,\pi)$. On the other hand, our approach partitions $(G,\pi)$ into vertex disjoint subgraphs such that automorphisms of the whole graph can be composed by automorphisms of each subgraph and isomorphisms between two subgraphs, and canonical labeling $C(G,\pi)$ can be easily achieved given canonical labeling of each subgraph.
As a consequence, our approach discovers the $k$-th minimum $(G,\pi)^\gamma$ value as the canonical labeling. Note that $k$ is not fixed and varies for different graphs.

The main idea underlying our approach is motivated by the following observation.
\begin{itemize}
\item Each automorphism of $(G, \pi)$ can be decomposed into a few isomorphisms, each is between two symmetric subgraphs and a few automorphisms of subgraphs.
\end{itemize}
For instance, in the example graph in Fig.~\ref{fig:expgraph}, automorphism $(4,5,6)$ is a single automorphism of subgraph induced by $\{4,5,6\}$. If we add a quadrangle with vertices $\{8,9,10,11\}$ and edges between vertex 7 and all new vertices, then permutation  $(0,8)(1,9)(2,10)(3,11)(4,5,6)$ is also an automorphism and can be decomposed into an isomorphism between subgraphs $\{0,1,2,3\}$ and $\{8,9,10,11\}$, and an automorphism of subgraph $\{4,5,6\}$.
Assume graph $(G, \pi)$ can be partitioned into subgraphs ${\mathcal S}=\{(S_1,\pi_1), \ldots, (S_k,\pi_k)\}$  s.t. any automorphism can be decomposed into isomorphisms between two symmetric subgraphs and automorphisms of subgraphs, and vertices $V_s$ that have no automorphic counterparts, then any graph $(G', \pi')$ that is isomorphic to $(G,\pi)$ can be partitioned into the same subgraphs and same $V_s'$ in the same manner. Note that all vertices have their IDs renamed by its structure information. Given a canonical labeling for each subgraph $(S_i, \pi_i)$, then $C(G, \pi)$, i.e., canonical labeling of $(G, \pi)$, can be easily achieved. The reasons are as follows.
First, since symmetric subgraphs are undistinguishable globally, partial order defined by canonical labeling of $\mathcal S$ and $V_s$ is capable to define the edges between $V(\mathcal S)$ and $V_s$ in $C(G, \pi)$. Second, canonical labeling of each subgraph is capable to define the edges among each subgraph in $C(G, \pi)$.
}

\comment{

As an overview, different from previous approaches that select the minimum $(G,\pi)^\gamma$ as the canonical labeling, our approach discovers a relabeled graph, shorted as relabeling, of $(G,\pi)$ that is uniquely determined by the structure of $G$ and the initial coloring $\pi$. Such a relabeling is the $k$-th minimum $(G,\pi)^\gamma$ in the relabeling set, here $k$ can be any possible number for $(G,\pi)$. Since the relabeling is uniquely determined, isomorphism graphs must share the same relabeling, which acts as a canonical labeling.
Following, we outline why such a canonical labeling is uniquely determined.

First consider an observation. With some equitable coloring $\pi$ uniquely determined by $(G,\pi)$, graph $(G,\pi)$ can be uniquely divided into vertex disjoint subgraphs $\{(S_1,\pi_1), \ldots, (S_k,\pi_k)\}$ Each subgraph is either a singleton subgraph containing a vertex in a singleton cell in $\pi$ or a maximal connected subgraph containing only vertices from several non-singleton cells. Given a canonical labeling for each subgraph $(S_i, \pi_i)$ with vertices $V(S_i)$ relabeled by $\pi_i$, subgraphs with the same canonical labeling are symmetric in $G$ and all subgraphs can be sorted by canonical labeling. Represent each subgraph by its canonical labeling, isomorphic graphs have the same sorted subgraph list, representative of a number of permutations generating the same relabeling acting as a canonical labeling.

Consider the example graph $(g, \pi)$ in Fig.~\ref{fig:axis}, with equitable coloring $\pi=[3,5,7,9,11,13|2,4,6,8,10,12|1]$ uniquely determined by $g$. Here, cells in $\pi$ are sorted by neighborhood structure, such as degrees. $\pi$ uniquely divides $g$ into three subgraphs $g_1$, $g_2$ and $g_3$.
To assign symmetric subgraphs in $g$ with the same canonical labeling, $V(g_i)$ for each $g_i$ are firstly relabeled such that vertices in $V_i \in \pi$ are continuous numbers starting from $\pi(V_i)$. Specifically,  $V(g_1)$ and $V(g_3)$, are relabeled by $[3,5,7,2,4,6] \rightarrow [1,2,3,7,8,9]$ and
$[9,11,13,8,10,12] \rightarrow [1,2,3,7,8,9]$, respectively.
With the same canonical labeling $cl_0=\{((1,7),(2,8),(3,9),(7,8),(7,9),(8,9)\}$, $g_1$ and $g_3$ are symmetric in $G$. Denote canonical labeling of $g_2$ as $cl_2=\{(13,13)\}$, sorted subgraph list for $g$ is $cl_0, cl_0, cl_2$. Easy to check, for any permutation $\gamma$ of $g$, $g^\gamma$ has the same sorted subgraph list. As a consequence, our approach generates canonical labeling $\{((1,7),(2,8),(3,9),(4,10),(5,11),(6,12),(7,8),(7,9),(7,13),
\\(8,9),(8,13),(9,13), (10,11),(10,12),(10,13),(11,12),(11,13),(12,13)\}$ for $g$ and $g^\gamma$.

}

\comment{
Revisit previous approaches, they enumerate all permutations in symmetry group $S_n$ to find a permutation $\gamma$ with global minimum $(G, \pi)^\gamma$ value as the canonical labeling, which is time-consuming for some difficult cases. In our approach, we exploit the following observation to find a canonical labeling for $(G,\pi)$ using only few local minimum $(s, \pi_s)^\gamma_s$, which are efficient for small subgraphs. With orbit coloring $\pi_o$, graph $G$ can be divided into a number of vertex disjoint subgraphs, either a singleton subgraph containing a vertex in a singleton cell in $\pi_o$ or a maximal subgraph containing vertices from several non-singleton cells. With canonical labeling

Revisit previous approaches, they enumerate all permutations in symmetry group $S_n$ to find a permutation with the minimum $(G, \pi)^\gamma$ value as the canonical labeling. One interesting point is that canonical labeling $\gamma$ is equivalent to $\gamma^{-1}$, a total order among vertices in $V$. One possible solution for the canonical labeling problem is to define a total order for a colored graph $(G, \pi)$ that can act as a canonical labeling.

Let $\gamma$ and $\gamma^{-1}$ denote the canonical labeling and total order finally discovered. We define a total order of $V$ in two steps. First, equitable coloring $\pi=[V_1 | V_2 | \ldots | V_k]$ of $(G, \pi)$ defines an interval of $v^\gamma$ for each vertex $v$, which leads to a total order among different cells. Specifically, for each vertex $v \in V_i$, $v^\gamma \in [\pi(v), \pi(v)+|V_i|)$, which implies that for any vertex $v \in V_i$ and $u \in V_j$ having $i<j$, $v^\gamma < u^\gamma$.
Second, we define a total order among vertices in each cell. Note that for any vertex $w$ in a singleton cell, both $w^\gamma$ and its adjacent edges in $(G, \pi)^\gamma$ are determined. Therefore, when defining ordering of each cell, edges adjacent to such vertices can be neglected, which can disconnect $G$ into a number of connected components $\{(S_1, \pi_1), \ldots, (S_k,\pi_k)\}$. Note that vertices in $S_i$ may come from different cells in $\pi$. Assume a total order of $V(S_i)$, which leads to a canonical labeling of $(S_i, \pi_i)$, is known, then the total order among $\{(S_1, \pi_1), \ldots, (S_k,\pi_k)\}$ is easily obtained by canonical labeling of each $(S_i, \pi_i)$ (ties are break arbitrarily), which in return defines a total order of each cell $V_i$ along with the total order of $V(S_i)$. As a consequence, we develop our approach by divide-and-conquer, and the partition of each $(S_i, \pi_i)$ construct an attributed tree, which we call {\sl AutoTree}. Another fascinating discovery is that all total orders, or equivalently permutations, result in the canonical labeling, share the same AutoTree.
}

Canonical labeling of each subgraph $(g_i, \pi_i)$ can be obtained in the same manner, which results in a tree index. In the tree, each node is associated with automorphism group and canonical labeling and child nodes of each non-leaf node are sorted by canonical labeling.   We call such an ordered tree an {\sl AutoTree}, denoted
as ${\mathcal {AT}}(G, \pi)$, for given graph $(G,
\pi)$. Such an AutoTree benefits to discovering the automorphism group
$Aut(G,\pi)$ and the canonical labeling $C(G, \pi)$.

\comment{
Generalize this idea to $Aut(G,\pi)$, vertices $v$ having
$v^\gamma=v$ for all $\gamma \in Aut(G,\pi)$ act as an axis to
preserve all symmetries in $Aut(G,\pi)$. Recall that in an equitable
coloring, vertices in singleton cells cannot be automorphic to any
other vertices, such vertices consist a subset of the axis that
preserves the symmetries of $Aut(G,\pi)$.
}

\comment{
We give an overview on our approach. We construct a tree-shape index,
called {\sl AutoTree}, denoted as ${\mathcal {AT}}(G, \pi)$, for the
given colored graph $(G, \pi)$, to discover the automorphism group
$Aut(G,\pi)$ and the canonical labeling $C(G, \pi)$. Here, $\pi$ is
assumed to be equitable below.
First, we give the main ideas behind the AutoTree ${\mathcal {AT}}(G,
\pi)$ to be constructed as follows.
\begin{itemize}
\item Two vertices, $u$ and $u'$, are automorphic in $G$, if there are
  two subgraphs, $g$ and $g'$, that are isomorphic and symmetric
  in $G$, where $u \in g$ and $u' \in  g'$.
\end{itemize}
}

\comment{
Consider an automorphism $\gamma$ that maps a subgraph $g_1$ to another subgraph $g_2$. Then $g_1$ and $g_2$ are symmetric in $G$ and vertices $v$ having $v^\gamma=v$ act as an axis.
Intuitively, let a vertex $v$ be a vertex in an axis, two vertices,
$u$, and $u'$, are symmetric in $G$, if there exists a one to one
mapping between the set of paths from $u$ to $v$ and the set of paths
from $u'$ to $v$, where every pair of corresponding paths are
isomorphic.  There are no constraints on the two subgraphs, $g$ and
$g'$, which may have overlapping.
We further explain our ideas below. Suppose $g$ is a subgraph of $G$
containing $u$, $\gamma$ is an automorphism of $(G,\pi)$ satisfying $u^\gamma=u'$. We can easily find a subgraph
$g'=g^\gamma$ with the following 3 properties: (1) $g'$ contains $u'$,
(2) $g'$ is isomorphic to $g$ and $u'$ corresponds to $u$ in the
isomorphic permutation (i.e., $g'=g^\gamma$ and $u'=u^\gamma$), and
(3) $g'$ and $g$ are symmetric following an axis.
This observation suggests that
we can recursively divide a colored graph $(G, \pi)$ into components,
$\{(g_1, \pi), (g_2, \pi), \cdots\}$, where $(G, \pi) = \cup (g_i,
\pi)$
such that automorphisms in $(G,\pi)$ can be discovered (i) among each
component and (ii) between two isomorphic components.
%
%
}

We explain key points of AutoTree  ${\mathcal {AT}}(g, \pi)$, using an
example in Fig.~\ref{fig:axis}. Here, we assume the coloring $\pi$
associated with the colored graph $(g, \pi)$ is equitable, and show how
the AutoTree is constructed for such a colored graph.  As shown in
Fig.~\ref{fig:axis}, the entire graph $g$ is represented by the root. There are 3 colors in the equitable coloring.  Here, two
vertices have the same color if they are in the same cell in $\pi$.
First, vertex $1$ in the singleton cell in $\pi$ acts as an axis for
$g$ and partitions $g$ into three subgraphs $g_1$ (left), $g_2$
(right), and $g_3$ (mid), where $g_2$ consists of a single vertex $1$.
Second, we construct sub-AutoTree rooted at $g_1$. We find that there is a complete subgraph, $g'
\subseteq g_1$, over all vertices $\{2, 4, 6\}$ that have the same
color. We observe that the automorphism group of $g_1$ will not be
affected without the edges in $g'$, and further divide $g_1$ into
$g_{11}$ (left), $g_{12}$ (mid), and $g_{13}$ (right).  Here, we
consider the set of $\{2, 4, 6\}$ as an additional axis ($a_{11}$) for
$g_1$.
Third, consider $g_{11}$ as an example, it will be divided into
another 2 subgraphs, each contains a vertex having unique color in
$g_{11}$.
In ${\mathcal AT}(g,\pi)$, two vertices, $2$ and $6$ are automorphic,
because $2 \in g_{112}$, $6 \in g_{132}$, and $g_{112}$ and $g_{132}$ are
isomorphic and  symmetric according to the axis $a_{11}$. Similarly, $2$ and $12$ are automorphic, because $2 \in g_{112}$, $12
\in g_{332}$, and $g_{112}$ and $g_{332}$ are isomorphic and are symmetric
according to the axis $a_{1}$.
%

We discuss the key property of AutoTree ${\mathcal {AT}}(G, \pi)$.
For any two automorphic vertices $u$ and $u'$, the axes recursively
divide, $u$ and $u'$, into a series of subgraph pairs $((g_1, g_1'),$
$(g_2, g_2'),$ $\ldots, (g_k,$ $g_k'))$ such that (1) $g_1 \supset g_2
\supset \ldots \supset g_k$ and $g_1' \supset g_2' \supset \ldots
\supset g_k'$, (2) $(g_k,\pi_k)$ and $(g_k',\pi_k')$ are leaf nodes in ${\mathcal
  {AT}}(G,\pi)$, and (3) $g_i$ and $g_i'$ are isomorphic and symmetric
in $G$.
For instance, the two automorphic vertices, 2 and 12, in
Fig.~\ref{fig:axis} are divided into subgraph pairs $((g_1, g_3),
(g_{11}, g_{33}), (g_{112},$ $g_{332}))$.  In other words, for any two
automorphic vertices, they must be in two leaf nodes in AutoTree,
whose corresponding subgraphs
have the
same canonical labeling and  are symmetric. As a consequence, automorphisms between vertices can be detected by comparing
canonical labeling of leaf nodes
containing these vertices.
As can be observed in the experiments, (1) most vertices in $G$ are in
singleton cells, (2) non-singleton leaf nodes in ${\mathcal {AT}}(G,
\pi)$ are small in size. By ${\mathcal {AT}}(G$, $\pi)$, automorphisms
between vertices can be efficiently detected.  It is worth mentioning
that, in the existing approaches, determining whether two vertices are
automorphic need to compare a set of permutations.
The generation of canonical labeling $C(G,\pi)$, as we will discuss,
can be done in a bottom-up manner, where the canonical labeling of a
non-leaf node in ${\mathcal {AT}}(G, \pi)$ can be done by
combining canonical labeling of its child nodes, which significantly
reduces the cost.

\comment{
Motivated by this observation, Algorithm
\DivideP constructs a tree rooted at $(G,\pi)$ by recursively
isolating singleton vertices in each non-singleton tree node.
Therefore, discovering automorphisms directly from $(G,\pi)$ is
equivalent to discovering automorphisms from the partial AutoTree
${\mathcal {AT}}(G, \pi)$ established by \DivideP.

Let $\pi_o$ be the {\sl orbit coloring} of
$(G, \pi)$, where each cell in an order coloring contains vertices
that are mutually automorphic if they have the same color.
Following $\pi_o$, in a colored graph $(G, \pi)$,
%
%
%
%
%
%

When $\pi_o$ contains no singleton cells, the idea of taking singleton
cells as the axis can be generalized to the idea of taking a set of
vertices as the axis. Here, the set of vertices can either be a single
cell or several cells in the orbit coloring $\pi_o$. For simplicity,
we discuss how to isolate a set of vertices for two cases, and other
cases can be easily generalized. First, when a single cell, denoted as
$V_i$, is taken as the axis, isolating $V_i$ means removing edges
among vertices in $V_i$, i.e., removing $E_i=\{(u,v) | u,v \in V_i\}$.
Second, when two cells, denoted as $V_i$ and $V_j$, are taken as the
axis, isolating $V_i \cup V_j$ is done by removing edges between $V_i$
and $V_j$, i.e., removing $E_{ij}=\{(u,v) | u \in V_i, v \in V_j\}$.
As we will prove in the following sections, removing $E_i$ and
$E_{ij}$ will not influence the automorphism group and the canonical
labeling of $(G,\pi)$. This observation motivate Algorithm \DivideS,
which attempts to construct a subtree rooted at a tree node
$(g,\pi_g)$ by simplifying the structure of $g$.
}

\comment{

\begin{figure*}[t]
\begin{center}
   \includegraphics[width=1.8\columnwidth,height=5cm]{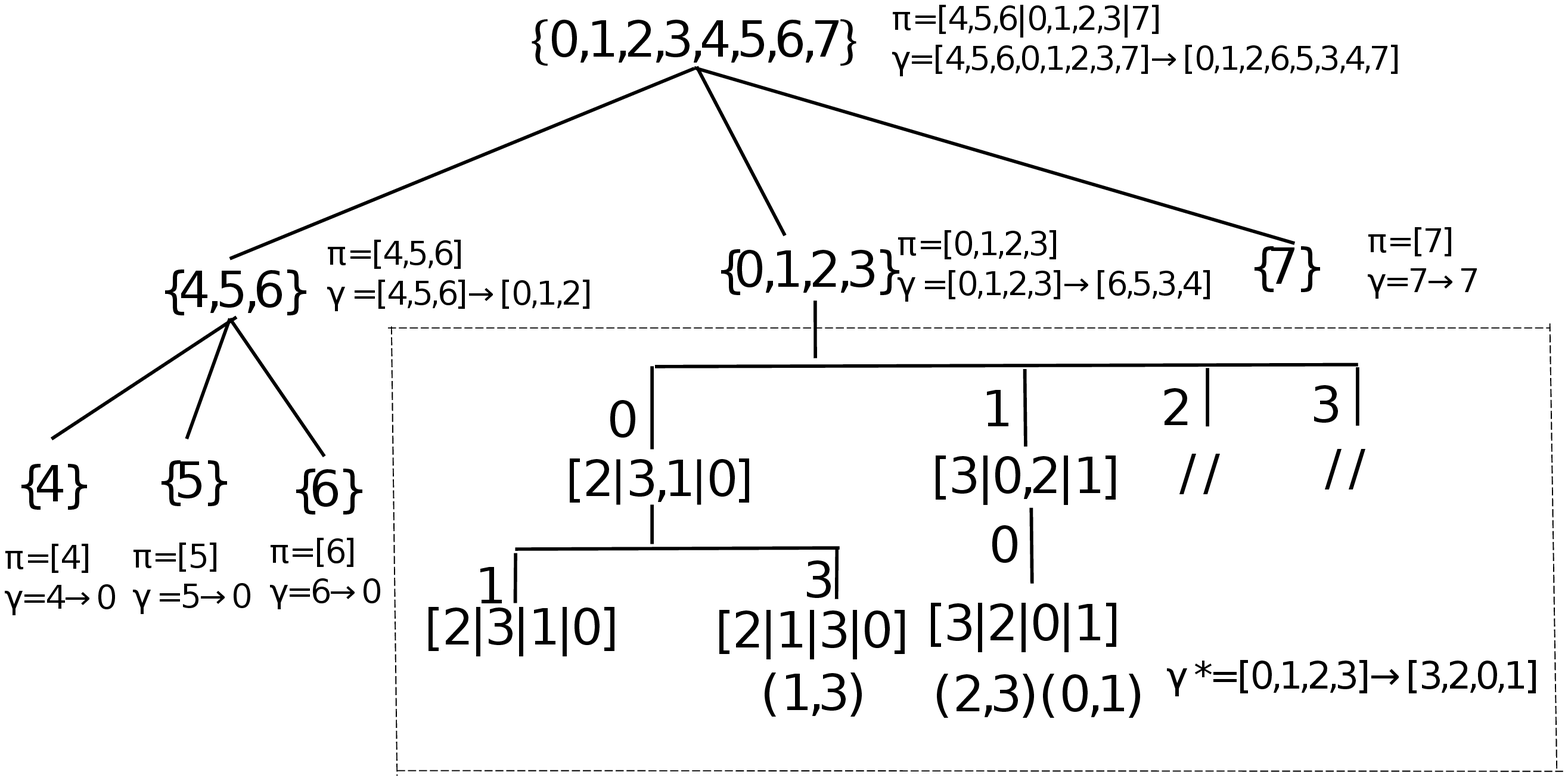}
\end{center}
\vspace{-0.6cm}
\caption{AutoTree for the graph $G$ in Fig.~\ref{fig:expgraph}.}
\label{fig:dectree}

\end{figure*}
}

${\mathcal {AT}}(G, \pi)$ is a sorted tree. In ${\mathcal {AT}}(G, \pi)$, the root represents $(G, \pi)$, and every node represents a
subgraph $(g, \pi_g)$. Here, $g$ is a subgraph of $G$ induced
by $V(g)$ and $\pi_g$ is the projection of $\pi$ on $V(g)$.  Note that
$\pi_g(v) = \pi(v)$ for any $v \in g$ and any $g \subset G$.  Each
node $(g,\pi_g)$ in ${\mathcal {AT}}(G, \pi)$ is associated with
canonical labeling $C(g, \pi_g)$, or equivalently,
a permutation $\gamma_g$ generating $C(g, \pi_g)$, i.e., $(g, \pi_g)^{\gamma_g}=C(g,\pi_g)$.
For any singleton
subgraph $g=\{v\}$, we define $C(g, \pi_g)=(v^{\gamma_g},v^{\gamma_g})=(\pi(v), \pi(v))$.
%
%
Permutation $\gamma_g$ can be generated for a node $(g,\pi_g)$ in
three cases: (a) $\gamma_g$ is trivially obtained for a singleton leaf
node, (b) $\gamma_g$ is generated with canonical labeling achieved by
any existing algorithm (e.g., \nauty, \bliss and \traces) for a
non-singleton leaf node, and (c) $\gamma_g$ is generated by combining
all canonical labeling of $g$'s children.
%
%
%
The canonical labeling of the root node is the one of
the given graph.
Automorphisms of $(G, \pi)$ can be discovered between two nodes with the same canonical labeling and automorphisms of each subgraph.

\comment{
  As discussed in Section~\ref{sec:previous}, state-of-the-art
  canonical labeling algorithms construct a backtrack search tree
  ${\mathcal T}(G,\pi)$ following the ``individualization-refinement''
  paradigm, and find the automorphism group $Aut(G, \pi)$ and the
  canonical labeling $C(G, \pi)$ from leaf nodes of the search tree.
  On the other hand, our approach \CL constructs a tree-shape index,
  called {\sl AutoTree}, for the given colored graph $(G, \pi)$.
  AutoTree, denoted as ${\mathcal {AT}}(G, \pi)$, contains symmetry
  information of $(G,\pi)$, from which the automorphism group and the
  canonical labeling can be easily derived. Specifically, the
  automorphism group can be discovered from both leaf nodes and
  non-leaf nodes, and the canonical labeling can be directly obtained
  from the root.  Each node of AutoTree ${\mathcal {AT}}(G, \pi)$ is a
  colored subgraph $(g, \pi_g)$ of $(G, \pi)$, and is associated with
  information such as a permutation $\gamma_g$ that can test
  isomorphism between two tree nodes, i.e., two subgraphs of $(G,
  \pi)$. Here, $\pi$ is an equitable coloring and $\pi_g$ is the
  projection of $\pi$ on $V(g)$. For the sake of simplicity, we reuse
  the notation permutation to represent $\gamma_g$, which is actually
  a injective mapping $V(g) \rightarrow V(G)$, and reuse the notation
  canonical labeling to represent $(g^{\gamma_g}, \pi_g^{\gamma_g})$.
}

\comment{
\begin{figure*}[t]
\begin{center}
\begin{tabular}[t]{c}
\hspace{-0.6cm}
    \subfigure[An AutoTree Example]{
         \includegraphics[width=1\columnwidth,height=4.5cm]{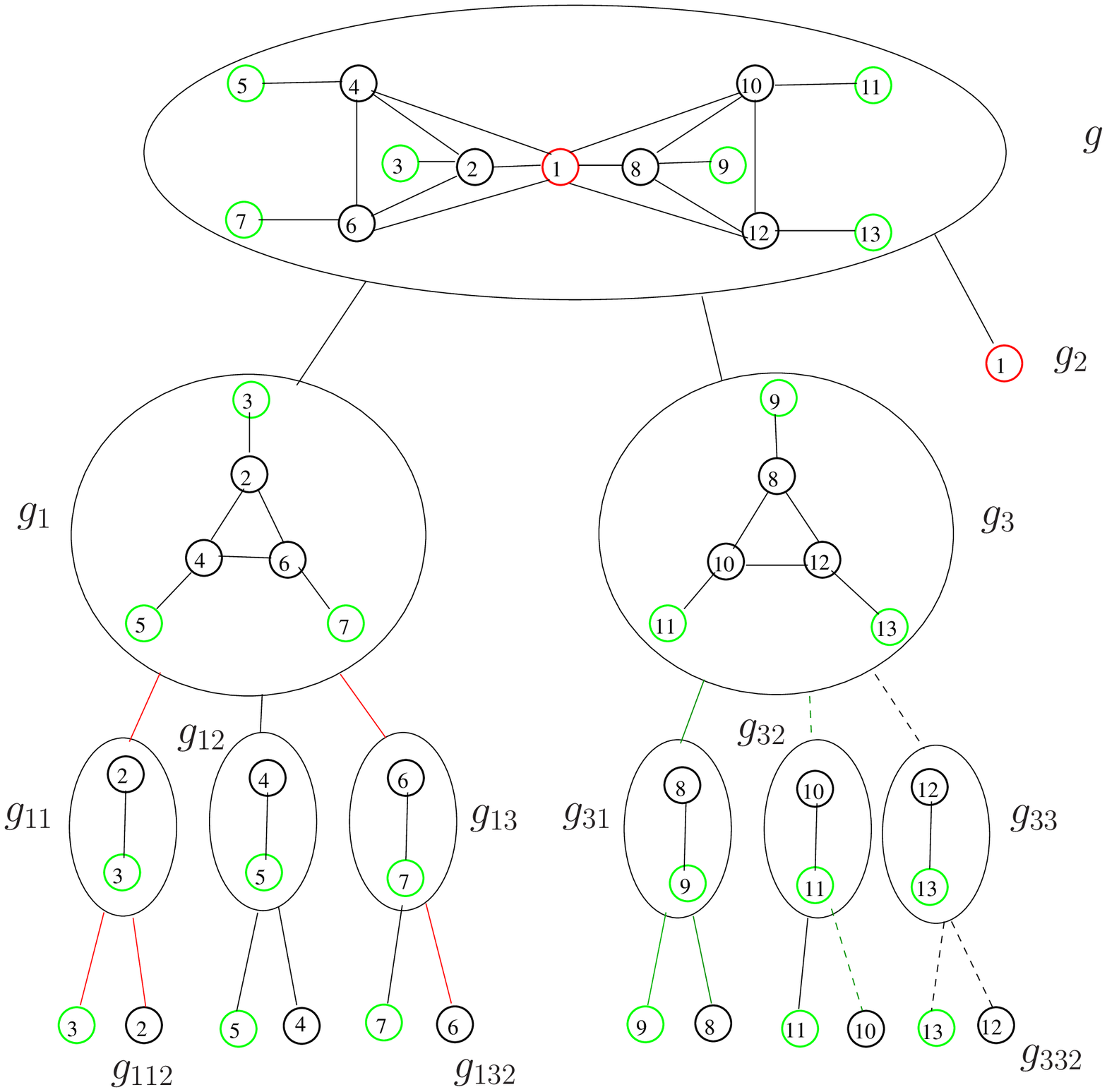}
        \label{fig:axis}
    }
    \hspace{0.3cm}
    \subfigure[AutoTree for the graph $G$ in Fig.~\ref{fig:expgraph}.]{
    \includegraphics[width=1\columnwidth,height=5cm]{figure/CRD-0}
   \label{fig:dectree}
    }
\end{tabular}
\end{center}
\vspace*{-0.4cm}
\caption{AutoTree example}
\end{figure*}
}

\begin{figure}[t]
\begin{center}
   \includegraphics[width=1\columnwidth,height=4.5cm]{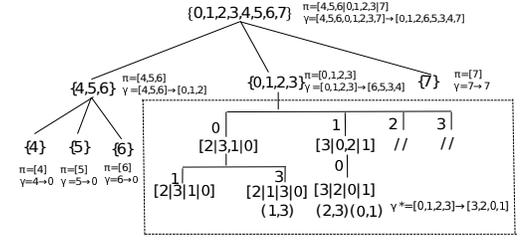}
\end{center}
\vspace*{-1.2cm}
\caption{AutoTree for the graph $G$ in Fig.~\ref{fig:expgraph}.}
\label{fig:dectree}
\vspace*{-0.6cm}
\end{figure}

Fig.~\ref{fig:dectree} shows the AutoTree ${\mathcal {AT}}(G, \pi)$
constructed for the graph $G$ in Fig.~\ref{fig:expgraph}. A node in
${\mathcal {AT}}(G, \pi)$ represents a subgraph $(g, \pi_g)$, by its
$V(g)$ together with its permutation $\gamma_g$.
Consider the three leaf nodes (singletons) from the left (i.e., the
three one-vertex subgraphs), $\{4\}$, $\{5\}$, and $\{6\}$, with
coloring $\pi_g=[4]$, $\pi_g=[5]$, $\pi_g=[6]$, respectively. The
permutations for the three  subgraphs are, $\gamma =
4 \rightarrow 0$, $\gamma = 5 \rightarrow 0$, and $\gamma = 6
\rightarrow 0$. Vertices 4, 5 and 6 are mutually automorphic since these three leaf nodes have the same canonical labeling.
The permutation for the parent of the three singletons
is $\gamma = [4, 5, 6] \rightarrow [0, 1, 2]$ by combining the
canonical labeling of the three singletons.
%
Subgraph $\{4,5,6\}$ does not have symmetric counterparts
since there exist no other nodes having the same canonical labeling.
The 4th leaf node from the left is non-singleton, since it cannot be
further divided. We use \bliss to obtain its
permutation, in dashed rectangle.
%

In an AutoTree, the permutation $\gamma_g$ for $(g, \pi_g)$, is done
as follows. First, $v^{\gamma_g}=\pi(v)$ is generated for a singleton
leaf node with $\{v\}$. For example, for the 2nd leaf node from the
left of $\{5\}$ in Fig.~\ref{fig:dectree}, $\pi(5)=0$ which
indicates the cell in the coloring where 5 exists.  Second,
$\gamma_g$ is generated by an existing algorithm for a non-singleton
node. For example, the 4th leaf node from the left in
Fig.~\ref{fig:dectree} is a non-singleton. its permutation $\gamma_g$
is obtained by a backtrack search tree constructed using an existing
algorithm.  Third, $\gamma_g$ for a non-leaf node is determined by those of its child nodes.


\comment{
We explain that $u$ and $u'$ are symmetric in the viewpoint of $v$ in two folds. First, let  $\gamma$ denote an automorphism satisfies $u^\gamma=u'$, then for any path $(v, w_1, \ldots, w_k, u)$ (or a colored path $((v,\pi(v)), (w_1,\pi(w_1)), \ldots, (w_k,\pi(w_k)), (u,\pi(u)))$),  there is another path $(v, w_1^\gamma, \ldots, w_k^\gamma, u')$. This means

Let $\pi_o$ be the {\sl orbit coloring} of $(G, \pi)$, in which each cell contains vertices that are mutually automorphic. Without loss of generality, we assume $\pi_o$ contains both singleton cells and non-singleton cells, i.e., in $(G, \pi)$, there are both vertices that are not automorphic to any other vertices and vertices that have automorphic counterparts.
Consider a singleton cell in $\pi_o$, and denote the vertex in this singleton cell as $v$. We explain that the whole graph is symmetric in the viewpoint of $v$. First, consider any two distinct automorphic vertices $u$ and $u'$.  Assume automorphism $\gamma$ satisfies $u^\gamma=u'$, then for any path $(v, w_1, \ldots, w_k, u)$ (or a colored path $((v,\pi(v)), (w_1,\pi(w_1)), \ldots, (w_k,\pi(w_k)), (u,\pi(u)))$),  there is another path $(v, w_1^\gamma, \ldots, w_k^\gamma, u')$, implying that any two automorphic vertices are symmetric taken any vertex in a singleton cell as an axis. Similarly, for any subgraph $g$ of $G$, $g$ and $g^\gamma$ are symmetric at the viewpoint
}


\stitle{The AutoTree ${\mathcal {AT}}(G, \pi)$ Construction}: We
design an algorithm \CL  to construct an AutoTree ${\mathcal {AT}}(G,
\pi)$ by divide-and-conquer.
In the divide phase, \CL
divides $(G, \pi)$ into a set of subgraphs $(g_i, \pi_{g_i})$, each
consists of a child node of   $(G, \pi)$ in ${\mathcal {AT}}(G,
\pi)$. \CL recursively construct AutoTree ${\mathcal {AT}}(g_i,
\pi_{g_i})$ rooted at $(g_i, \pi_{g_i})$.  In the combine phase, \CL
determines the canonical labeling of $(G, \pi)$ by the canonical labeling of its child nodes $(g_i,\pi_{g_i})$.
In the divide phase, two algorithms are used to divide $(g, \pi_g)$,
namely, \DivideP and \DivideS, by removing edges in $g$ that have no
influence in determining the automorphism group and the canonical
labeling of $(g, \pi_g)$. Consider Fig.~\ref{fig:axis}. \DivideP is to
remove edges by finding singleton cells in $\pi_g$ (e.g., the vertex 1 in $g$), whereas
\DivideS is to remove edges by complete subgraphs or complete bipartite subgraphs (e.g., the complete
subgraph in $g_1$).
A leaf node in ${\mathcal {AT}}(G, \pi)$ is a node that cannot by
divided by \DivideP or \DivideS.

\begin{figure}[t]
\begin{center}
  \includegraphics[width=0.8\columnwidth,height=2.8cm]{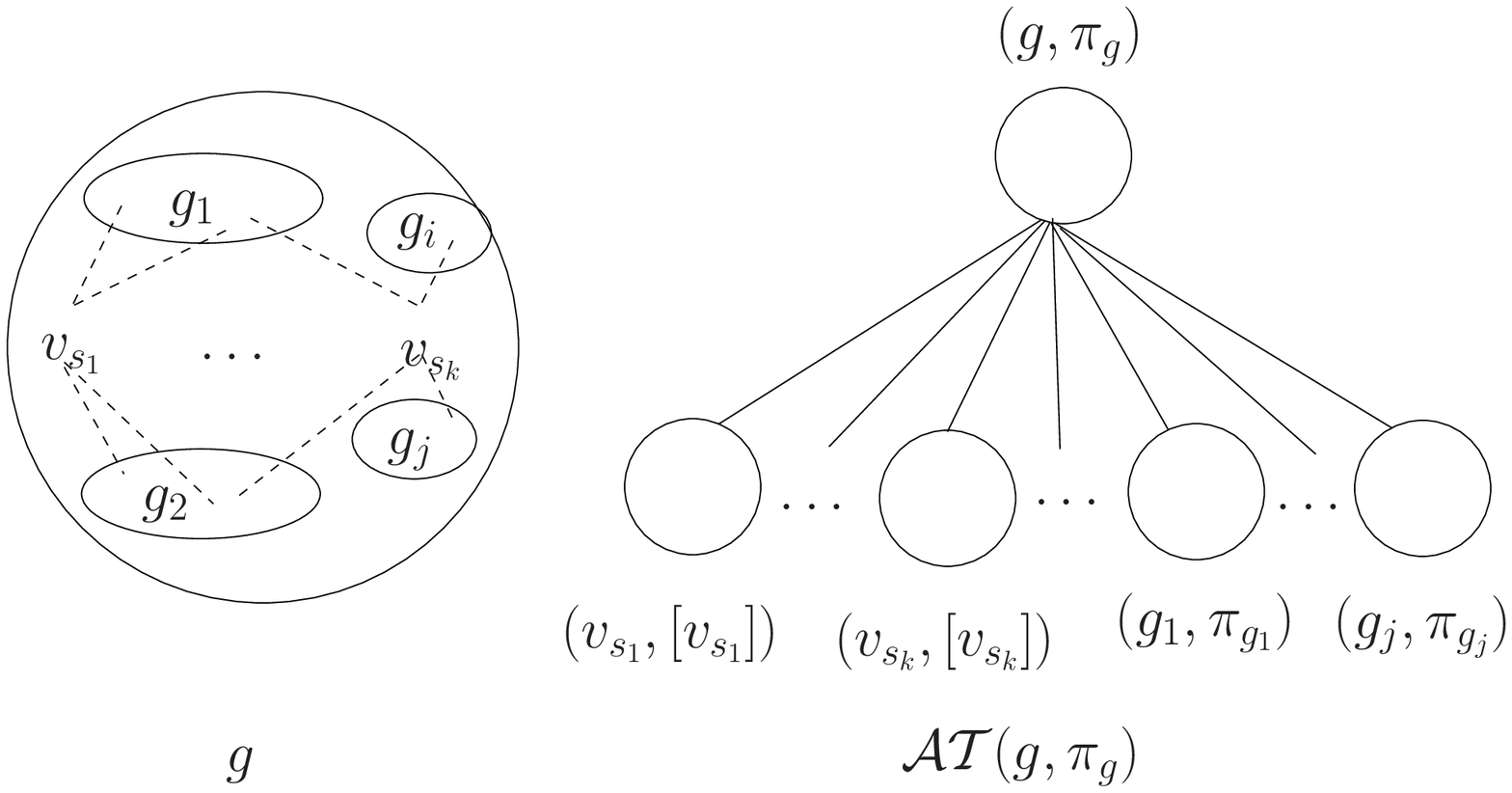}
\end{center}
\vspace*{-0.4cm}
\caption{The Overview of Algorithm \DivideP}
\label{fig:dividep}
\vspace*{-0.4cm}
\end{figure}

\begin{figure}[t]
\begin{center}
  \includegraphics[width=0.75\columnwidth,height=2.8cm]{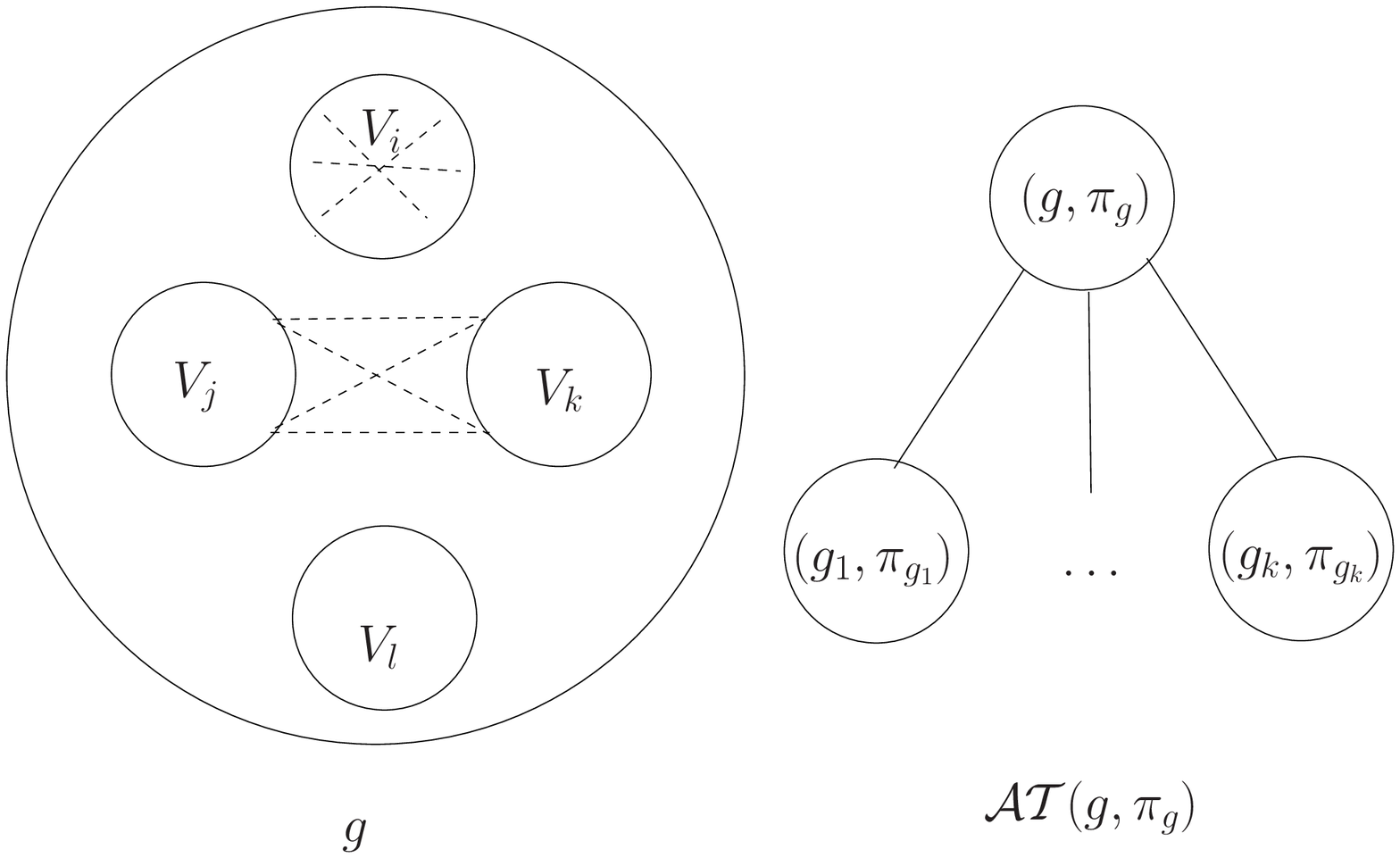}
\end{center}
\vspace{-0.4cm}
\caption{The Overview of Algorithm \DivideS}
\label{fig:divides}
\vspace{-0.4cm}
\end{figure}

An overview is shown in Fig.~\ref{fig:dividep} for \DivideP.  In
Fig.~\ref{fig:dividep}, the left shows a tree node  $(g,
\pi_g)$ where vertices $v_{s_i}$ are in singleton cells in $\pi_g$,
and the right shows the child nodes constructed for $(g,\pi_g)$ by
\DivideP.  Isolating singleton cells in $\pi_g$ is to remove dashed
edges in $g$ and partition $g$ into a set of connected components
$g_i$.

{\small
\begin{displaymath}
 (g, \pi_g) \rightarrow \bigcup_i (v_{s_i}, [v_{s_i}]) \cup \bigcup_j
  (g_j, \pi_{g_j})
\end{displaymath}
 }
 \vspace{-0.1cm}
Here, each $(v_{s_i}, [v_{s_i}])$ represents a one-vertex colored
subgraph as a result from a singleton cell in
$\pi_g$, and each $(g_j, \pi_{g_j})$ is a connected component of
$(g,\pi_g)$.

An overview is shown in Fig.~\ref{fig:divides} for \DivideS. In
Fig.~\ref{fig:divides}, the left shows a subgraph $(g, \pi_g)$ whose
vertices are in 4 different cells,
$V_i$, $V_j$, $V_k$, and $V_l$.
The right shows the
child nodes constructed for the node that represents $(g,\pi_g)$ by
\DivideS.  \DivideS removes edges for 2 cases.  First, \DivideS
removes all edges from the induced subgraph over the cell $V_i$ if
it is a complete subgraph. Second, \DivideS removes all edges between
2 different colors $V_j$ and
$V_k$ if there is a complete bipartite subgraph between all
vertices in $V_j$ and all vertices in $V_k$. Removing such
edges does not affect the   automorphism group $Aut(g,\pi_g)$.
By removing such edges, $(g,\pi_g)$ can be possibly divided into
several disconnected components $(g_i,\pi_i)$. We have

\vspace{-0.2cm}
{\small
\begin{displaymath}
(g, \pi_g) \rightarrow \bigcup_k (g_k, \pi_{g_k})
\end{displaymath}
}
\vspace{-0.2cm}
%
%
\comment{
In the left of Fig.~\ref{fig:divides}, dashed lines are the edges to
be removed by \DivideS. There are two cases. First,
in $V_i$
that consist of a clique or between $V_{j_1}$ and $V_{j_2}$ that
consist of a biclique.
}
%
%

In the combine phase, \CL generates the permutation $\gamma_g$ for the
node $(g, \pi_g)$ in AutoTree. Note that permutation $\gamma_g$ is the
one that produces the canonical labeling $C(g, \pi_g)$.  First,
consider the base case when $(g, \pi_g)$ is a leaf node in
${\mathcal {AT}}(G, \pi)$.
If $(g, \pi_g)$ is a singleton leaf node (e.g., $g=\{v\}$), we define
$g^{\gamma_g}=\pi(v)$.
If $(g,\pi_g)$ is a non-singleton leaf node, we obtain $\gamma_g$ by
\CombineCL. Here, \CombineCL first applies an existing approach to
generate a canonical labeling $\gamma^*$ for $(g, \pi_g)$. With
$\gamma^*$, vertices sharing the same color, i.e., in the same cell in
$\pi_g$, are differentiated by the ordering introduced by $\gamma^*$.
Second, consider the case when $(g, \pi_g)$ is a non-leaf
node. \CombineST exploits the canonical labeling of the child nodes
of $(g, \pi_g)$ (i.e., $\gamma_{g_i}$ and $(g_i,
\pi_{g_i})^{\gamma_{g_i}}$ of $(g_i, \pi_{g_i})$) to determine an
ordering that can differentiate vertices in the same cell in $\pi_g$,
and obtains $\gamma_g$ in a similar manner.

\comment{

\section{An Overview of Our Approach}
\label{sec:overview}

As discussed in Section~\ref{sec:previous}, state-of-the-art approaches follow the ``individualization-refinement'' schema and construct search trees based on concepts and theorems in group theory. On the other hand, our approach
solves the canonical representative discovery problem, and consequently
the graph isomorphism problem,
by constructing a AutoTree of the given graph from the graph theory perspective. With the AutoTree, both the automorphism group (or equivalently, the orbit partition) and the canonical representative can be easily derived. The structure of the AutoTree is as follows. The root is the entire graph; each leaf node is a subgraph, 
either a single vertex or a subgraph where no vertex can be easily distinguished from the others;
and for each non-leaf node,
it is divided into a set of non-overlapping subgraphs, each consists a child node.
In addition, each node in the AutoTree is associate with a certificate, which can help to find isomorphisms between two disjoint subgraphs.
And for each non-leaf node, its child nodes are sorted by certificates such that isomorphic subgraphs are grouped together.
Note that in this paper, a node in the AutoTree is equivalent to a subgraph of the given graph $G$.
Fig.~\ref{fig:dectree} shows the AutoTree constructed by our approach for the example graph in Fig.~\ref{fig:expgraph}. In the AutoTree, each node is represented by the vertices that induce the subgraph, and the subtree in the dashed rectangle is the search tree achieved by \bliss for the unique non-singleton leaf node $\{0,1,2,3\}$, to obtain its canonical representative.

Our approach \CL constructs the AutoTree by divide and
conquer. Consider a graph $g$, here $g$ can be either the given graph
$G$ or a subgraph of $G$ extracted in the algorithm. \CL, or its main
component \cl, generates a partition of $V(g)$, explicitly or
implicitly. With the partition, $g$ is divided into a set of
non-overlapping subgraphs, constructing tree edges in the AutoTree,
and each subgraph is recursively divided until the base cases. Each
base case, which is a leaf node in the AutoTree, occurs when the
vertices in the subgraph cannot be easily distinguished. For each leaf
node, \CL invokes existing algorithms to obtain
a canonical representative and the orbit partition for the subgraph, further generating its certificate. For each non-leaf node, its certificate is achieved by combining the certificates of its child nodes.
As a consequence, the certificate of the root node can be used as a canonical representative of the given graph.

The main components of our approach \CL is as follows.
\begin{itemize}
\item Algorithm \Partition generates an equitable ordered partition $\pi$ of $V(G)$. The partition $\pi$ will be frequently reused in the other algorithms. In Fig.~\ref{fig:dectree}, $\pi=[0,1,2,3,4,5,6|7]$ since vertex 7 has a distinct degree from the others while its not easy to distinguish the remaining vertices with neighborhood information;
\item Algorithm \cl constructs the subtree rooted at the input tree node $g$ following the ``divide and conquer'' paradigm. In addition, \cl generates the certificate $c(g)$ of $g$;
\item Algorithms \DivideP and \DivideS, invoked by \cl, divide a subgraph $g$ into a set of smaller subgraphs. In Fig.~\ref{fig:dectree}, \DivideP separates vertex 7 from the others in the root node and \DivideS removes edges in tree node $\{4,5,6\}$, dividing it into three singleton subgraphs;
\item Algorithms \CombineCL and \CombineST, utilized in \cl, generate the certificates of tree nodes in the AutoTree. Specifically, \CombineCL generates the certificate of a leaf node by its canonical representative, while \CombineST generates the certificate of a non-leaf node, combining the certificates of its child nodes. In Fig.~\ref{fig:dectree}, \CombineCL generates the certificate of node $\{0,1,2,3\}$  while \CombineST generates the certificate of node $\{4,5,6\}$ and the root node;
\end{itemize}
}

\comment{

\subsection{Optimizations by Structural Equivalence}

\begin{figure}[t]
\begin{center}
\begin{tabular}[t]{c}
\hspace{-0.6cm}
    \subfigure[simplified graph $G_s$]{
         \includegraphics[scale=0.32]{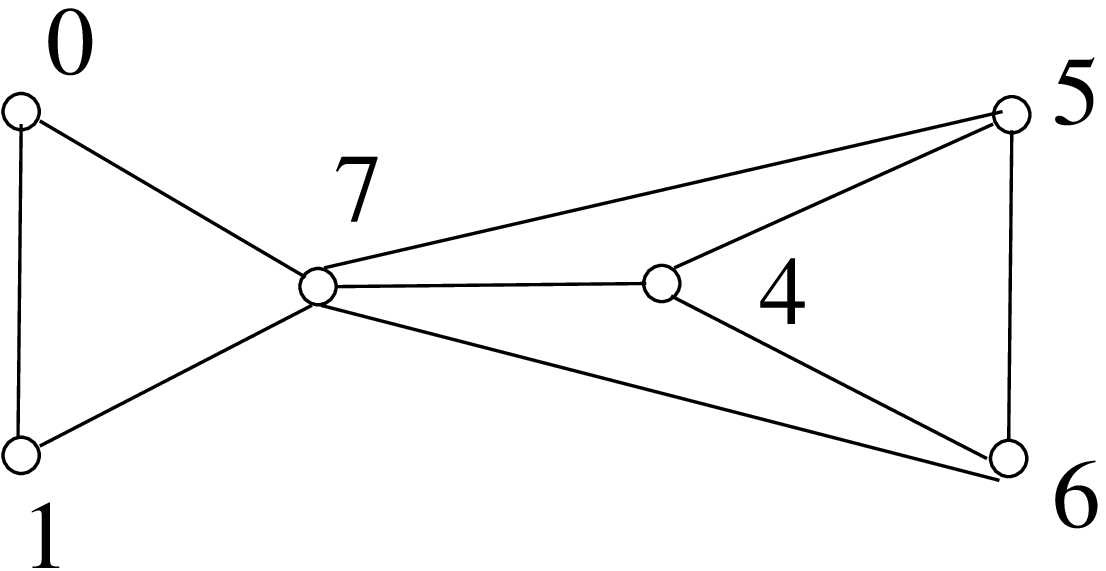}
        \label{fig:simgraph}
    }
    \hspace{0.3cm}
    \subfigure[AutoTree of $G_s$]{
    \includegraphics[scale=0.18]{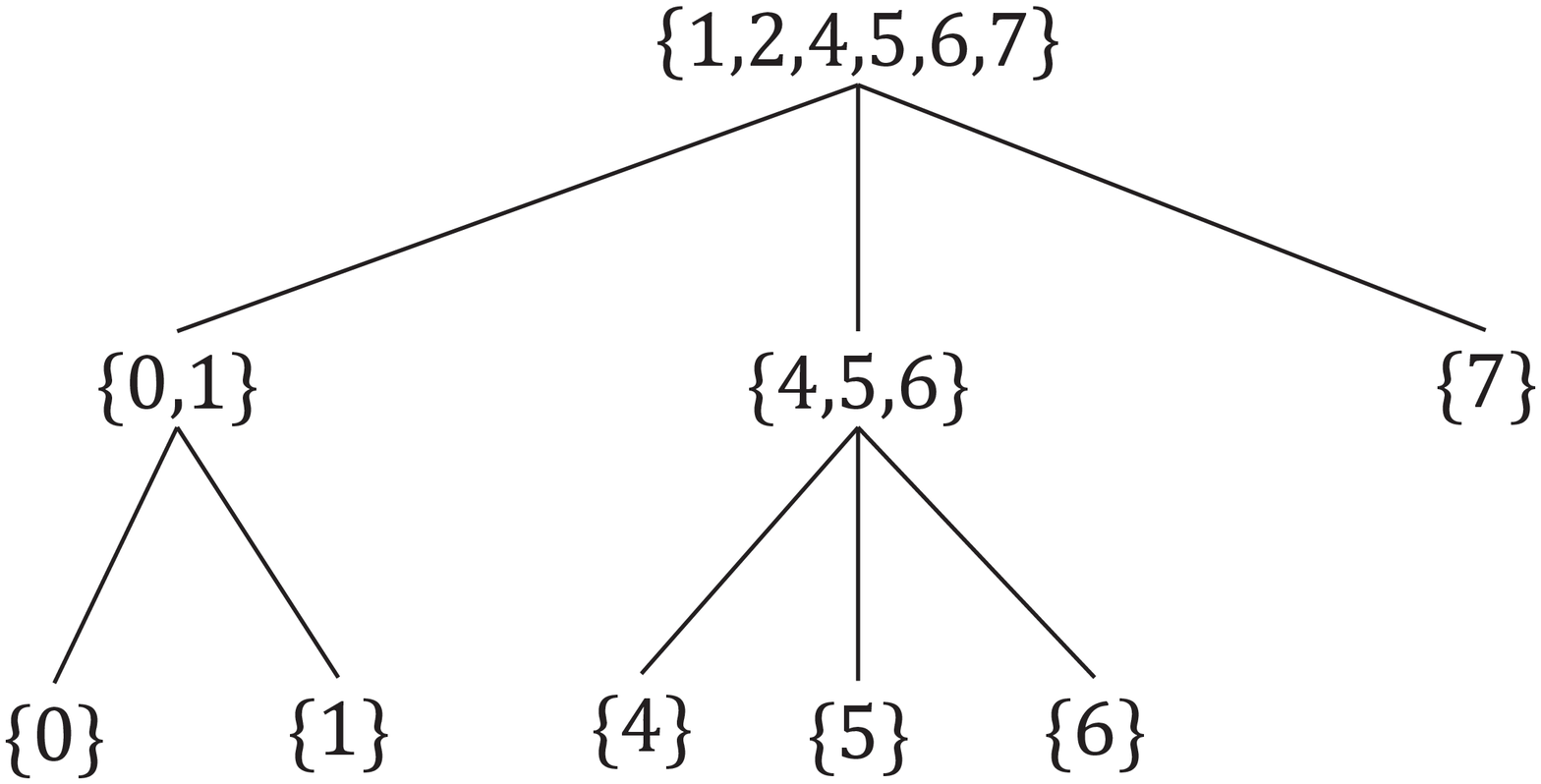}
   \label{fig:dectrees}
    }
\end{tabular}
\end{center}
\vspace*{-0.4cm}
\caption{Simplified graph and its AutoTree}
\vspace*{-0.6cm}
\label{fig:sgraph}
\end{figure}

\begin{figure*}[t]
\begin{center}
  \includegraphics[width=1.8\columnwidth,height=6.8cm]{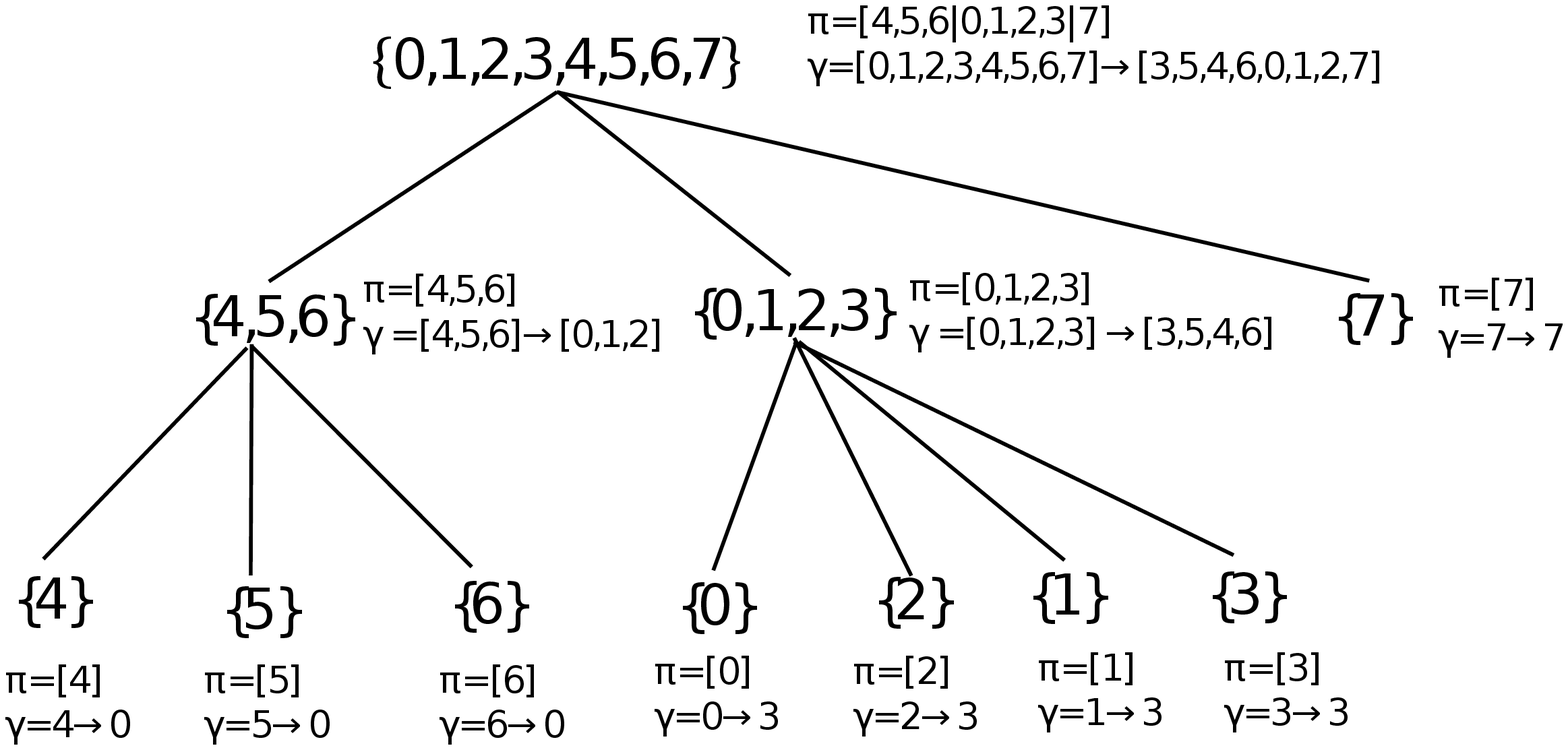}
\end{center}
\vspace{-0.6cm}
\caption{AutoTree constructed by our approach for the Graph
  in Fig.~\ref{fig:expgraph} based on the simplified graph and its
  AutoTree in Fig.~\ref{fig:sgraph}.}
\label{fig:dectree_se}
\vspace{-0.4cm}
\end{figure*}

Recall that two structural equivalent vertices must be automorphic equivalent. Such property can be applied to simplify $G$ and improve the performance of \CL. Specifically, vertices in $V$ are partitioned into a number of structural equivalent subsets. Vertices in each non-singleton subsets are replaced by an arbitrary vertex $v$ in the subset, and the graph $G$ and the ordered partition $\pi$ are simplified accordingly.
When constructing the AutoTree, the leaf node containing $v$ is extended either by a number of sibling leaf nodes, each containing a vertex in $s$, or by adding vertices in $s$ back to the subgraph representing the leaf node, retaining the structure of the given graph.

Fig.~\ref{fig:simgraph} and Fig.~\ref{fig:dectrees} illustrate the simplified graph $G_s$ of the example graph $G$ in Fig.~\ref{fig:expgraph} and its AutoTree, respectively. Note that the AutoTree in Fig.~\ref{fig:dectrees} contains the tree structure without any information about node certificates or orbit partitions on each tree node. In the example graph $G$, shown in Fig.~\ref{fig:expgraph}, there are two non-singleton structural equivalent subsets,  $\{0,2\}$ and $\{1,3\}$. Therefore, in the simplified graph $G_s$, shown in Fig.~\ref{fig:simgraph}, vertices 2 and 3 along with their adjacent edges  are removed, and the cell $\{0,1,2,3,4,5,6\}$ in $\pi$ is simplified as $\{0,1,4,5,6\}$ and partitioned into two subgraphs according to connectivity.  Based on the simplified graph and its decomposition, the AutoTree $G$ is constructed by extending leaf nodes containing vertices 0 and 1, shown in Fig.~\ref{fig:dectree_se}. It is worth noting that different approaches, or even different implementations, can generate different canonical representatives, while each approach, or implementation, will generate same canonical representation for isomorphic graphs. For instance, the certificates of the root nodes in Fig.~\ref{fig:dectree} and Fig.~\ref{fig:dectree_se} are different.
}

\comment{
From another viewpoint, \CL divides vertex set $V$ into four disjoint subsets $V=V_s \cup V_a \cup V_e \cup V_{ne}$.
\begin{itemize}
\item $V_s$: Vertices in $V_s$ cannot be automorphic equivalent to any other vertices, thus each vertex in $V_s$ consists a leaf node of the root. Vertices in $V \setminus V_s$, i.e.,$V_a \cup V_e \cup V_{ne}$, are grouped into disjoint subgraphs, each is a connected component in the subgraph induced by $V \setminus V_s$.
\item $V_a$: Vertices in $V_a$ are automorphic equivalent to some vertices, and the automorphic relations for each vertex in $V_a$ are all discovered. Each vertex in $V_a$ consists a leaf node in the AutoTree.
\item $V_{ne}$: Vertices in $V_{ne}$
\end{itemize}

\begin{itemize}
\item $V_s$: Vertices in $V_s$ cannot be automorphic equivalent to any other vertices, thus each vertex in $V_s$ consists a leaf node in the AutoTree.
    Vertices in $V \setminus V_s$, i.e.,$V_a \cup V_e \cup V_{ne}$, are grouped into disjoint subgraphs, each is a connected component in the subgraph induced by $V \setminus V_s$.
\item $V_a$: Vertices in $V_a$  are automorphic equivalent to some vertices, and the automorphic relations among $V_a$ are easy to determine. Similar to $V_s$, each vertex in $V_a$ consists a leaf node in the AutoTree.
\item $V_e$: vertices in $V_e$ are divided into several subsets, and none
\item $V_{ne}$: the labeling of vertices in $V_{ne}$ can be easily derived utilizing the labeling of vertices in $V_e$.
\end{itemize}
Detecting all symmetries among vertices in $V_e$, canonical labeling of the whole graph can be easily determined.
Since $|V_e| \ll |V|$ holds in most real-life massive graphs, our approach \CL succeeds in transferring the problem of canonical labeling for a massive graph to symmetry detection in several small subgraphs, improving the efficiency and scalability significantly. We discuss procedures proposed to divide $V$ into $V_s$, $V_a$, $V_{ne}$ and $V_e$ below.

\stitle{Partition}: Algorithm \Partition extracts a vertex subset $V_s \subset V$ by achieving an ordered partition $\pi$ for $V$.
\begin{displaymath}
\begin{aligned}
V &\rightarrow \pi=[V_1 | V_2 | \ldots | V_k] \\
(V, \pi) & \rightarrow V_s \cup \overline V_s
\end{aligned}
\end{displaymath}
Ordered partition $\pi$ is achieved by grouping vertices with the same signature into a cell. Here, vertex signature encodes the neighborhood structure for each vertex. As proved by Theorem~\ref{the:unique}, $\pi$ is equitable and only vertices in the same cell can probably be automorphic equivalent. $V_s$ contains vertices  in singleton cells and $\overline V_s = V \setminus V_s$ contains vertices in non-singleton cells. Easy to see, for any vertex $v \in V_s$, it cannot be automorphic equivalent to any other vertex $u \in V$.

\stitle{Preprocessing}: Procedure Preprocessing divides vertices $\overline V_s$ into a set of subgraphs $\mathcal S$=$\{S_1, S_2, \ldots, S_k\}$, and extracts a vertex subset $V_a$ when it simplifies each subgraph $S_i \in \mathcal S$.
\begin{displaymath}
\begin{aligned}
\overline V_s &\rightarrow {\mathcal S} ={\mathcal S}_1 \cup {\mathcal S}_2 \cup {\mathcal S}_3 \\
{\mathcal S}_1 &\rightarrow {\mathcal S}_1' \cup {\mathcal S}_1''  \\
{\mathcal S}_1'' \cup {\mathcal S}_3 &\rightarrow {\mathcal S} \\
{\mathcal S}_1' \cup {\mathcal S}_2 &\rightarrow V_a,  \quad
{\mathcal S}_1'' \cup {\mathcal S}_3 \rightarrow \overline V_a = \overline V_s \setminus V_a
\end{aligned}
\end{displaymath}
Here, each subgraph $S_i \in \mathcal S$ is a connected component of the subgraph induced by $\overline V_s$, and is associated with a subgraph signature encoding its vertex set. Subgraphs in $\mathcal S$ are divided into three subsets. ${\mathcal S}_1$ contains subgraphs with unique signature, each subgraph  $S$ in ${\mathcal S}_2$ has non-unique signature while all vertices in $S$ have distinct labels, and ${\mathcal S}_3$ contains the remaining subgraphs, i.e., each subgraph $S$ in ${\mathcal S}_3$ has non-unique signature while some vertices in $S$ have the same label. Procedure Preprocessing simplifies subgraphs in ${\mathcal S}_1$ and ${\mathcal S}_2$. For each subgraph $S \in {\mathcal S}_1$, Preprocessing recursively removes edges based on Theorem~\ref{the:subgraph_intra}, and all the resulting connected components consist two subsets, ${\mathcal S}_1'$ contains singleton components, where symmetries can be determined, and ${\mathcal S}_1''$ contains non-singleton components.  For subgraphs in ${\mathcal S}_2$, symmetries between vertices can be easily determined according to Theorem~\ref{the:subgraph_direct}.  Procedure Preprocessing returns ${\mathcal S}_1'' \cup {\mathcal S}_3$ as the new $\mathcal S$, and vertex subset $V_a$ consists vertices in ${\mathcal S}_1' \cup {\mathcal S}_2$.

\stitle{Condense}: Procedure Condense constructs a tree $T_i$ for each subgraph $S_i \in \mathcal S$. Each $T_i$ illustrates the symmetry hierarchy between vertices and subgraphs in $S_i$, and the tree set $\mathcal T$ divides $\overline V_a$ into two subsets $V_e$ and $V_{ne}$.
\begin{displaymath}
\begin{aligned}
S_i &\rightarrow T_i \\
{\mathcal T} &\rightarrow {\mathcal S}_{s} \cup {\mathcal S}_{ns} \\
{\mathcal S}_{ns} &\rightarrow V_e, \quad {\mathcal S}_{s} \rightarrow V_{ne}
\end{aligned}
\end{displaymath}
Here, each tree $T_i$ is constructed when Algorithm \Preprocessing recursively removes edges in $S_i$. Each node in $T_i$ is a subgraph of $S_i$ and leaf nodes of $T_i$ are pair-wise disjoint subsets whose union is $S_i$. For each tree $T_i \in \mathcal T$, leaf nodes containing only one vertex are maintained in ${\mathcal S}_s$ and the remaining, i.e., non-singleton leaf nodes, are maintained in ${\mathcal S}_{ns}$.
Vertices in ${\mathcal S}_{ns}$ consist vertex subset $V_e$ and vertices in ${\mathcal S}_{s}$ consist vertex subset $V_{ne}$.

\stitle{Symmetry Hierarchy Tree}: As can be seen, when our approach \CL applies the above mentioned procedures to find a canonical labeling for graph $G$, we also rearrange both vertices and subgraphs of $G$ in the form of a tree, representing the symmetry hierarchy in $G$, similar to each tree $T_i$ constructed for subgraph $S_i$ in Procedure Condense. We denote the tree as $T$ and call it a symmetry hierarchy tree of $G$. We clarify the structure of $T$ in detail. Each node of $T$ represents a subgraph of $G$, where the root node represents the whole graph $G$ and each leaf node represents either a vertex or a subgraph of $G$. It is worth noting that each vertex $v \in V$ exists in one and only one leaf node of $T$. For each non-leaf node $p$, its child nodes represent the connected components obtained when our approach \CL applies procedures to simplify the subgraph represented by node $p$, and subtrees that represent isomorphic subgraphs are placed together.
With symmetry hierarchy tree $T$, we can develop a surjective vertex labeling function $l$ such that $l(u)=l(v)$ iff $u \sim v$, and a depth-first traversal can easily assign a partial order among automorphic equivalent vertices, which results in the canonical labeling $C(G)$. With such vertex labeling function $l$, each node in $T$ can be assigned with a signature in the same manner mentioned in procedure Preprocessing, and two subgraphs are isomorphic if the two corresponding nodes have the same signature.
}

\begin{figure}[t]
\begin{center}
\begin{tabular}[t]{c}
\hspace{-0.6cm}
    \subfigure[simplified graph $G_s$]{
         \includegraphics[scale=0.32]{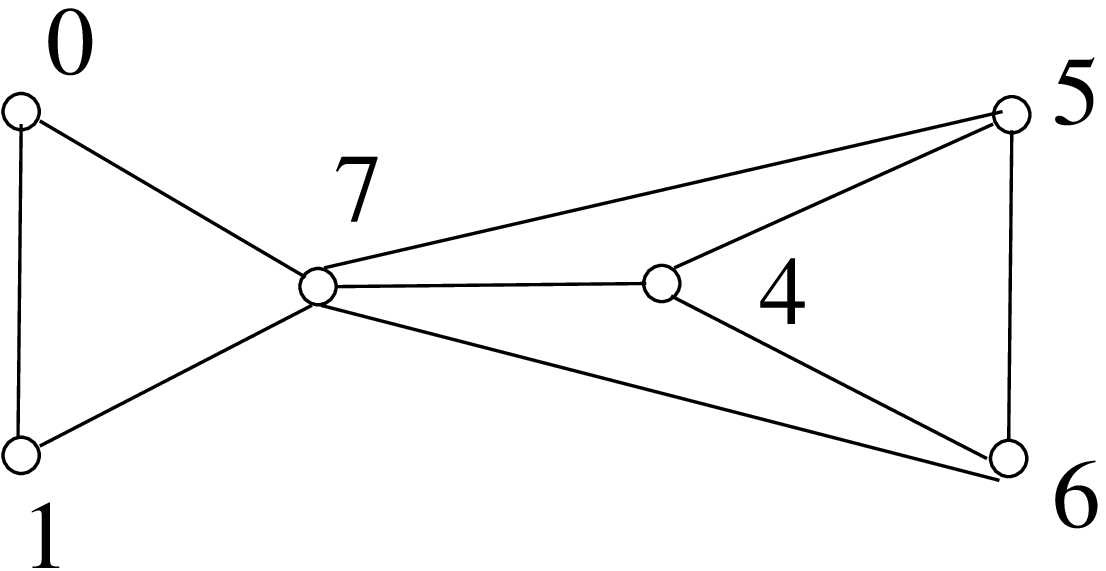}
        \label{fig:simgraph}
    }
    \hspace{0.3cm}
    \subfigure[AutoTree of $G_s$]{
    \includegraphics[scale=0.18]{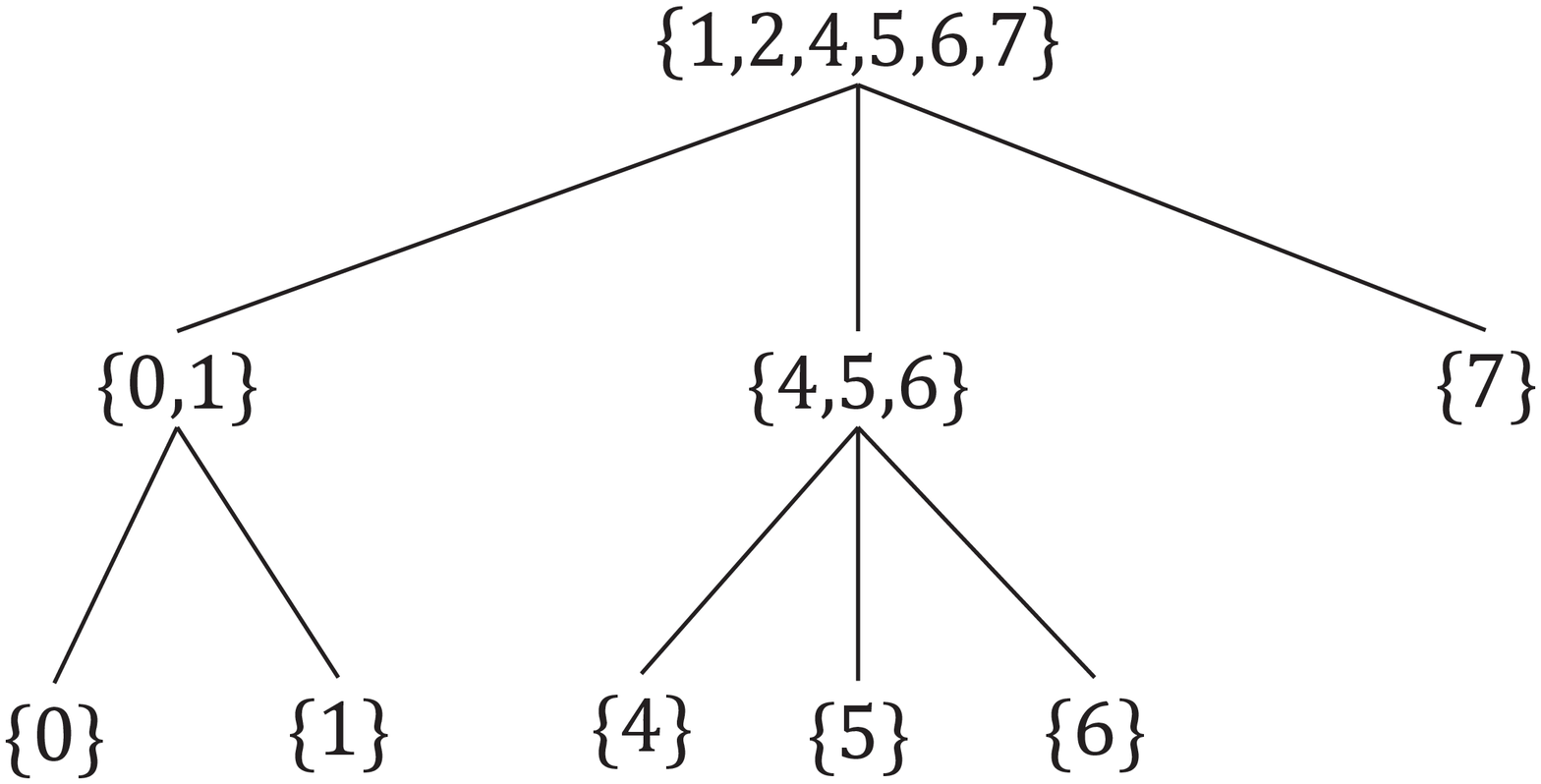}
   \label{fig:dectrees}
    }
\end{tabular}
\end{center}
\vspace*{-0.4cm}
\caption{Simplified graph and its AutoTree}
\vspace*{-0.4cm}
\label{fig:sgraph}
\end{figure}

\comment{
\begin{figure*}[t]
\begin{center}
  \includegraphics[width=1.8\columnwidth,height=6.8cm]{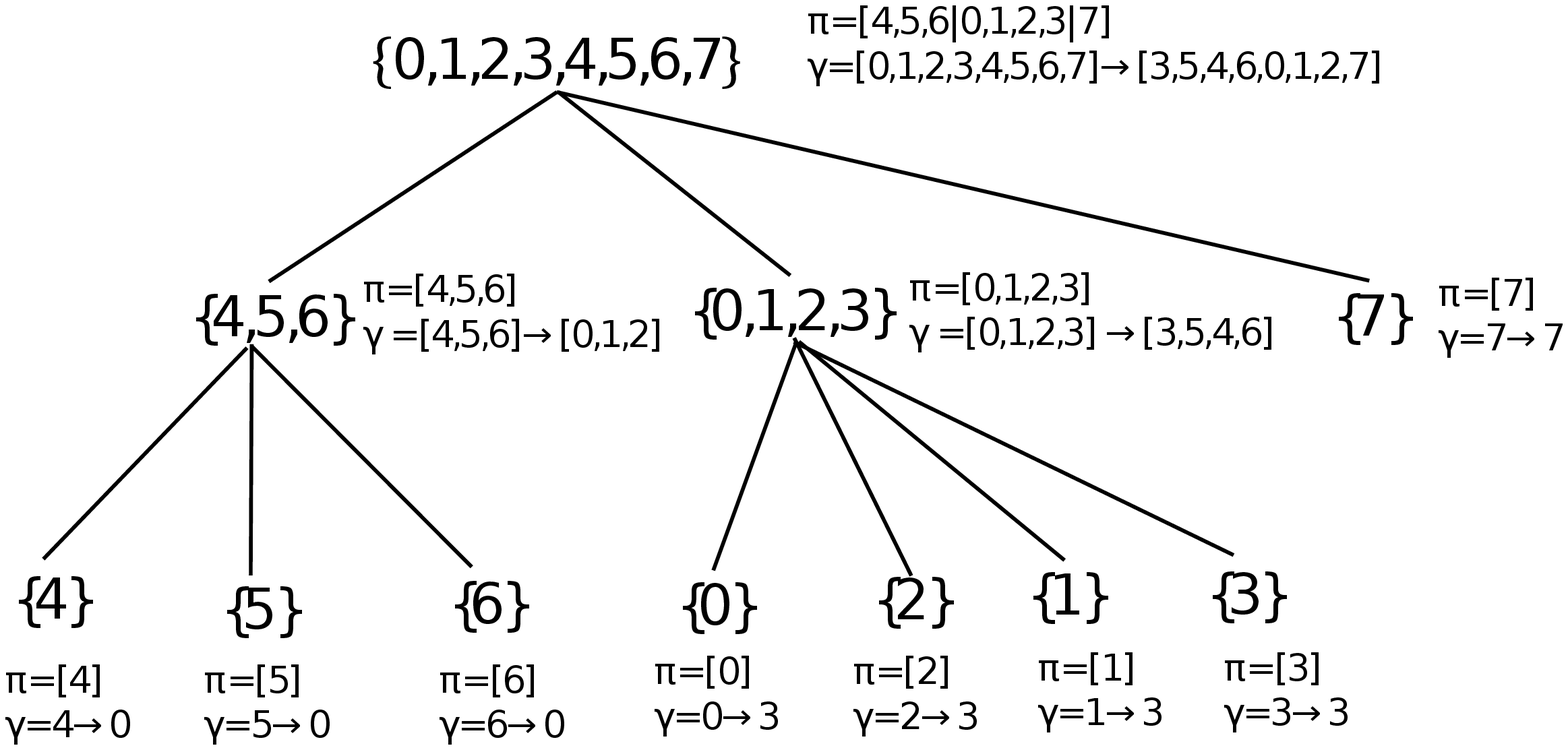}
\end{center}
\vspace{-0.6cm}
\caption{AutoTree constructed by \CL for $G$ in
  Fig.~\ref{fig:expgraph} based on Fig.~\ref{fig:sgraph}(a)(b).}
\label{fig:dectree_se}

\end{figure*}
}

\begin{figure}[t]
\begin{center}
  \includegraphics[width=1.2\columnwidth,height=5cm]{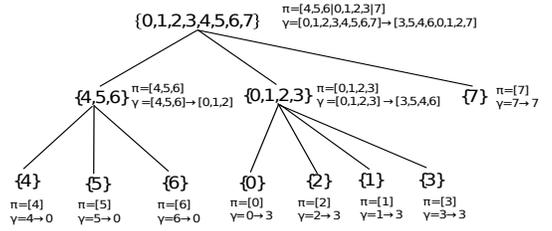}
\end{center}
\vspace{-1.8cm}
\caption{AutoTree by \CL for $G$ (Fig.~\ref{fig:expgraph}) on
  Fig.~\ref{fig:sgraph}(a)(b).}

\label{fig:dectree_se}
\vspace{-0.4cm}
\end{figure}

\section{The New Approach }
\label{sec:cl}

We give our \CL algorithm in Algorithm~\ref{alg:CL}. Given a colored
graph $(G, \pi)$, \CL constructs an AutoTree ${\mathcal
  {AT}}(G, \pi)$ for $(G, \pi)$ (Line~4). Note that the canonical
labeling  $C(G,\pi)$ at the root of
${\mathcal {AT}}(G, \pi)$ acts as the canonical labeling of the given
graph.  In \CL, the given coloring $\pi$ is refined to be
equitable by a refinement function $R$, for instance,
Weisfeiler-Lehman algorithm \cite{weisfeiler2006construction}, and is
further exploited to assign each vertex $v$ with color $\pi(v)$
(Line~1-2).  Then, \CL applies procedure \cl to construct AutoTree
${\mathcal {AT}}(G, \pi)$(Line~3).  We discuss Procedure \cl in
detail.  Procedure \cl constructs ${\mathcal {AT}}(g, \pi_g)$
rooted at $(g, \pi_g)$, for a colored subgraph $(g, \pi_g)
\subset (G, \pi)$ following the divide-and-conquer
paradigm. ${\mathcal {AT}}(g, \pi_g)$ is initialized with root node
$(g, \pi_g)$ (Line~6).  \cl divides $(g, \pi_g)$ into a set of
subgraphs $(g_i, \pi_{g_i})$, each consists of a child node of $(g,
\pi_g)$, utilizing Algorithm \DivideP (Algorithm~\ref{alg:DivideP})
and Algorithm \DivideS (Algorithm~\ref{alg:DivideS}) (Line~11-12).  \cl
recursively constructs subtrees ${\mathcal {AT}}(g_i, \pi_{g_i})$ rooted at
each $(g_i, \pi_{g_i})$ (Line~13-14) and identifies the canonical
labeling   $C(g, \pi_g)$ for the root node $(g, \pi_g)$
utilizing Algorithm
\CombineST (Algorithm~\ref{alg:CombineST}) (Line~15). The base cases
occur when either  $g$ contains a single vertex (Line~7-8) or
$(g,\pi_g)$ cannot be disconnected by  \DivideP or
\DivideS (Line~9-10).  For the former case, obtaining
$C(g, \pi_g)$ is trivial. For the latter case, $C(g,
\pi_g)$ can be achieved by applying Algorithm \CombineCL
(Algorithm~\ref{alg:CombineCL}), which exploits the canonical labeling $\gamma^*$ by existing algorithms like \bliss.

\comment{
In \CL, we apply Weisfeiler-Lehman algorithm  as the refinement function $R$. As proved by \cite{weisfeiler2006construction}, only vertices in the same cell in the equitable coloring $\pi$ can probably be automorphic equivalent.
We prove that $\pi_g$, the projection of $\pi$ on $V(g)$, inherits the properties of $\pi$. Specifically, (1) only vertices in the same cell in $\pi_g$ can probably be automorphic equivalent. (2) $\pi_g$ is equitable with respect to $g$. The first property can be proved trivially. We focus on the second property. We prove the claim based on the  mathematical induction. Assume $g$ is a connected  component in $g'$ that emerges due to either \DivideP or \DivideS, and $\pi_{g'}$ satisfies the second property.
In either case, the edges removed are those between two cells in $\pi_{g'}$. Then for any two vertices in the same cell in $\pi_g$, they either retain all neighbors or remove all neighbors in any cell in $\pi_g$, i.e., $\pi_g$ is equitable with respect to $g$.
}

\comment{
Given an equitable coloring $\pi$ achieved by Weisfeiler-Lehman algorithm, we give properties of $\pi_g$, which is the projection of  $\pi$ on a subgraph $g \subset G$.
\begin{lemma}
\label{lem:equitable}
$\pi_g$ is equitable and only vertices in the same cell in $\pi_g$ can probably be automorphic.
\end{lemma}

\proofsketch
First, as required by Weisfeiler-Lehman algorithm, $\pi$ is equitable and only vertices in the same cell in $\pi$ can probably be automorphic.
\eop
}


\subsection{Optimized by Structural Equivalence}

Recall that structural equivalent vertices must be automorphic
equivalent. Such property can be applied to simplify $(G,\pi)$ and improve
the performance of \CL. Specifically, vertices in $V$ are partitioned
into a number of structural equivalent subsets. Vertices in each
non-singleton subset $S$ are simplified by retaining only one vertex $v$ in the
subset, and the colored graph $(G,\pi)$ is simplified
accordingly.
When constructing AutoTree, the leaf node containing $v$ is extended
either by adding a number of sibling leaf nodes, each contains a
vertex in $S$ if the leaf node containing $v$ is singleton, or by
adding vertices in $S$ to the subgraph of the leaf node otherwise.

Fig.~\ref{fig:simgraph} and Fig.~\ref{fig:dectrees} illustrate the
simplified graph $G_s$ of the graph $G$ in Fig.~\ref{fig:expgraph} and
its AutoTree, respectively. For simplicity, AutoTree in
Fig.~\ref{fig:dectrees} contains the tree structure without any
information such as canonical labeling on each tree node. In the
example graph $G$, shown in Fig.~\ref{fig:expgraph}, there are two
non-singleton structural equivalent subsets, $\{0,2\}$ and
$\{1,3\}$. Therefore, in the simplified graph $G_s$, shown in
Fig.~\ref{fig:simgraph}, vertices 2 and 3 along with their adjacent
edges are removed.
Based on the simplified graph and its
AutoTree ${\mathcal {AT} (G_s, \pi_s)}$, the AutoTree ${\mathcal {AT}(G , \pi)}$ of $(G,\pi)$ is constructed by extending leaf nodes
containing vertices 0 and 1, shown in Fig.~\ref{fig:dectree_se}. It is worth noting that different approaches, or even
different implementations, can generate different canonical labeling,
while each approach, or implementation, will generate the same canonical
labeling for isomorphic graphs. For instance, the canonical labeling
of the root nodes in Fig.~\ref{fig:dectree} and
Fig.~\ref{fig:dectree_se} are different.


\comment{

We give our \CL (Canonical Labeling) algorithm in Algorithm~\ref{alg:cl}. In Algorithm~\ref{alg:cl}, Algorithm \Partition first achieves an ordered partition $\pi$ (Line~1), which divides $V$ into two subsets $V_s$, containing vertices in singleton cells, and $\overline V_s =V \setminus V_s$. Since only vertices in the same cell in $\pi$ can probably be automorphic equivalent, we concentrate on detecting symmetries among vertices in $\overline V_s$ in the following.
For simplicity, we assign each vertex $v$ with labeling $l(v)$ to simplify vertex signature $s(v)$ (Line~2-4). Here the mapping from $s(v)$, or equivalently the cell $V_i$ containing $v$, to $l(v)$ is bijective.
To detect symmetries in $\overline V_s$, Algorithm \CSG first divides vertices in $\overline V_s$ into a set of subgraphs $\mathcal S$ =$\{S_1, S_2, \ldots, S_k\}$, where each subgraph $S_i$ is a connected component of the subgraph induced by $\overline V_s$ (Line~5). Next, Algorithm \Preprocessing simplifies $\mathcal S$ by removing some specific edges that have little influence on the canonical labeling of $G$, and extracts a subset $V_a \subset \overline V_s$ s.t., the symmetries between vertices in $V_a$ can be easily determined (Line~6). For the resulting $\mathcal S$ returned by Algorithm \Preprocessing, none subgraph $S_i$ can be easily simplified and more sophisticated algorithms are required to detect symmetries both among vertices in $S_i$ and between $S_i$ and another subgraph $S_j$ sharing the same signature with $S_i$. To detect symmetries among vertices in each subgraph $S_i \in \mathcal S$, Algorithm \Condense deals with $ \mathcal S$, constructs a tree $T_i$ representing the symmetry hierarchy of vertices for each  $S_i$ and extracts the most essential subgraphs in each $S_i$ for canonical labeling, consisting ${\mathcal S}_{ns}$ (Line~7).
To detect symmetries between two subgraphs, Algorithm \CanonicalLabeling first generates canonical labeling for each subgraph $S_i \in {\mathcal S}_{ns}$ (Line~8-10), and the labeling for ${\mathcal S}_{ns}$ along with tree $T_i$  derives canonical labeling for each subgraph $S_i \in \mathcal S$, applying Algorithm \Labeling (Line~11-13). It is worth noting that Algorithm \CanonicalLabeling can be any previous approaches proposed for canonical labeling problem, for instance, naucy, saucy, Bliss, conauto and Traces. Until now, each subgraph in $ \mathcal S$ is associated with a canonical labeling and two subgraphs are symmetric in graph $G$ iff they have the same canonical labeling. To obtain the canonical labeling for $G$, Algorithm \SymmetryAndLabeling finds all symmetries between subgraphs in $\mathcal S$ and assigns each vertex $v$ with a unique label $l(v)$ (Line~14). Algorithm \CL defines the permutation $\gamma$ (Line~15-17) and returns canonical labeling $G^\gamma$ (Line~18).
}

\comment{
We give our \CL (Canonical Labeling) algorithm in Algorithm~\ref{alg:cl}. In Algorithm~\ref{alg:cl}, we first achieve an ordered partition $\pi$ using an iterative algorithm \Partition (Line~1). In Algorithm \Partition, every vertex is associate with a series of hash functions, where the $i$-th function maps the vertex to a signature, representing the most distinctive information in its $i$-hop neighborhood. Sorting these signatures, \Partition iteratively refines an ordered partition. We will prove that the resulting partition $\pi$ is equitable and only vertices in the same cell can probably be automorphic equivalent, i.e., $\pi$ divides $V$ into $V_s$ consists of vertices in singleton cells and the remaining consist $V_e \cup V_{ne}$.
With partition $\pi$, each vertex $v$ is initialized with a labeling $l(v)$ (Line~2-4). Here, labeling $l$ is an approximation of the permutation $\gamma$ in the canonical labeling, and for any vertex $v$ in a singleton cell, its labeling keeps unchanged, i.e., $v^\gamma=l(v)$. In the following, \CL focuses on vertices in non-singleton cells and detects automorphism relationships between these vertices.
Second, vertices in non-singleton cells are divided into a set of connected subgraphs $\mathcal S$ =$\{S_1, S_2, \ldots, S_k\}$, applying BFS from these vertices while ignoring vertices in singleton cells (Line~5). It is worth noting that each subgraph $S_i$ is associated with a signature consists of $l(v)$ in non-descending order of $l$ for all $v \in S_i$, i.e., $s(S_i)=<l(v_1^i), l(v_2^i), \ldots, l(v_k^i)>$, here $l(v_1^i) \leq l(v_2^i) \leq \ldots \leq l(v_k^i)$. With subgraph signature, Algorithm \Preprocessing detects a number of symmetries between subgraphs in $\mathcal S$ directly(Line~6). For instance, if there are two subgraphs $S_p$ and $S_q$ satisfying $s(S_p)= s(S_q)$ and $l(v_1^p) < l(v_2^p) < \ldots <l(v_k^p)$, then $v_1^p \sim v_1^q$, $v_2^p \sim v_2^q$, \ldots, $v_k^p \sim v_k^q$. To further reduce search space, Algorithm \Condense

}

\comment{
\begin{algorithm}[t]
\caption{\CL($G$) }
\label{alg:cl}
\begin{algorithmic}[1]
\STATE $\pi \leftarrow $\Partition($G$);
\FOR {every vertex $v \in V$}
\STATE $l(v) \leftarrow \sum_{0 < k <i} |V_k|$, here $v \in V_i$;
\ENDFOR
\STATE $\mathcal S \leftarrow $\CSG($G, \pi$);
\STATE $\mathcal S \leftarrow $\Preprocessing($\mathcal S$);
\STATE $({\mathcal S}_{ns}, \mathcal T) \leftarrow $\Condense($G, \mathcal S$);
\FOR {every subgraph $S_i \in {\mathcal S}_{ns}$}
\STATE \CanonicalLabeling($S_i$);
\ENDFOR
\FOR {every subgraph $S_i \in \mathcal S$}
\STATE \Labeling($ S_i, T_i, {\mathcal S}_{ns}$);
\ENDFOR
\STATE \SymmetryAndLabeling ($\mathcal S$);
\FOR {every vertex $v \in V$}
\STATE $v^\gamma=l(v)$;
\ENDFOR
\RETURN $G^\gamma$;
\end{algorithmic}
\end{algorithm}
}

\begin{algorithm}[t]
\scriptsize
\caption{\CL($G,\pi$) }
\label{alg:CL}
$\pi=[ V_1 | V_2 | \ldots | V_k ] \leftarrow $\Partition($G,\pi$)\;
$\pi(v) \leftarrow \sum_{0<j<i} |V_j|$ for each $v \in V$\;
${\mathcal {AT}}(G, \pi) \leftarrow \cl(G, \pi)$\;
{\bf return} ${\mathcal {AT}}(G, \pi)$\;

\vspace*{0.2cm}

{\bf Procedure} {\cl($g,\pi_g$) } \\
initialize ${\mathcal {AT}}(g, \pi_g)$ with root $(g, \pi_g)$\;
\If{$g=\{v\}$}{
    $v^{\gamma_g} \leftarrow \pi(v)$, $C(g,\pi_g) \leftarrow ( v^{\gamma_g}, v^{\gamma_g})$; return ${\mathcal {AT}}(g, \pi_g)$\;
}
\If{neither \DivideP nor \DivideS can disconnect $(g, \pi_g)$}{
  $ C(g,\pi_g) \leftarrow \CombineCL(g,\pi_g)$;
  {\bf return} ${\mathcal {AT}}(g, \pi_g)$\;
}
$\bigcup_{1\leq i\leq k}(g_i, \pi_{g_i}) \leftarrow \DivideP(g, \pi_g) ( \DivideS(g, \pi_g)$)\;
construct tree edges $((g,\pi_g), (g_i, \pi_{g_i}))$ for all $i$\;
\For{$i$ from 1 to $k$}{
  ${\mathcal AT}(g_i, \pi_{g_i}) \leftarrow \cl(g_i,\pi_{g_i})$;
}
 $ C(g,\pi_g) \leftarrow \CombineST(g,\pi_g)$\;
{\bf return} ${\mathcal {AT}}(g, \pi_g)$;
\end{algorithm}
\setlength{\textfloatsep}{2pt}

\comment{
\subsection{\Partition}
\label{sec:partition}

\begin{algorithm}[t]
\caption{\Partition($G$) }
\label{alg:Partition}
\begin{algorithmic}[1]
\STATE Initialize vertex signature $s(v) \leftarrow d(v)$ for each $v \in V$;
\STATE sort vertices in $V$ in non-descending order of $s(v)$;
\STATE $\pi \leftarrow$ an ordered partition of $V$ where each cell contains vertices with the same $s(v)$ value;
\FOR {every vertex $v \in V$}
\STATE $i(v) \leftarrow \sum_{0 < k <i} |V_k|$, here $v \in V_i$;
\ENDFOR
\STATE $\pi' \leftarrow [ V ]$;
\WHILE { $\pi \neq \pi'$}
\STATE $\pi' \leftarrow \pi$;
\FOR {every non-singleton cell $V_i \in \pi'$}
\STATE $s(v) \leftarrow ( i(v_1), i(v_2), \ldots , i(v_{d(v)}) )$ for every $v \in V_i$;
\STATE replace $V_i$ with an ordered partition $\pi_i$;
\ENDFOR
\STATE refine $\pi$ by replacing each $V_i$ by $\pi_i$;
\STATE refine $i(v)$ using new $\pi$;
\ENDWHILE
\RETURN $\pi$;
\end{algorithmic}
\end{algorithm}

Algorithm \Partition (Algorithm~\ref{alg:Partition}) takes a graph $G$ as input and returns an ordered partition $\pi$ of vertex set $V$ (Line~17). In Algorithm \Partition, each vertex $v$ is associated with a signature $s(v)$, representing its neighborhood structure from local to global, and vertices with the same signature consist a cell in $\pi$. First, $s(v)$ for each vertex $v$ is initialized as degree $d(v)$ and the initial $\pi$ groups vertices with the same degree into a cell and all cells are sorted in non-descending order of degrees (Line~1-3). Next, Algorithm \Partition iteratively refines $\pi$, distinguishing vertices in the same non-singleton cell by more sophisticated vertex signatures (Line~8-16). Specifically, in the $i$-th iteration, for any vertex $v$ in a non-singleton cell, $s(v)$ encodes the most distinctive information in its $i$-hop neighborhood. Here, each $s(v)$ in the $i$-th iteration is composed by concatenating $s(u)$ in the $(i-1)$-th iterations for $u \in N(v)$ in non-descending order of $s(u)$ (Line~10-13). It is worth noting that, if a vertex $v$ is partitioned into a singleton cell in some iteration, its signature $s(v)$ will keep unchanged in later iterations.
Since for any two vertices $u$ and $v$, Algorithm \Partition only cares about whether $s(u)$ equals $s(v)$, we design an index $i(v)$ for each vertex $v$ such that the mapping from signature $s(v)$ (or equivalently cell $V_i$) to index $i(v)$ is bijective (Line~4-6). Therefore, index $i(v)$, which is a scalar, can be utilized to replace signature $s(u)$, which is a vector, without introducing any errors, simplifying the representation of $s(v)$ and improving the efficiency of Algorithm \Partition. The while loop terminates when $\pi$ cannot be further refined.
We give the properties of the ordered partition $\pi$ achieved by Algorithm \Partition in Theorem~\ref{the:unique}.

\begin{theorem}
\label{the:unique}
The ordered partition $\pi$ returned by Algorithm \Partition is equitable and only vertices in the same cell can probably be automorphic equivalent.
\end{theorem}

\proofsketch
We first prove $\pi$ is equitable. Consider partitions $\pi$ (Line~14) and $\pi'$ (Line~9) in the last iteration that terminate the while loop. Then $\pi = \pi'$, which means that in the last iteration, vertices in every non-singleton cell $V_i \in \pi'$ have the same signature. Since the mapping from cell $V_i$ to $i(v)$ is bijective for
each vertex $v \in V_i$. Two vertices with the same signature must have the same number of neighbors in each cell $V_i$, i.e., $\pi$ is equitable.

Next, we consider two automorphic equivalent vertices, denoted as $u$ and $w$, and prove that they must be in the same cell in each $\pi$ obtained in each iteration in Algorithm \Partition. For simplicity, we denote the neighbor set of $u$ as $N(u)=\{ v_1^u, v_2^u, \ldots, v_k^u \}$ and the neighbor set of $w$ as $N(w)=\{v_1^w, v_2^w, \ldots, v_k^w \}$, here $k=d(u)=d(w)$. We prove the claim by mathematical induction on each $\pi$ obtained.
\begin{enumerate}
\item For the first $\pi$, i.e., the initialized $\pi$ (Line~1-3), vertices with the same degree are partitioned into the same cell. Since $d(u)= d(w)$, $u$ and $w$ are in the same cell.
\item Assume $u$ and $w$ are in the same cell $V_j$ in the $i$-th $\pi$. In order to prove $u$ and $w$ are in the same cell in the $(i+1)$-th $\pi$, it is equivalent to prove signatures $s(u)=s(w)$ when refining $V_j$. Since $u \sim w$, there is a coordination between $N(u)$ and $N(w)$ s.t. $v_1^u \sim v_1^w$, $v_2^u \sim v_2^w$,
    $\ldots$, $v_k^u \sim v_k^w$. According to assumption, $v_r^u$ and
  $v_r^w$ are in the same cell in the $i$-th $\pi$, indicating
  $i(v_r^u) = i(v_r^w)$ for $1 \leq r \leq k$. Therefore,
  $s(u)=(i(v_1^u), i(v_2^u), \ldots , i(v_k^u)) = (i(v_1^w), i(v_2^w),
  \ldots , i(v_k^w) ) =s(w)$, i.e., $u$ and $v$ are in the same cell
  in  the $(i+1)$-th $\pi$. \eop
\end{enumerate}

\stitle{Time complexity of Algorithm \Partition}: Easy to see, the most time-consuming component in Algorithm~\ref{alg:Partition} is the while loop (Line~8-16).
In each iteration, it costs $\sum d_i ln d_i$, which is bounded by
$mlnm$, to encode vertices' signatures and costs $\sum |V_i|ln |V_i|$,
bounded by $nlnn$, to sort vertices in non-singleton cells. Therefore,
the time complexity of each iteration is $O(mlnm)$. For the number of
iterations, the while loop terminates when no cell can be further
refined. Since each non-singleton cell $V_i$ can be refined at most
$|V_i|-1$ times, the number of iterations is bounded by $n$. In
conclusion, the time complexity of  Algorithm \Partition is
$O(nmlnm)$.

\comment{
According to Theorem~\ref{the:unique}, for each vertex $v\in V$, Algorithm \Partition extracts a set of candidate vertices that can probably be automorphic equivalent to $v$. For any vertex in a singleton cell, its candidate set is empty and determining its label is trivial. We denote vertices in singleton cells as a subset $V_s \subset V$ and concentrating on finding symmetries between vertices in $V \setminus V_s$ in the following. Note that we can also extract a subset, where all vertices in the subset are structural equivalent, in each non-singleton cell and replace the subset using a representative vertex in it.
}

\subsection{\cl}

\comment{
\begin{algorithm}[t]
\caption{\cl($g$, $\pi$) }
\label{alg:cl}
\begin{algorithmic}[1]
\IF {$g$ contains only one vertex $v$}
\STATE $c(g) \leftarrow i(v)$;
\RETURN;
\ENDIF
\STATE ${\mathcal S}^p=\{ S_1^p, S_2^p, \ldots, S_k^p\} \leftarrow $ \DivideP($g, \pi$);
\IF {$|{\mathcal S}^p|=1$}
\STATE ${\mathcal S}^s=\{ S_1^s, S_2^s, \ldots, S_k^s\} \leftarrow $ \DivideS($g, \pi$);
\IF{ $|{\mathcal S}^s|=1$}
\STATE generate $g$'s canonical labeling $C(g)$, canonical labeling $\gamma$ and orbit partition $\pi_{orbit}$;
\STATE $c(g) \leftarrow \CombineCL(C(g), \gamma,\pi_{orbit})$;
\RETURN;
\ENDIF
\FOR {each subgraph $S_i^s \in {\mathcal S}^s$}
\STATE construct a tree node $S_i^s$ as a child node of $g$;
\STATE \cl($S_i^s, \pi$);
\ENDFOR
\STATE sort child nodes of $g$ and $c(g) \leftarrow $ \CombineST($g$);
\RETURN ;
\ENDIF
\FOR {each subgraph $S_i^p \in {\mathcal S}^p$}
\STATE construct a tree node $S_i^p$ as a child node of $g$;
\IF {$S_i^p$ contain only one vertex}
\STATE \cl($S_i^p, \pi$);
\ELSE
\STATE ${\mathcal S}^s=\{ S_1^s, S_2^s, \ldots, S_k^s\} \leftarrow $ \DivideS($S_i^p, \pi$);
\FOR {each subgraph $S_i^s \in {\mathcal S}^s$}
\STATE construct a tree node $S_i^s$ as a child node of $S_i^s$;
\STATE \cl($S_i^s, \pi$);
\ENDFOR
\STATE sort child nodes of $S_i^p$ and  $c(S_i^p) \leftarrow $\CombineST($S_i^p$);
\ENDIF
\ENDFOR
\STATE sort child nodes of $g$ and  $c(g)\leftarrow$ \CombineST($g$);
\RETURN ;
\end{algorithmic}
\end{algorithm}

}

We introduce Algorithm \cl, shown in Algorithm~\ref{alg:cl}. Algorithm \cl follows the ``divide and conquer'' paradigm, it constructs the subtree rooted at $g$ and generates the certificate $c(g)$ for node $g$ in the AutoTree.
The base cases occur either when $g$ contains only one vertex $v$ (Line~1-4) or when the vertices in $g$ cannot be easily distinguished from each other, i.e., Algorithms \DivideP and \DivideS cannot divide $g$ into $\geq 2$  subgraphs (Line~6-12). In both cases, node $g$ is a leaf node in AutoTree and \cl generates the certificate $c(g)$ for $g$ directly. For the former case, $c(g)$ is set as $i(v)$ (Line~2), recall $i(v)$ is the index of vertex $v$ by partition $\pi$. For the latter case, $c(g)$ is based on the canonical representative $C(g)$, the  canonical labeling $\gamma$ and the orbit partition $\pi_{orbit}$ obtained by existing approaches, such as \nauty-based approaches (Line~9-10).
Otherwise, $g$ is divided into a number of subgraphs by Algorithms \DivideP and \DivideS (Line~5, Line~7, Line~25). Algorithm \cl recursively constructs subtrees rooted at each subgraph and generates certificates for each subgraph (Line~13-16, Line~20-32). To construct the subtree rooted at $g$, child nodes of $g$ are sorted by their certificates and $c(g)$, certificate of $g$, is achieved by Algorithm \CombineST (Line~16, Line~32), combining $c(s)$ for each subgraph $s$ of $g$ based on the structure of $g$.
}

\subsection{\DivideP and \DivideS}

We show \DivideP and \DivideS in Algorithm~\ref{alg:DivideP}
and Algorithm~\ref{alg:DivideS}.  Both algorithms take a
colored graph $(g, \pi_g)$ as input and attempt to divide $(g, \pi_g)$
into a set of subgraphs $(g_i, \pi_{g_i})$.  \DivideP isolates each
singleton cell $\{v_{s_i}\}$ in $\pi_g$ as a colored subgraph
$(v_{s_i}, [v_{s_i}])$ of $(g, \pi_g)$ (Line~2-3). Each connected
component $g_i$ due to the isolation results in a colored subgraph
$(g_i, \pi_{g_i})$ of $(g, \pi_g)$ (Line~4-5). On the other hand,
\DivideS divides $(g, \pi_g)$ based on Theorem~\ref{the:simplify}
(Line~1-6). Similar to \DivideP, each connected component $g_i$
results in a colored subgraph $(g_i, \pi_{g_i})$ of $(g, \pi_g)$
(Line~8-9).

We first discuss properties of refinement function \Partition.
In \CL, we apply Weisfeiler-Lehman algorithm \cite{weisfeiler2006construction} as the refinement function $R$. As proved by \cite{weisfeiler2006construction}, only vertices in the same cell in the resulting equitable coloring $\pi$ can probably be automorphic equivalent. In \CL, only the coloring $\pi$ for $G$ is achieved by the refinement function $R$, all the other colorings, i.e., $\pi_g$ for subgraphs $g$, are obtained by projecting $\pi$ on $V(g)$. The following theorem proves the equivalence between projecting $\pi$ on $V(g)$ and applying $R$ on $(g, \pi_g)$.

\begin{theorem}
\label{the:reuse}
$\pi_g$, the projection of $\pi$ on $V(g)$ by \DivideP and \DivideS, inherits the properties of $\pi$. Specifically, (1) only vertices in the same cell in $\pi_g$ can  be automorphic equivalent. (2) $\pi_g$ is equitable with respect to $g$.
\end{theorem}

\proofsketch The first property can be proved trivially. We focus on
the second property, and prove the claim based on the mathematical
induction. Assume $g$ is a connected component in $g'$ that emerges
due to either \DivideP or \DivideS, and $\pi_{g'}$ satisfies the
second property.  In either case, the edges removed are those between
two cells in $\pi_{g'}$. Then for any two vertices in the same cell in
$\pi_g$, they either retain all neighbors or remove all neighbors in
any cell in $\pi_{g'}$, i.e., $\pi_g$ is equitable with respect to $g$.
 \eop

\begin{lemma}
\label{lem:clique}
Given a graph $(g,\pi_g)$. For any cell $V_i \in \pi_g$, if the subgraph induced by $V_i$ is a clique, removing edges among vertices in $V_i$, i.e., $E_i=\{(u,v)| u, v \in V_i, u\neq v\} \cap E(g)$, will not influence the automorphism group of $(g,\pi_g)$.
\end{lemma}

\proofsketch: Let $g'$ denote the graph
after removing edges $E_i$ from $g$, and $Aut(g',\pi_{g'})$ denote its
automorphism group.
By Theorem~\ref{the:reuse}, $\pi_{g'}=\pi_g$. For simplicity, we will use $\pi_g$ for $\pi_{g'}$ below.

Consider automorphisms $\gamma \in Aut(g, \pi_g)$,
$\gamma' \in Aut(g', \pi_g)$.
We prove $\gamma \in Aut(g', \pi_g)$ and $\gamma' \in Aut(g, \pi_g)$, respectively.
We prove $\gamma' \in Aut(g, \pi_g)$, and $\gamma \in Aut(g',\pi_g)$ can be proved in the similar manner.
Consider  $v \in V_i$. Since $v$ and $v^{\gamma'}$ are automorphic,
$v$ and $v^{\gamma'}$ must be in the same cell in $\pi_g$, i.e., $v^{\gamma'} \in
V_i$, implying that $V_i^{\gamma'}=V_i$. As a consequence, for any
edge $(u,v) \in E_i$, $(u^{\gamma'}, v^{\gamma'}) \in E_i$. Therefore,
$(g, \pi_g)^{\gamma'} = (g,\pi_g)$, i.e., $\gamma' \in Aut(g, \pi_g)$.  \eop

\begin{lemma}
\label{lem:biclique}
Given a colored graph $(g,\pi_g)$. For any two cells $V_i$ and $V_j$ in $\pi_g$,
let $E_{ij}$ denotes the edges between $V_i$ and $V_j$, i.e., $E_{ij}=\{(u,v)| u \in V_i, v \in V_j\} \cap E(g)$.
If the subgraph $(V_i\cup V_j, E_{ij})$ is a complete bipartite graph, removing edges $E_{ij}$ will not influence the automorphism group of $(g,\pi_g)$.
\end{lemma}

\proofsketch
The proof is similar to that of Lemma~\ref{lem:clique}.
\eop

Note that \DivideP is a special case of Lemma~\ref{lem:biclique}.

\begin{theorem}
\label{the:simplify}
Given a colored graph $(g,\pi_g)$, applying \DivideP and \DivideS on  $(g,\pi_g)$ retains the automorphism group of $(g,\pi_g)$. In other words, removing the following two classes
of edges will not influence   $Aut(g,\pi_g)$:
(1) edges among vertices in $V_i$, i.e., $E_i=\{(u,v)| u, v \in V_i,
u\neq v\} \cap E(g)$, if the subgraph induced by $V_i$ is a clique.
(2) edges between $V_i$ and $V_j$, i.e., $E_{ij}=\{(u,v)| u \in V_i, v
\in V_j\} \cap E(g)$, if the subgraph $(V_i\cup V_j, E_{ij})$ is a complete
bipartite graph.
\end{theorem}

\proofsketch
It can be proved by Lemma~\ref{lem:clique}, Lemma~\ref{lem:biclique}.

\begin{lemma}
\label{lem:tree_node}
Given two isomorphic graphs $(G , \pi )$ and $(G', \pi')$, if they are simplified by either \DivideP or \DivideS, then the remaining graphs are isomorphic. Specifically, each remaining graph can be partitioned into a subgraph set, i.e., $\{(g_i, \pi_{g_i})\}$ for $(G, \pi)$ and $\{(g'_i, \pi'_{g_i})\}$ for $(G', \pi')$, and the two subgraph sets
can be sorted such that $(g_i, \pi_{g_i}) \cong (g'_i, \pi'_{g_i})$.
\end{lemma}

\proofsketch Similar to the proof of automorphism retainment, i.e.,
Lemma~\ref{lem:clique} and Lemma~\ref{lem:biclique}, we prove that
each edge set removal will retain any isomorphism between $(G, \pi )$
and $(G', \pi')$. Without loss of generality, we prove the case when removing
edges $E_i$ from $(G,\pi)$ and removing $E'_i$ from $(G',\pi')$
simultaneously.
Here, $E_i$ and $E'_i$ are the same as defined in Lemma~\ref{lem:clique}, and  vertices in the corresponding cells $V_i$ and $V'_i$ have the same color.
Such property can be easily extended to prove Lemma~\ref{lem:tree_node}.

Let $(g,\pi_g)$ and $(g', \pi_{g'})$ denote the remaining graph after
removing $E_i$ from $(G, \pi)$ and $E'_i$ on $(G, \pi)$ and $(G',
\pi')$, respectively.
Here, $\pi_g = \pi$ and $\pi_{g'} =\pi'$ by Theorem~\ref{the:reuse}.
Denote $\gamma$ as an arbitrary isomorphism
between $(G,\pi)$ and $(G', \pi')$, i.e., $(G,\pi)^\gamma=(G', \pi')$.
We prove that $(g, \pi )^\gamma= (g', \pi')$.  First, by
isomorphism, we have $V_i^\gamma=V'_i$. Since both $V_i$ and $V'_i$
induce complete subgraphs, for any edge $(u,v) \in E_i$, $(u,v)^\gamma
\in E'_i$. As a consequence, $E_i^\gamma=E'_i$.  Second, since $(g,
\pi )^\gamma =  ((G^\gamma \setminus E_i^\gamma , \pi')$, $(g', \pi_{g'}) = ((G'\setminus E'_i , \pi')$ we have $(g, \pi )^\gamma = (g', \pi')$.
 \eop

\begin{algorithm}[t]
\scriptsize
\caption{\DivideP($g, \pi_g$) }
\label{alg:DivideP}
${\mathcal S} \leftarrow \emptyset$\;
\For {each singleton cell $\{v_{s_i}\}$ in $\pi_g$}{
${\mathcal S} \leftarrow {\mathcal S} \cup \{(v_{s_i}, [v_{s_i}])\}$;
$g \leftarrow g \setminus v_{s_i}$\;
}
\For {each connected component $g_i$ in $g$}{
 ${\mathcal S} \leftarrow {\mathcal S} \cup \{(g_i, \pi_{g_i})\}$\;
}
{\bf return} ${\mathcal S}$;
\end{algorithm}
\setlength{\textfloatsep}{2pt}

\begin{algorithm}[t]
\scriptsize
\caption{\DivideS($g, \pi_g$) }
\label{alg:DivideS}
\For {each cell $V_i \in \pi_g$}{
\If {$V_i$ induces a clique}{
 remove all edges between vertices in $V_i$\;
}
}
\For {any two distinct cells $V_i$ and $V_j$ in $\pi_g$}{
\If{edges between $V_i$ and $V_j$ consist a complete bipartite graph}{
 remove all edges between $V_i$ and $V_j$\;
}
}
${\mathcal S} \leftarrow \emptyset$\;
\For {each connected component $g_i$ in $g$ after removing edges}{
 ${\mathcal S} \leftarrow {\mathcal S} \cup \{ (g_i, \pi_{g_i})\}$\;
}
{\bf return} ${\mathcal S}$;
\end{algorithm}
\setlength{\textfloatsep}{2pt}

\comment{
\proofsketch
We make the following denotations.
\begin{small}
\begin{itemize}
\item $\gamma$: an isomorphism between $(G,\pi)$ and $(G',\pi')$, i.e., $(G,\pi)^\gamma=(G', \pi')$ \\
\item $(g,\pi_g)$: remaining graph applying \DivideP (\DivideS) on $(G, \pi)$; \\
\item $(g', \pi_{g'})$: remaining graph applying \DivideP (\DivideS) on $(G', \pi')$; \\
\item $(G^*,\pi^*)$: a graph consists of $(G, \pi)$ and $(G', \pi')$; \\
\item $(g^*, \pi_{g^*})$: a graph consists of $(g,\pi_g)$ and $(g', \pi_{g'})$;\\
\end{itemize}
\end{small}
\comment{
Let $\gamma$ be one isomorphism between $(G,\pi)$ and $(G',\pi')$, i.e., $(G,\pi)^\gamma=(G', \pi')$.  Denote $(g,\pi_g)$ as the remaining graph applying \DivideP (\DivideS) on $(G, \pi)$, and $(g', \pi_{g'})$ for $(G', \pi')$. Let $(G^*,\pi^*)$ denote the graph consists of $(G, \pi)$ and $(G', \pi')$, and $(g^*, \pi_{g^*})$ denote the graph consists of $(g,\pi_g)$ and $(g', \pi_{g'})$.
}

Then $\{(g_i, \pi_{g_i})\}$ are connected components in
$(g,\pi_g)$, $\{(g'_i, \pi'_{g_i})\}$ are connected components in $(g', \pi_{g'})$. Proving the lemma is equivalent to proving $\gamma$ is an isomorphism between
 $(g,\pi_g)$ and $(g', \pi_{g'})$.

Consider two automorphisms $\gamma_1 \in Aut(G, \pi)$, $\gamma_2 \in Aut(G', \pi')$ having $\gamma_2= \gamma_1 \gamma$. By Theorem~\ref{the:simplify}, $\gamma_1 \in Aut(g,\pi_g)$ and $\gamma_2 \in Aut(g', \pi_{g'})$.
According to the definition of automorphism, $\gamma_1, \gamma_2, \gamma \in Aut(G^*, \pi^*)$, and $\gamma_1, \gamma_2 \in Aut(g^*, \pi_{g*})$. Since $\gamma=\gamma_1^{-1} \gamma_2$, $\gamma \in Aut(g^*, \pi_{g^*})$, i.e., $\gamma$ is an isomorphism between $(g, \pi_g)$ and $(g', \pi_{g'})$.
\eop
}

\begin{theorem}
\label{the:tree}
Given two isomorphic graphs $(G , \pi )$, $(G', \pi')$, the
structure of  ${\mathcal AT}(G, \pi)$ and ${\mathcal
  AT}(G', \pi')$ are the same. Here, the structure of an AutoTree is a
tree without any labels.
\end{theorem}

\proofsketch
The proof can be constructed by mathematical induction on each tree node with Lemma~\ref{lem:tree_node}.
\eop

\comment{
We discuss Algorithms \DivideP (Algorithm~\ref{alg:DivideP}) and \DivideS (Algorithm~\ref{alg:DivideS}), which may probably divide a subgraph $g$ of $G$ into several smaller subgraphs.
For each subgraph $g$, Algorithm \DivideP first divides $g$ into a set of smaller subgraphs 
and for each non-singleton subgraph $s \in \mathcal S$, Algorithm \DivideS further simplifies the structure of $s$, 
dividing $s$ into a number of smaller subgraphs similarly.
Both of \DivideP and \DivideS take a graph $g$ and an ordered partition $\pi$ as inputs and return a set of non-overlapping subgraphs ${\mathcal S}$, which consists the input graph $g$.
\DivideP is based on a special case of Lemma.~\ref{lem:biclique} where at least one of $V_i$ and $V_j$ is a singleton cell. It isolates each singleton cell in $\pi_g$  as a subgraph in the resulting $\mathcal S$ (Line~1-6) and groups the remaining vertices into a number of subgraphs(Line~7-9).  Here, $\pi_g$ is obtained by projecting $\pi$ on $V(g)$. Specifically, each cell of $\pi_g$ is obtained by restricting the corresponding cell in $\pi$ contain only vertices in $V(g)$. 
\DivideS is based on Theorem~\ref{the:simplify}. It simplifies each non-singleton subgraph $g$ by removing edges that have no influence on the automorphism group of $g$ (Line~1-11). The removed edges can probably disconnect $g$, and each connected component of $g$ is a subgraph in $\mathcal S$ (Line~12-15).


Revisit \cl, \DivideP and \DivideS, it is worth noting that, $pi_g$, an ordered partition of $V(g)$, is achieved by simply projecting $\pi$ on $V(g)$ rather than invoking Algorithm \Partition on $g$. Here, $\pi$ is the ordered partition of the input graph $G$ (Line~2 in Algorithm~\ref{alg:CL}). Easy to see, reusing $\pi$ improves the efficiency of \cl significantly. The correctness is given in Theorem~\ref{the:reuse}.
}

\comment{
\subsection{Refinement \Partition}

We discuss properties of refinement function \Partition.
In \CL, we apply Weisfeiler-Lehman algorithm \cite{weisfeiler2006construction} as the refinement function $R$. As proved by \cite{weisfeiler2006construction}, only vertices in the same cell in the resulting equitable coloring $\pi$ can probably be automorphic equivalent. In \CL, only the coloring $\pi$ for $G$ is achieved by the refinement function $R$, all the other colorings, i.e., $\pi_g$ for subgraphs $g$, are obtained by projecting $\pi$ on $V(g)$. The following theorem proves the equivalence between projecting $\pi$ on $V(g)$ and applying $R$ on $g$.

\begin{theorem}
\label{the:reuse}
$\pi_g$, the projection of $\pi$ on $V(g)$ by \DivideP and \DivideS, inherits the properties of $\pi$. Specifically, (1) only vertices in the same cell in $\pi_g$ can  be automorphic equivalent. (2) $\pi_g$ is equitable with respect to $g$.
\end{theorem}

\proofsketch The first property can be proved trivially. We focus on the second
property, and prove the claim based on the  mathematical
induction. Assume $g$ is a connected  component in $g'$ that emerges
due to either \DivideP or \DivideS, and $\pi_{g'}$ satisfies the
second property.
In either case, the edges removed are those between two cells in $\pi_{g'}$. Then for any two vertices in the same cell in $\pi_g$, they either retain all neighbors or remove all neighbors in any cell
in $\pi_g$, i.e., $\pi_g$ is equitable with respect to $g$.
\eop
}

\stitle{Time complexity of \DivideP and \DivideS}:
Easy to see, the time complexity of \DivideP is $O(m)$ as each component of \DivideP costs $O(m)$.
We focus on the time complexity of \DivideS.
Recall that a coloring $\pi$ is equitable with respected to a graph $G$, if for all vertices $v_i, v_2 \in V_i$, they have the same number of neighbors in $V_j$. Such property can be utilized to accelerate \DivideS.
Specifically, \DivideS assigns each cell $V_i$ with a vector, where each element maintains the number of neighbors of each vertex $v \in V_i$ in $V_j$. Then checking if $V_i$ consists a clique is equivalent to checking if the $i$-th element in the vector equals $|V_i|-1$. Checking whether $V_i$ and $V_j$ consists a biclique is equivalent to checking if the $j$-th element in the vector equals $|V_j|$. Therefore, each component of \DivideS also costs $O(m)$, i.e., \DivideS costs $O(m)$.

\begin{algorithm}[t]
\scriptsize
\caption{\CombineCL($g, \pi_g$) }
\label{alg:CombineCL}
 $\gamma^* \leftarrow \bliss(g, \pi_g)$\;
\For {each vertex $v \in V(g)$}{
 $v^{\gamma_g} \leftarrow \pi(v)+|\{ u | \pi_g(u)=\pi_g(v), u^{\gamma^*} < v^{\gamma^*} \}|$\;
}
 $C(g,\pi_g)=(g, \pi_g)^{\gamma_g}$\;
{\bf return} $C(g,\pi_g)$;
\end{algorithm}
\setlength{\textfloatsep}{2pt}

\subsection{\CombineCL and \CombineST}

We discuss algorithms \CombineCL and \CombineST, which generate the
canonical labeling $C(g, \pi_g)$ for the input colored
graph $(g, \pi_g)$.  \CombineCL, shown in
Algorithm~\ref{alg:CombineCL}, generates $\gamma_g$ for a
non-singleton leaf node exploiting the canonical labeling $\gamma^*$
obtained by existing approaches (Line~1).  $\gamma^*$ introduces a total order
among vertices in the same cell in $\pi_g$, resulting in the
canonical labeling $\gamma_g$, along with vertex color $\pi(v)$ due to
$\pi$ (Line~2-3). Canonical labeling $C(g,\pi_g)$ can be trivially
obtained as $(g,\pi_g)^{\gamma_g}$ (Line~4).
On the other hand, \CombineST, shown in Algorithm~\ref{alg:CombineST},
generates $\gamma_g$ for a non-leaf node by combining canonical
labeling of its child nodes $(g_i, \pi_{g_i})$.
Canonical labeling $C(g_i, \pi_{g_i})$ introduces a total order between vertices in different subgraphs (Line~1-2) and
canonical labeling $\gamma_{g_i}$ introduce a total order among vertices in the same subgraph $(g_i, \pi_{g_i})$ (Line~3). These two orders determines a total order between vertices in the same cell in $\pi_g$, resulting in the canonical labeling $\gamma_g$ for $(g, \pi_g)$ (Line~4-5), in the similar manner. Canonical labeling $C(g, \pi_g)$ can be obtained by as $(g,\pi_g)^{\gamma_g}$ (Line~6).

\comment{
We discuss Algorithms \CombineCL and \CombineST, which generate a certificate $c(g)$ and the orbit partition $\pi_{orbit}$ of $g$ by combining existing information.
Both of \CombineCL and \CombineST first determine a vertex label $L(v)$ for each vertex $v \in V(g)$ (Line~1-3 in Algorithm~\ref{alg:CombineCL}, Line~1-9 in Algorithm~\ref{alg:CombineST}), and then generate the certificate $c(g)$ as the lexicographically sorted edge list under the mapping of $L$ (Line~5 in Algorithm~\ref{alg:CombineCL}, Line~10 in Algorithm~\ref{alg:CombineST}).

For Algorithm \CombineCL, shown in Algorithm~\ref{alg:CombineCL}, the canonical labeling $\gamma$ corresponds to the canonical representative $C(g)$  introduces a total order of vertices in $V(g)$. With such a total order, vertices with the same index, i.e., in the same cell in $\pi$, are assigned with distinct labels $L(v)$ (Line~1-3), which are further used to update orbit partition $\pi_{orbit}$ of $g$ (Line~4) and generate certificate $c(g)$ (Line~5).

Algorithm \CombineST, shown in Algorithm~\ref{alg:CombineST}, first generates the orbit partition $\pi_{orbit}$ for $g$ (Line~1-7).
Each cell in the combined orbit partition $\pi_{orbit}$ is either one derived by concatenating corresponding cells in the partitions of isomorphic subgraphs (Line~3-5) or an original cell in a partition if the subgraph is not isomorphic to any other subgraphs (Line~6). Note that vertices in each cell in $\pi_{orbit}$ are sorted, which is inherited from the orbit partitions of subgraphs.
The ordering is utilized to assign vertex label $L(v)$ for each vertex $v \in V(g)$ (Line~7-9).
}

\begin{lemma}
\label{lem:leaf}
For two leaf nodes $(g_1,\pi_{g_1})$ and $(g_2, \pi_{g_2})$ in AutoTree, if they are symmetric in $(G,\pi)$, i.e., these is a permutation $\gamma \in Aut(G,\pi)$ such that $(g_1, \pi_{g_1})^\gamma =(g_2, \pi_{g_2})$,   $C(g_1,\pi_{g_1})=C(g_2, \pi_{g_2})$ by Algorithm \CombineCL.
\end{lemma}

\proofsketch
The proof is trivial when $(g_1,\pi_{g_1})$ and $(g_2,\pi_{g_2})$ are singleton, since the vertices are in the same cell in $\pi$.

We focus on the case when $(g_1,\pi_{g_1})$ and $(g_2,\pi_{g_2})$ are non-singleton. For ease of discussion, we assume vertices in $g_1$ and $g_2$ are relabeled from 1 to $k$, respectively. Here $k=|V(g_1)|=|V(g_2)|$. Since $(g_1,\pi_{g_1})$ and $(g_2, \pi_{g_2})$ are symmetric in $(G,\pi)$, they are isomorphic, i.e.,
$(g_1,\pi_{g_1})^{\gamma_1}= (g_2,\pi_{g_2})^{\gamma_2}$.
Here, $\gamma_1$ and $\gamma_2$ are the corresponding permutations by \bliss.  Let $v \in g_1$ and $u \in g_2$ be two vertices having $v^{\gamma_1}=u^{\gamma_2}$, we prove $v^{\gamma_{g_1}}=u^{\gamma_{g_2}}$.
Let $v_1 \in g_1$ and $u_1 \in g_2$ be two vertices having $v_1^{\gamma_1} = u_1^{\gamma_2}$.
If $\pi(v_1)= \pi(v)$ and $v_1^{\gamma_1} < v^{\gamma_1}$, then $\pi(u_1) = \pi(v_1) = \pi(v) =\pi(u)$ and $u_1^{\gamma_2}  < u^{\gamma_2} $. The reverse also holds, implying that
$v_1$ and $u_1$ have the same influence on $v^{\gamma_{g_1}}$ and $u^{\gamma_{g_2}}$.
If  $\pi(v_1) \neq \pi(v)$, then $\pi(u_1) \neq \pi(u)$, i.e., $v_1$ and $u_1$ have no, which is also the same, influence on $v^{\gamma_{g_1}}$ and $u^{\gamma_{g_2}}$.
As a consequence, $v^{\gamma_{g_1}} = u^{\gamma_{g_2}}$, i.e., $C(g_1, \pi_{g_1}) = C(g_2, \pi_{g_2})$.
\eop

\begin{lemma}
\label{lem:non-leaf}
For two non-leaf nodes $(g_1,\pi_{g_1})$ and $(g_2, \pi_{g_2})$, if $(g_1,\pi_{g_1})$ and $(g_2, \pi_{g_2})$ are symmetric in $(G, \pi)$, i.e., these is a permutation $\gamma \in Aut(G, \pi)$ such that $(g_1,\pi_{g_1})^\gamma =(g_2, \pi_{g_2})$, then $C(g_1,\pi_{g_1})=C(g_2, \pi_{g_2}) $ by \CombineST.
\end{lemma}

\comment{
\proofsketch
By Lemma~\ref{lem:tree_node}, child nodes of $(g_1, \pi_{g_1})$ and $(g_2, \pi_{g_2})$ can be sorted such that each pair $(g_i, \pi_{g_i})$, $(g_j, \pi_{g_j})$ are isomorphic.
For any vertex pair $v \in g_i$ and $u \in g_j$ having $v^{\gamma_{g_i}} = u^{\gamma_{g_j}}$, similar to the proof of Lemma~\ref{lem:leaf}, we can prove $v^{\gamma_{g_1}} = u^{\gamma_{g_2}}$, implying that $C(g_1, \pi_{g_1}) =C(g_2, \pi_{g_2})$.
\eop
}

\proofsketch
We prove the base case, i.e., child nodes of $(g_1, \pi_{g_1})$ and $(g_2, \pi_{g_2})$ are leaf nodes, the other cases can be proved by mathematical induction.

First, by Lemma~\ref{lem:tree_node}, child nodes of $(g_1, \pi_{g_1})$ and $(g_2, \pi_{g_2})$ can be sorted such that each pair $(g_i, \pi_{g_i})$ and $(g_j, \pi_{g_j})$ are isomorphic.
Second, by Lemma~\ref{lem:leaf}, $C(g_i, \pi_{g_i}) = C(g_j, \pi_{g_j})$.
Then for any vertex pairs $v \in g_i$ and $u \in g_j$ having $v^{\gamma_{g_i}} = u^{\gamma_{g_j}}$, we have (1) there are the same number of subgraph pair $(g'_i, \pi'_{g_i})$ and $(g'_j, \pi'_{g_j})$ that are isomorphic and share the same canonical labeling with $(g_i, \pi_{g_i})$ and $(g_j, \pi_{g_j})$, while $(g'_i, \pi'_{g_i})$ is sorted before $(g_i, \pi_{g_i})$ and $(g'_j, \pi'_{g_j})$ is sorted before
$(g_j, \pi_{g_j})$;
(2) there are the same number of vertex pairs $v' \in g_i$ and $u' \in g_j$ with $v'^{\gamma_{g_i}} = u'^{\gamma_{g_j}}$, having $v'^{\gamma_{g_i}} < v^{\gamma_{g_i}}$ and $u'^{\gamma_{g_j}} < u^{\gamma_{g_j}}$.
As a consequence, $v^{\gamma_{g_1}} = u^{\gamma_{g_2}}$, in other words, $C(g_1, \pi_{g_1}) =C(g_2, \pi_{g_2})$.

\comment{
\proofsketch
Since $g_1$ and $g_2$ are non-leaf nodes, Algorithms \DivideP and \DivideS divide them into two sets of subgraphs, denoted as ${\mathcal S}_1=\{s_1^1, s_2^1, \ldots, s_{k_1}^1\}$ and ${\mathcal S}_2=\{s_1^2, s_2^2, \ldots, s_{k_2}^2\}$. Without loss of generality, we assume subgraphs in ${\mathcal S}_1$ and ${\mathcal S}_2$ are sorted by certificates, and ties are broken by permutation $\gamma$.
Since $g_1$ are $g_2$ are symmetric in $G$, and both \DivideP and \DivideS are deterministic, it is easy to prove, $k_1= k_2$ and $s_i^1$ and $s_i^2$ are symmetric in $G$, for $1 \leq i \leq k$, here $k=k_1=k_2$.

We prove the claim in the way certificates of tree nodes are generated, i.e., in  post-order traversal manner.

First, when all subgraphs in ${\mathcal S}_1$ and ${\mathcal S}_2$ are leaf nodes.
Consider two subgraphs $s_i^1 \in {\mathcal S}_1$ and $s_i^2 \in {\mathcal S}_2$ that are symmetric in $G$, and two vertices $v \in s_i^1$ and $u \in s_i^2$ having $v^\gamma =u$. According to Lemma~\ref{lem:leaf} and its proof, we have $L(v)=L(u)$. Note that $L(v)$ and $L(u)$ refer to the vertex labels obtained in leaf nodes $s_i^1$ and $s_i^2$. For clarity, we denote them as $L(s_i^1, v)$ and $L(s_i^2, u)$, respectively. We prove $L(g_1, v)=L(g_2, v)$. According to Algorithm~\ref{alg:CombineST}, we rewrite $L(g_1, v)$ and $L(g_2, u)$ as follows
\begin{align}
L(g_1, v)&= L(s_i^1,v)+ |\{p| i(p)=i(v), p \in s_j^1, j<i \}| \\
L(g_2, u)&= L(s_i^2,u)+|\{q| i(q)=i(u), q \in s_j^2, j<i\}|
\end{align}
Similar to the proof of Lemma~\ref{lem:leaf}, we can prove that for each vertex pair $p$ and $q$ having $p^\gamma=q$, $p$ satisfies (3) iff $q$ satisfies (4). Therefore, $L(g_1, v)=L(g_2, v)$, implying that $c(g_1)=c(g_2)$.
Second, when some subgraphs in ${\mathcal S}_1$ and ${\mathcal S}_2$ are non-leaf nodes, the proof is similar.
As a consequence, for any two symmetric non-leaf node $g_1$ and $g_2$, $c(g_1)=c(g_2)$.

}

\comment{
\proofsketch
Since $(g_1,\pi_{g_1})$ and $(g_2, \pi_{g_2})$ are non-leaf nodes, Algorithms \DivideP and \DivideS divide them into two sets of subgraphs, denoted as $(g_{1i}, \pi_{g_{1i}})$ and $(g_{2i}, \pi_{g_{2i}})$. Without loss of generality, we assume subgraphs in $(g_{1i}, \pi_{g_{1i}})$ and $(g_{2i}, \pi_{g_{2i}})$  are sorted by canonical labeling, and ties are broken by permutation $\gamma$.
Since $(g_1,\pi_{g_1})$ and $(g_2, \pi_{g_2})$ are symmetric in $(G,\pi)$, and both \DivideP and \DivideS are deterministic, it is easy to prove, $|(g_{1i}, \pi_{g_{1i}})| = |(g_{2i}, \pi_{g_{2i}})|$ and $(g_{1i}, \pi_{g_{1i}})$ and $(g_{2i}, \pi_{g_{2i}})$ are symmetric in $(G,\pi)$ for $1 \leq i \leq k$, here $k=|(g_{1i}, \pi_{g_{1i}})|$.

We prove the claim in the way canonical labeling of tree nodes are generated, i.e., in bottom-up manner.

First, when all subgraphs in $(g_{1i}, \pi_{g_{1i}})$ and $(g_{2i}, \pi_{g_{2i}})$ are leaf nodes.
Consider two subgraphs $(g_{1i}, \pi_{g_{1i}})$ and $(g_{2i}, \pi_{g_{2i}})$  that are symmetric in $(G, \pi)$, and two vertices $v \in (g_{1i}, \pi_{g_{1i}})$ and $u \in (g_{2i}, \pi_{g_{2i}})$ having $v^\gamma =u$. According to Lemma~\ref{lem:leaf} and its proof, we have $v^{\gamma_{g_1}}=u^{\gamma_{g_2}}$. Note that $v^{\gamma_{g_1}}=u^{\gamma_{g_2}}$ refer to the vertex labels obtained in leaf nodes $s_i^1$ and $s_i^2$. For clarity, we denote them as $L(s_i^1, v)$ and $L(s_i^2, u)$, respectively. We prove $L(g_1, v)=L(g_2, v)$. According to Algorithm~\ref{alg:CombineST}, we rewrite $L(g_1, v)$ and $L(g_2, u)$ as follows
\begin{align}
L(g_1, v)&= L(s_i^1,v)+ |\{p| i(p)=i(v), p \in s_j^1, j<i \}| \\
L(g_2, u)&= L(s_i^2,u)+|\{q| i(q)=i(u), q \in s_j^2, j<i\}|
\end{align}
Similar to the proof of Lemma~\ref{lem:leaf}, we can prove that for each vertex pair $p$ and $q$ having $p^\gamma=q$, $p$ satisfies (3) iff $q$ satisfies (4). Therefore, $L(g_1, v)=L(g_2, v)$, implying that $c(g_1)=c(g_2)$.
Second, when some subgraphs in ${\mathcal S}_1$ and ${\mathcal S}_2$ are non-leaf nodes, the proof is similar.
As a consequence, for any two symmetric non-leaf node $g_1$ and $g_2$, $c(g_1)=c(g_2)$.
\eop
}

\begin{algorithm}[t]
\scriptsize
\caption{\CombineST($g, \pi_g$) }
\label{alg:CombineST}
 sort child nodes $(g_i, \pi_{g_i})$ of $(g, \pi_g)$ in non-descending order of $C(g_i, \pi_{g_i})$\;
 sort vertices in each cell in $\pi_g$, s.t., $u$ is before $v$ if $u \in g_i, v \in g_j, i<j$\;
 sort vertices in each cell in $\pi_{g_i}$, s.t., $u$ is before $v$ if $ u^{\gamma_{g_i}}<v^{\gamma_{g_i}}$\;
\For {each vertex $v \in V(g)$}{
$v^{\gamma_g} \leftarrow \pi(v)+ |\{ u |\pi_g(u)=\pi_g(v), u $ is before $v \}|$\;
}
 $C(g,\pi_g)=(g, \pi_g)^{\gamma_g}$\;
{\bf return}  $C(g,\pi_g)$;
\end{algorithm}
\setlength{\textfloatsep}{2pt}

The following theorem gives the correctness of \CL.

\begin{theorem}
\label{the:correctness}
Given two graphs $(G_1,\pi_1)$ and $(G_2, \pi_2)$, they are isomorphic
iff the canonical labeling $C(G_1, \pi_1)$ and $C(G_2,$ $\pi_2)$ by
\CL satisfy $C(G_1,\pi_1)=C(G_2,\pi_2)$.
\end{theorem}

\proofsketch
We construct an auxiliary graph $G$ containing $G_1$, $G_2$ and an
vertex $u$ connecting to every vertex in $G_1$ and $G_2$. Easy to see,
$u$ is distinct from any other vertices in $G$, and $G_1$ and $G_2$
are symmetric in $G$. Therefore, the root of the AutoTree ${\mathcal AT}(G, \pi)$ has
three child nodes, $(u, \pi_u)$, $(G_1,\pi_1)$ and $(G_2,\pi_2)$.
According to Lemma~\ref{lem:leaf} and Lemma~\ref{lem:non-leaf}, $C(G_1,\pi_1)=C(G_2,\pi_2)$.
\eop

\begin{theorem}
\label{the:tree_sym}
In  $(G, \pi)$, if  two vertices are symmetric,   they are in two leaf nodes in ${\mathcal AT}(G,\pi)$ sharing the same canonical labeling.
\end{theorem}
\proofsketch
It can be proved by Lemma~\ref{lem:leaf} and Lemma~\ref{lem:non-leaf}, since a leaf node cannot be isomorphic to a non-leaf node.
\eop

We revisit previous algorithms, e.g., \nauty, \bliss, \traces as well as our approach \CL. As mentioned, previous algorithms
enumerate all possible permutations and select the minimum $(G,\pi)^\gamma$ as the canonical labeling. On the other hand, our approach \CL constructs a tree index AutoTree ${\mathcal AT}$ that recursively partitions the given
graph $(G,\pi)$ into subgraphs.
By partition, \CL exploits properties of $(s, \pi_s)$ that
enable canonical labeling computation from combining without enumeration.
Canonical labeling for each node $(s, \pi_s)$ in ${\mathcal AT}(G,\pi)$ is either the minimum $(s ,\pi_s)^\gamma$, for a leaf node, or the $k$-th minimum $(s ,\pi_s)^\gamma$ obtained by combining the canonical labeling of child nodes, for a non-leaf node.
Note that for different tree nodes, the $k$ values are different.
As a consequence,   \CL returns the $k$-th minimum $(G, \pi)^\gamma$ as the canonical labeling and ensures that $k$ is the same for isomorphic graphs.

\stitle{Time complexity of \CombineCL and \CombineST}:  For \CombineCL, easy to see, the most time-consuming parts are invoking existing canonical labeling algorithms to generate $\gamma^*$ (Line~1) and generating the canonical labeling $C(G, \pi)$ (Line~4). Therefore, the time complexity of \CombineCL is $O(X+|E(g)|ln(|E(g)|))$ where $X$ is the time complexity of canonical labeling algorithms. Similarly, the most time-consuming parts of \CombineST are  determining total order between  different child nodes of $(g,\pi_g)$ (Line~1) and generating the canonical labeling (Line~6), since the other parts either cost $O(|V(g)|log(|V(g)|))$ or cost $O(|V(g)|)$. Therefore, the time complexity of \CombineST is $O(|E(g)|ln(|E(g)|))$.

\subsection{Symmetric Subgraph Matching}
\label{sec:ssm}

 \comment{

We propose Algorithm \SSM (Algorithm~\ref{alg:SSM}) for SSM, following
divide-and-conquer paradigm. \SSM is designed by utilizing the
properties of AutoTree $\mathcal AT$.
Specifically, two tree nodes with the same
canonical labeling implies that the two corresponding subgraphs in $G$
are symmetric.
%
An SSM query graph $q$ is divided into vertex disjoint subgraphs
$\{s_1, \ldots, s_{k_1}\}$ where all vertices in $s_i$ are contained in
a leaf node $l_i$ in $\mathcal AT$ (Line~1).  \SSM   finds mosaic
subgraphs that are symmetric to each $s_i$ (Line~3-7) and reconstructs
the resulting subgraphs symmetric to $q$ using these mosaics
(Line~8-12).  For each $s_i$, its symmetric subgraphs can be found by
first locating at $l_j^i$, a leaf node in $\mathcal AT$ that is
symmetric to $l_i$ (Line~5) and then mapping  subgraphs in $R_i=SM(s_i,
l_i)$, i.e., subgraphs symmetric to $s_i$ in $l_i$, to subgraphs in $l_j^i$
(Line~6-7). This mapping can be achieved with the isomorphism between $l_i$ and $l_j^i$ easily.  To reconstruct subgraphs symmetric to
$q$, any subtree matching of $T$  helps to identify valid
combinations. Here, $T$ is the minimum subtree in $\mathcal AT$
containing $\{s_1, \ldots, s_{k_1}\}$.  In \SSM, the most time consuming part is $SM(s_i, l_i)$
(Line~4) and $SM(T, {\mathcal AT})$ (Line~9), where SM  is an algorithm for
subgraph matching. For $SM(s_i, l_i)$, (1) both $s_i$ and $l_i$ are
much smaller than $q$ and $G$, and in most cases, $s_i$ and $l_i$ are singleton; (2) coloring of $l_i$ prunes
significant unnecessary searching; (3) in most cases, $s_i = l_i$.  For $SM(T, {\mathcal AT})$, since both $T$
and $\mathcal AT$ are trees, SM on trees are much easier than on general graphs.
In conclusion, Algorithm \SSM transfers the problem of symmetric subgraph matching to several subgraph matchings in leaf nodes in ${\mathcal AT}$, where the majority are trivial since most leaf nodes in ${\mathcal AT}$ are singleton.
Owe to the symmetries maintained by ${\mathcal AT}$, each symmetric subgraph of $q$ can be discovered with polynomial delay. On the other hand, existing subgraph matching algorithms have several drawbacks. (1) the time complexity is not bounded. (2) they will find much more candidate matchings than the result. (3) the verification of symmetry between a matching $g$ and the query graph $q$ is not trivial. (4) there is no guarantee to find all symmetric subgraph matchings.

\begin{example}
Consider the AutoTree ${\mathcal AT}$ in Fig.~\ref{fig:axis}, where
all the leaf nodes are singleton, and leaf nodes with the same
canonical labeling correspond to the vertices with the same color.
Consider an SSM query $q$, 3-2-6, on $g$.  The
three subgraphs correspond to three leaf nodes in ${\mathcal AT}$, and
the minimum subtree $T$ of ${\mathcal AT}$ containing these leaf nodes
is the one with red tree edges. Two of total 11 SSM matchings are 9-8-10, derived by the matching $T_1$ of $T$, in green tree edges, and 13-12-10, derived by $T_2$ of $T$, in dashed tree edges.
\end{example}

\begin{algorithm}[t]
\scriptsize
\caption{\SSM($G,q,{\mathcal AT}$) }
\label{alg:SSM}
 partition $q$ into vertex disjoint subgraphs $\{s_1, \ldots, s_{k_1}\}$,  $s_i \in l_i$\;
 construct the minimum subtree $T$ of $\mathcal AT$ containing $\{l_1, \ldots, l_{k_1}\}$\;
\For {each $(s_i, l_i)$}{
 $R_i \leftarrow SM(s_i, l_i)$\;
 $L_i \leftarrow \{l_1^i, \ldots, l_{k_2}^i\}$, where $l_j^i \cong l_i$\;
\For {each $l_j^i \in L_i$}{
 mapping each subgraph in $R_i$ to a subgraph in $l_j^i$, resulting in $R_j^i$\;
}
}
 ${\mathcal R} \leftarrow \emptyset$\;
\For {any $T' \in SM(T, \mathcal AT)$}{
 extracts the leaf nodes of $T'$, denoted as $(l_{i1}^1, l_{i2}^2, \ldots, l_{ik}^k)$\;
 ${\mathcal R} \leftarrow {\mathcal R} \cup R_{i1}^1 \times \ldots \times R_{ik}^k$\;
}
 ${\mathcal S} \leftarrow G[{\mathcal R}]$\;
{\bf return} $\mathcal S$;
\end{algorithm}
 }

\begin{algorithm}[t]
\scriptsize
\caption{\SSM($G,q,{\mathcal AT}(g)$) }
\label{alg:SSM}
find tree node $n_q \in {\mathcal AT}(g)$ with max depth that contains $q$\;
\If{$n_q$ is a leaf node or $n_q=q$}{
${\mathcal S} \leftarrow SM(n_q, q)$ or ${\mathcal S} \leftarrow n_q$\;
}
\Else{
divide $q$ into subgraphs $\{q_1, \ldots, q_k\}$, contained in children $\{n_1, \ldots, n_k\}$ of $n_q$\;
\For{each $(q_i, n_i)$}{
 ${\mathcal S_i} \leftarrow$ \SSM$(G, q_i, {\mathcal AT}(n_i))$\;
}
\For{each child $n_j$ of $n_q$ that shares the same canonical labeling with $n_i$}{
${\mathcal S_j} \leftarrow {\mathcal S_i}^{\gamma_{ij}}$\;
}
${\mathcal S} \leftarrow \emptyset$\;
\For{each $\{n_{1'}, \ldots, n_{k'}\}$}{
${\mathcal S} \leftarrow {\mathcal S} \cup  {\mathcal S_{1'}} \times \ldots \times {\mathcal S_{k'}}     $\;
}
}
\For {each $n_{q'}$ sharing the same canonical labeling with $n_q$}{
${\mathcal S}  \leftarrow {\mathcal S} \cup {\mathcal S}^{\gamma_{qq'}}$\;
}
{\bf return} $\mathcal S$;
\end{algorithm}

We propose Algorithm \SSM (Algorithm~\ref{alg:SSM}) for SSM, following divide-and-conquer paradigm. \SSM is designed by the properties of AutoTree $\mathcal AT$.
Specifically, two tree nodes sharing the same
canonical labeling implies that the two corresponding subgraphs in $G$ are symmetric, and one isomorphism between these two subgraphs can be easily obtained. \SSM$(G, q, {\mathcal AT}(g))$ finds all symmetric subgraphs of $q$ in the subtree of ${\mathcal AT}$ that rooted at $g$, i.e., in a subgraph $g$ of $G$. \SSM first finds the minimal subgraph, a tree node $n_q$ in ${\mathcal AT}$, that contains $q$ (Line~1). Then, symmetric subgraphs of $q$ in $n_q$ can be extended to those in subgraphs $n_{q'}$ that are symmetric to $n_q$ by an isomorphism $\gamma_{qq'}$ from $n_q$ to $n_{q'}$, consisting the symmetric subgraphs of $q$ in $G$ (Line~13-14). Symmetric subgraphs of $q$ in $n_q$ can be found by divide-and-conquer.
The basic cases occur  when $n_q$ is a leaf node or $n_q=q$, then an existing subgraph isomorphism algorithm SM can be applied or returns $n_q$ as the result (Line~2-3). Otherwise, $q$ is divided into subgraphs $\{q_1, \ldots, q_k\}$ by the children of $n_q$, where $q_i$ is contained in $n_i$ (Line~5). Symmetric subgraphs of $q_i$ in $n_i$ can be found recursively by \SSM$(G, q_i, {\mathcal AT}(n_i))$, and mapped to those in $n_j$ that is symmetric to $n_i$ (Line~6-9). As a consequence, each symmetric subgraph of $q$ in $n_q$ can be composed by mosaic subgraphs in $\{n_{1'}, \ldots, n_{k'}\}$ where $n_{i'}$ is $n_i$ or is a sibling node  symmetric to $n_i$ (Line~11-12).
Since the majority of leaf nodes are singleton, \SSM is efficient and robust.  On the other hand, existing subgraph matching algorithms have several drawbacks. (1) the time complexity is not bounded. (2) they will find much more candidate matchings than the result. (3) the verification of symmetry between a matching $g$ and the query graph $q$ is not trivial. (4) there is no guarantee to find all symmetric subgraph matchings.

\begin{example}
Consider the AutoTree ${\mathcal AT}$ in Fig.~\ref{fig:axis}, where
all the leaf nodes are singleton, and leaf nodes with the same
canonical labeling correspond to the vertices with the same color.
Consider an SSM query $q$, 3-2-6, on $g$. We find symmetric subgraphs of $q$ in $g_1$, and those in $g_3$ can be extended by isomorphism $\gamma=(3,9)(2,8)(5,11)(4,10)(7,13)(6,12)$. Symmetric subgraphs of $q$ in $g_1$ are divided into $q_1$, 3-2, in $g_{11}$ and $q_2$, 6 in $g_{13}$. ${\mathcal S}_1= g_{11} $, which can be extended to ${\mathcal S}_2= g_{12} $ and ${\mathcal S}_3= g_{13}$. Those for $q_2$ can be obtained similarly. As a consequence, symmetric subgraphs of $q$ in $g_1$ can be composed as ${\mathcal S}=\{$3-2-4, 3-2-6, 5-4-2, 5-4-6, 7-6-2, 7-6-4$\}$. Those in $g_3$ can be obtained by ${\mathcal S}^{\gamma}$.
\end{example}

\comment{
\subsection{Procedure Preprocessing}
\label{sec:preprocessing}


As we mention in Section~\ref{sec:overview}, vertices in $V \setminus V_s$ are grouped into a set of subgraphs $\mathcal S$. Each subgraph $S_i \in \mathcal S$ is assigned with a signature $s(S_i)$ encoding its vertex set.
We discuss Algorithm \Preprocessing in Algorithm~\ref{alg:preprocessing}. As Theorem~\ref{the:subgraph_inter} claims, only vertices in subgraphs with the same signature can probably be automorphic equivalent,  Algorithm \Preprocessing groups subgraphs with the same signature into a subset and deals with each subset $s \subset \mathcal S$ to find symmetries that can be easily determined (Line~1).
As a result, subgraphs in $\mathcal S$ are further partitioned into components, consisting ${\mathcal S}_p$ and ${\mathcal S}_r$, and Algorithm \Preprocessing returns ${\mathcal S}_r$, for which more  sophisticated algorithms are required to detect symmetries. On the other hand, ${\mathcal S}_p$ consists of components for which symmetries can be easily detected and labels for vertices in each subgraph in ${\mathcal S}_p$ is given in Algorithm \Preprocessing (Line~22).
Note that vertices in ${\mathcal S}_p$ consist $V_a$ mentioned in Section~\ref{sec:overview}. Both of ${\mathcal S}_p$ and ${\mathcal S}_r$ are initialized as $\emptyset$ (Line~2).
Consider each subset $s \subset \mathcal S$ and a subgraph $S_i \in s$, there are three cases: (1) $s$ contains only one subgraph; (2) $s$ contains more than one subgraph, and all vertices in $S_i$ have different $l(v)$ values; (3) $s$ contains more than one subgraph, and some vertices in $S_i$ have same $l(v)$ values. For case (1), we simplify $S_i$ by removing edges according to Theorem~\ref{the:subgraph_intra} (Line~7). Consider the simplified $S_i$. If $S_i$ is still connected, then $S_i$ cannot be simplified and more sophisticated algorithms are needed to detect symmetries in $S_i$ (Line~8-9). Otherwise, labels of vertices in singleton connected components can be easily determined (Line~11) and we simplify the remaining components in a similar manner (Line~12). For case (2), as Theorem~\ref{the:subgraph_direct} claims, symmetries between vertices in subgraphs in $s$ can be easily determined (Line~15-17). For case (3), we can neither simplify $S_i$ without introducing errors nor determine symmetries between vertices in $S_i$ (Line~18-19).

\begin{algorithm}[t]
\caption{\PBBFS($u, v$) }
\label{alg:pbbfs}
\begin{algorithmic}[1]
\STATE $ S_i \leftarrow \emptyset$, $S_j \leftarrow \emptyset$, $Q_i \leftarrow \{ u \}$, $Q_j \leftarrow \{v \}$;
\WHILE { $Q_i \neq \emptyset$}
\STATE $u' \leftarrow Q_i.dequeue()$, $v' \leftarrow Q_j.dequeue()$;
\STATE $S_i \leftarrow S_i \cup \{u'\}$, $S_j \leftarrow S_j \cup \{v'\}$;
\STATE order $N(u')$ and $N(v')$ such that $u_1'^\gamma= v_1'$, $u_2'^\gamma=v_2'$, \ldots, $u_d'^\gamma=v_d'$, here $d=d(u')=d(v')$;
\FOR {$k$ from 1 to $d$}
\IF {$u_k' \notin S_i$ and $u_k'$ is contained in a non-singleton cell in $\pi$}
\STATE $Q_i.enqueue( u_k')$;
\ENDIF
\IF {$v_k' \notin S_i$ and $v_k'$ is contained in a non-singleton cell in $\pi$}
\STATE $Q_j.enqueue(v_k')$;
\ENDIF
\ENDFOR
\ENDWHILE
\RETURN $(S_i, S_j)$;
\end{algorithmic}
\end{algorithm}

\begin{lemma}
\label{lemma:pbbfs}
Conditions in Line~7 and Line~10 in Algorithm~\ref{alg:pbbfs} hold simultaneously.
\end{lemma}

\proofsketch
We prove that if conditions in Line~7 hold, conditions in Line~10 also hold. The reverse is similar.

Assume $u_k' \neq v_k'$, otherwise the lemma holds trivially. Since $u_k'^\gamma = v_k'$, then $u_k' \sim v_k'$, $u_k'$ and $v_k'$ are in the same cell in $\pi$ according to Theorem~\ref{the:unique}.
Assume $v_k' \in S_i$, then there are vertices $u''$ and $v''$ such that $u''^\gamma = v''$, $u_k' \notin N(u'')$ while $v_k' \in N(v'')$, contradicting the claim that permutation $\gamma$ is an automorphism of $G$.
\eop

\begin{theorem}
\label{the:subgraph_inter}
For any two vertices $u \in S_i$ and $v \in S_j$, if they are automorphic equivalent, then $s(S_i)=s(S_j)$, i.e., one of the following conditions  must hold
\begin{enumerate}
\item $i = j$
\item $i \neq j$ and $s(S_i) = s(S_j)$
\end{enumerate}
\end{theorem}

\proofsketch
Let permutation $\gamma$ be an automorphism of $G$ that maps $u$ to $v$, i.e., $u^\gamma =v$. We prove the claim by constructing $S_i$ and $S_j$, conducted by Algorithm \PBBFS (Parallel Pruned BFS), shown in Algorithm~\ref{alg:pbbfs}. Consider queues $Q_i$ and $Q_j$. (1) Initially, we enqueue $u$ and $v$ into $Q_i$ and $Q_j$, having $u^\gamma = v$. (2) In each iteration,  vertices $u_k'$ and $v_k'$ enqueued into $Q_i$ and $Q_j$ having $u_k' \sim v_k'$ according to Lemma~\ref{lemma:pbbfs}. Then the vertices at the same position in $Q_i$ and $Q_j$ are automorphic equivalent. Similarly, we can prove vertices $u'$ and $v'$ dequeued from $Q_i$ and $Q_j$ (Line~3) in each iteration are automorphic equivalent. Therefore, $s(S_i)=s(S_j)$.
\eop

\begin{theorem}
\label{the:subgraph_intra}
Consider a subgraph $S_i \in \mathcal S$ satisfying $s(S_i) \neq s(S_j)$ for all $j \neq i$,
then the following two sets of edges in $S_i$ can be removed without introducing errors when detecting symmetries in $S_i$.
\begin{itemize}
\item If vertices labeled $l_1$ consist a complete bipartite subgraph $b$ with vertices labeled $l_2$, then edges in $b$ can be removed
\item If vertices labeled $l_1$ consist a complete subgraph $c$, then edges in $c$ can be removed
\end{itemize}
\end{theorem}

\proofsketch
We prove
\eop

\comment{
\begin{theorem}
\label{the:singleton}
Consider a simplified $S_i$
\end{theorem}
}

\begin{algorithm}[t]
\caption{\Preprocessing($\mathcal S$) }
\label{alg:preprocessing}
\begin{algorithmic}[1]
\STATE group subgraphs in $\mathcal S$ into subsets, each contains subgraphs with the same signature;
\STATE ${\mathcal S}_p \leftarrow \emptyset$, ${\mathcal S}_r \leftarrow \emptyset$;
\FOR {each subset $s \subset \mathcal S$}
\STATE $\mathcal S \leftarrow \mathcal S$$ \setminus s$, denote one subgraph in  $s$ as $S_i$;
\SWITCH {states of $s$ and $S_i$}
\CASE {$|s|=1$}
\STATE remove edges in $S_i$ according to Theorem~\ref{the:subgraph_intra}
\IF { $S_i$ contains only one connected component}
\STATE ${\mathcal S}_r \leftarrow {\mathcal S}_r \cup \{ S_i\}$;
\ELSE
\STATE add connected components with size $1$ into ${\mathcal S}_p$
\STATE group connected components with size $>1$, and add these components into $\mathcal S$;
\ENDIF
\STATE {\bf break};
\ENDCASE
\CASE {$|s|>1$, all vertices in $S_i$ have different $l(v)$ values}
\STATE ${\mathcal S}_p \leftarrow {\mathcal S}_p \cup s$;
\STATE {\bf break};
\ENDCASE
\CASE {$|s|>1$, some vertices in $S_i$ have same $l(v)$ values}
\STATE ${\mathcal S}_r \leftarrow {\mathcal S}_r \cup s$;
\ENDCASE
\ENDSWITCH
\ENDFOR
\STATE determine labels for vertices in ${\mathcal S}_p$;
\RETURN ${\mathcal S}_r$;
\end{algorithmic}
\end{algorithm}

\comment{
\begin{algorithm}[t]
\caption{\Preprocessing($\mathcal S$) }
\label{alg:preprocessing}
\begin{algorithmic}[1]
\STATE group subgraphs in $\mathcal S$ into subsets, each contains subgraphs with the same signature;
\STATE ${\mathcal S}_p \leftarrow \emptyset$, ${\mathcal S}_r \leftarrow \emptyset$;
\FOR {each subset $s \subset \mathcal S$}
\STATE $\mathcal S \leftarrow \mathcal S$$ \setminus s$, denote one subgraph in  $s$ as $S_i$;
\SWITCH {states of $s$ and $S_i$}
\CASE {$|s|=1$}
\STATE let $L_i$ denotes the set of $l(v)$ values for all $v \in S_i$;
\FOR {every label $l_1 \in L_i$}
\IF {vertices with label $l_1$ consist a complete subgraph $c$}
\STATE remove edges in $c$;
\ENDIF
\ENDFOR
\FOR {every two different labels $l_1, l_2 \in L_i$}
\IF {vertices with label $l_1$ consist a complete bipartite subgraph $b$ with those labeled $l_2$}
\STATE remove edges in $b$;
\ENDIF
\ENDFOR
\IF { $S_i$ contains only one connected component}
\STATE ${\mathcal S}_r \leftarrow {\mathcal S}_r \cup \{ S_i\}$;
\ELSE
\STATE add connected components with size $1$ into ${\mathcal S}_p$
\STATE group connected components with size $>1$, and add these components into $\mathcal S$;
\ENDIF
\STATE {\bf break};
\ENDCASE
\CASE {$|s|>1$, all vertices in $S_i$ have different $l(v)$ values}
\STATE ${\mathcal S}_p \leftarrow {\mathcal S}_p \cup s$;
\STATE {\bf break};
\ENDCASE
\CASE {$|s|>1$, some vertices in $S_i$ have same $l(v)$ values}
\STATE ${\mathcal S}_r \leftarrow {\mathcal S}_r \cup s$;
\ENDCASE
\ENDSWITCH
\ENDFOR
\STATE determine labels for vertices in ${\mathcal S}_p$;
\RETURN ${\mathcal S}_r$;
\end{algorithmic}
\end{algorithm}
}

\comment{
\begin{algorithm}[t]
\caption{\Preprocessing($\mathcal S$) }
\label{alg:preprocessing}
\begin{algorithmic}[1]
\STATE group subgraphs in $\mathcal S$ into subsets, each contains subgraphs with the same signature;
\STATE $\mathcal S' \leftarrow \emptyset$;
\FOR {each subset $s \subset \mathcal S$}
\STATE denote one subgraph in $s$ as $S_i$;
\IF {$s$ contains only one subgraph}
\STATE let $L_i$ denotes the set of $l(v)$ values for all $v \in S_i$;
\FOR {every label $l_1 \in L_i$}
\IF {vertices with label $l_1$ consist a complete subgraph $c$}
\STATE remove edges in $c$;
\ENDIF
\ENDFOR
\FOR {every two different labels $l_1, l_2 \in L_i$}
\IF {vertices with label $l_1$ consist a complete bipartite subgraph $b$ with those labeled $l_2$}
\STATE remove edges in $b$;
\ENDIF
\ENDFOR
\ELSIF {all vertices in $S_i$ have different $l(v)$ values}
\STATE $\mathcal S \leftarrow \mathcal S \setminus $$ s$;
\ENDIF
\ENDFOR
\RETURN $\mathcal S$;
\end{algorithmic}
\end{algorithm}
}

\comment{
\begin{algorithm}[t]
\caption{\Preprocessing($\mathcal S$) }
\label{alg:preprocessing}
\begin{algorithmic}[1]
\STATE group subgraphs in $\mathcal S$ into subsets, each contains subgraphs with the same signature;
\FOR {each subset $s \subset \mathcal S$}
\IF {$s$ contains more than one subgraph, denote one subgraph as $S_i$}
\IF {all vertices in $S_i$ have different $l(v)$ values}
\STATE $\mathcal S \leftarrow \mathcal S \setminus $$ s$;
\ENDIF
\ELSE
\STATE let $L_i$ denotes the set of $l(v)$ values for all $v \in S_i$;
\FOR {every label $l_1 \in L_i$}
\IF {vertices with label $l_1$ consist a complete subgraph $c$}
\STATE remove edges in $c$;
\ENDIF
\ENDFOR
\FOR {every two different labels $l_1, l_2 \in L_i$}
\IF {vertices with label $l_1$ consist a complete bipartite subgraph $b$ with those labeled $l_2$}
\STATE remove edges in $b$;
\ENDIF
\ENDFOR
\ENDIF
\ENDFOR
\RETURN $\mathcal S$;
\end{algorithmic}
\end{algorithm}
}

\begin{theorem}
\label{the:subgraph_direct}
Given two subgraphs $S_i =\{v_1^i, v_2^i, \ldots, v_k^i \}$ and $S_j=\{ v_1^j, v_2^j, \ldots, v_k^j\}$ with the same signature, if all vertices in $S_i$ ($S_j$ ) have different $l(v)$ values, then $S_i$ and $S_j$ are isomorphic, and the symmetry between vertices in $S_i$ and $S_j$ can be easily determined.
\end{theorem}

\proofsketch
We prove the claim by achieving the canonical labeling $C(S_i)$ for $S_i$ and $C(S_j)$ for $S_j$, respectively. As a consequence, if $C(S_i) = C(S_j)$, then $S_i$ and $S_j$ are isomorphic.

Without loss of generality, we assume $l(v_1^i) = l(v_1^j)$, $l(v_2^i) = l(v_2^j)$, $\ldots$, $l(v_k^i)=l(v_k^j)$, and consider two vertices $v_r^i \in S_i$ and $v_r^j \in S_j$.
For simplicity, we use $S$ to represent $S_i$ and $S_j$ and use $v_r$ to represent $v_r^i$ and $v_r^j$.
In order to achieve $C(S)$, it is sufficient to restrict $N(v_r)$ on the subset that exists in $S$, i.e., $N'(v_r) \leftarrow N(v_r)\cup S$.
Since both of the mappings from $s(v)$ to $V_k$ and from $V_k$ to $l(v)$ is bijective, $s(v_r^i) = s(v_r^j)$.
Let $u$ be a vertex in a singleton cell in $\pi$, then $l(v)$ is unique. If $u \in N(v_r^i)$, then $l(u) \in s(v_r^i)$ and $l(u) \in s(v_r^j)$, implying $u \in N(v_r^j)$. Removing each such $u$ from both $N(v_r)$ and $s(v_r)$, then $N'(v_r^i)= N'(v_r^j)$ and $s'(v_r^i) = s'(v_r^j)$. Since $C(S)$ can be easily composed using $s'(v_r)$, $C(S_i) =C(S_j)$, i.e., $S_i$ and $S_j$ are isomorphic equivalent. Applying Theorem~\ref{the:subgraph_inter}, $v_1^i \sim v_1^j$, $v_2^i \sim v_2^j$, $\ldots$, $v_k^i \sim v_k^j$.
\eop

\subsection{Procedure Condense}
\label{sec:condense}

In this section, we discuss Algorithm \Condense, illustrated in Algorithm~\ref{alg:condense}.
As an overview,
Algorithm \Condense works on subgraph set $\mathcal S$, constructs a tree $T_i$ representing the symmetry hierarchy for each $S_i \in \mathcal S$. Here the leaf nodes of $T_i$ are pair-wise disjoint subsets whose union is $S_i$. Each leaf node is a connected component of $S_i$ after removing edges according to Theorem~\ref{the:subgraph_intra}, and a leaf node $N$ represents a subgraph in which the symmetries are not fully detected if $N$ is non-singleton, i.e., $|N| >1$.  As a consequence, Algorithm \Condense returns a set of trees $\mathcal T$, containing $T_i$ for each $S_i \in \mathcal S$ and a set of subgraphs $\mathcal S'$, containing non-singleton leaf nodes in each $T_i \in \mathcal T$. In Algorithm \Condense, sets $\mathcal S'$ and $\mathcal T$ are initialized as $\emptyset$ (Line~1). For each $S_i \in \mathcal S$, Algorithm \Condense applies Procedure \DecTree to construct a tree $T_i$ (Line~3-4). All non-singleton leaf nodes of $T_i$ are added into $\mathcal S'$ and $T_i$ is added into $\mathcal T$ (Line~5-10). Next, we discuss Procedure \DecTree. Procedure \DecTree applies divide-and-conquer paradigm to generate a subtree of tree $T$ rooted at node $S$.
Procedure \DecTree removes edges in $S$ according the Theorem~\ref{the:subgraph_intra}, dividing the problem of constructing the subtree rooted at $S$ into a number of sub-problems, each constructs a subtree rooted at $S'$ invoking $\DecTree(T, S')$. The recursion terminates when $S$ cannot be divided, either $S$ contains only one vertex (Line~14-16) or $S$ contains only one connected component (Line~18-20).

\comment{
\begin{algorithm}[t]
\caption{\DecTree($T, N$) }
\label{alg:dectree}
\begin{algorithmic}[1]
\IF {$|N|=1$}
\RETURN ;
\ENDIF
\STATE let $L$ denotes the set of $l(v)$ values for all $v \in N$;
\FOR {every label $l_1 \in L$}
\IF {vertices with label $l_1$ consist a complete subgraph $c$}
\STATE remove edges in $c$;
\ENDIF
\ENDFOR
\FOR {every two different labels $l_1, l_2 \in L$}
\IF {vertices with label $l_1$ consist a complete bipartite subgraph $b$ with those labeled $l_2$}
\STATE remove edges in $b$;
\ENDIF
\ENDFOR
\end{algorithmic}
\end{algorithm}
}

\begin{algorithm}[t]
\caption{\Condense($\mathcal S$) }
\label{alg:condense}
\begin{algorithmic}[1]
\STATE $\mathcal S' \leftarrow \emptyset$, $\mathcal T \leftarrow \emptyset$;
\FOR {each $S_i \in \mathcal S$}
\STATE construct a tree $T_i$ rooted at node $S_i$;
\STATE \DecTree($T_i, S_i$);
\FOR {every leaf node $N$ of $T_i$}
\IF {$|N| >1$}
\STATE $\mathcal S' \leftarrow \mathcal S' \cup \{$ $N \}$;
\ENDIF
\ENDFOR
\STATE $\mathcal T \leftarrow \mathcal T \cup \{$ $T_i \}$;
\ENDFOR
\RETURN $(\mathcal S', \mathcal T)$;

\vspace*{0.4cm}

\STATE {\bf Procedure} \DecTree($T, S$)
\IF {$|S|=1$}
\RETURN ;
\ENDIF
\STATE remove edges in $S$ according to Theorem~\ref{the:subgraph_intra}
\IF {$S$ contains only one connected component}
\RETURN ;
\ENDIF
\FOR {every connected component $S'$ in $S$}
\STATE add a tree node $S'$ and a tree edge $(S, S'$);
\STATE \DecTree($T, S'$);
\ENDFOR
\RETURN ;
\end{algorithmic}
\end{algorithm}

}

\begin{table}[t]
{\scriptsize 
    \begin{center}
   \begin{tabular}{|l@{}|r|r|r@{}|r@{}|r@{}|r@{}|} \hline
       {\bf Graph}  & { $|V|$}  & { $|E|$} &{$d_{max}$} & {$d_{avg}$} & {$  cells$} & {$  singleton$}  \\       \hline\hline
        Amazon & 403,394& 2,443,408 &2,752 &12.11 &396,034 &390,706  \\ \hline
        BerkStan &685,230 & 6,649,470 &84,230 &19.41 &387,172 &316,162  \\ \hline
        Epinions & 75,879 &405,740 &3,044 &10.69 &53,067 & 45,552
  \\ \hline
        Gnutella &62,586 &147,892 &95 &4.73 &46,098  &38,216  \\ \hline
        Google  &875,713 &4,322,051 &6,332 &9.87 &525,232  &424,563
  \\ \hline
        LiveJournal &4,036,538  &34,681,189 &14,815 &17.18  &3,703,527 &3,518,490 \\ \hline
        NotreDame &325,729 &1,090,108 &10,721  &6.69 &115,038 &89,791   \\ \hline
        Pokec &1,632,803 &22,301,964 &14,854  &27.32  &1,586,176 &1,561,671  \\ \hline
        Slashdot0811 &77,360  &469,180 &2,539 &12.13   &61,457 &56,219
   \\ \hline
        Slashdot0902 &82,168  &504,229 &2,552  &12.27  &65,264
 &59,384  \\ \hline
        Stanford  &281,903  &1,992,636 &38,625  &14.14  &168,967  &133,992   \\ \hline
        WikiTalk  &2,394,385  &4,659,563  &100,029 &3.89 &553,199  &498,161  \\ \hline
        wikivote  &7,115  &100,762 &1,065  &28.32 &5,789  &5,283
  \\ \hline
        Youtube   &1,138,499  &2,990,443 &28,754  &5.25 &684,471  &585,349  \\ \hline
        Orkut    &3,072,627   &117,185,083 &33,313 &11.19  &3,042,918  &3,028,961  \\ \hline
        BuzzNet  &101,163  &2,763,066 &64,289  &54.63  &77,588 &76,758  \\ \hline
        Delicious &536,408 &1,366,136  &3,216  &5.09 &263,961 &221,669  \\ \hline
        Digg  &771,229 &5,907,413  &17,643 &15.32  &445,181 &400,605  \\ \hline
        Flixster &2,523,386  &7,918,801 &1,474  &6.28  &1,047,509 &928,445  \\ \hline
        Foursquare  &639,014  &3,214,986  &106,218  &10.06  &364,447 &315,108   \\ \hline
        Friendster &5,689,498  &14,067,887  &4,423  &4.95  &2,135,136 &1,973,584   \\ \hline
        Lastfm  &1,191,812   &4,519,340 &5,150   &7.58  &675,962  &609,605  \\ \hline
    \end{tabular}
    \end{center}
     \vspace*{-0.4cm}
  \caption{Summarization of real graphs}
  \label{tbl:summarization}
}
\end{table}

\begin{table}[t]
{\scriptsize 
    \begin{center}
   \begin{tabular}{|l|@{}r|r|r|r|r|r|} \hline
       {\bf Graph}  & { $|V|$}  & { $|E|$} &{$d_{max}$} & {$d_{avg}$} & {$  cells$} & {$  singleton$}  \\       \hline\hline
        ag2-49 & 4,851 &120,050 &50 &49.49 &2  &0   \\ \hline
        cfi-200 &2,000 & 3,000 &3 &3 &800 &0  \\ \hline
        difp-21-0-wal-rcr & 16,927 &44,188 &1,526 &5.22 &16,215 & 15,755
  \\ \hline
        fpga11-20-uns-rcr  &5,100 &9,240  &21 &3.62  &3,531  &2,418
  \\ \hline
        grid-w-3-20 &8,000 &24,000  &6  &6  &1 &0  \\ \hline
        had-256  &1,024  &131,584 &257 &257 &1 &0  \\ \hline
        mz-aug-50 &1,000  &2,300  &6  &4.6  &250
 &0 \\ \hline
        pg2-49  &4,902  &122,550 &50  &50  &1  &0   \\ \hline
        s3-3-3-10    &12,974  &23,798  &26  &3.67  &9,146  &5,318 \\ \hline

    \end{tabular}
    \end{center}
    \vspace*{-0.4cm}
  \caption{Summarization of benchmark graphs}
\label{tbl:sum_bm}

}
\end{table}

\begin{table}[h]
{\scriptsize 
    \begin{center}
   \begin{tabular}{|l|r|r|r|r|r|} \hline
       {\bf Graph}  & { $|V({\mathcal {AT}})| $}  & { singleton}  & { non-singleton } & {avg size} & {depth }  \\       \hline\hline
        Amazon & 407,032 & 403,388 &1 &6 &3   \\ \hline
        BerkStan &709,702 &681,680 &118  &30.08 &5   \\ \hline
        Epinions & 76,919 &75,879 &0  & 0 &3   \\ \hline
        Gnutella &62,598 &62,586 &0  &0 &2  \\ \hline
        Google  &910,617 &874,908 &71  &11.34 &5  \\ \hline
        LiveJournal &4,064,750 &4,036,533 &1 &5  &3 \\ \hline
        NotreDame &328,259 &318,204 &46  &163.59 &5  \\ \hline
        Pokec &1,633,602 &1,632,803 &0  &0  & 3 \\ \hline
        Slashdot0811 &77,809  &77,360 &0  &0   &3   \\ \hline
        Slashdot0902 &82,661 &82,168 &0 &0  &3 \\ \hline
        Stanford  &291,006  &279,912 &55  &36.2   &5\\ \hline
        WikiTalk  &2,398,843 &2,394,385 &0  &0 &3   \\ \hline
        wikivote  &7,139  &7,115  &0  &0 & 2  \\ \hline
        Youtube   &1,161,551  &1,138,499 &0   &0 &3 \\ \hline
        Orkut    &3,073,414 &3,072,627 &0  &0  & 3\\ \hline
        BuzzNet  &101,179 &101,163 &0  &0  & 2\\ \hline
        Delicious &537,831  &533,507  &339   &8.56  &3 \\ \hline
        Digg  &771,879 &771,229 &0 &0 &3  \\ \hline
        Flixster &2,524,659  &2,523,386 &0  &0  &3\\ \hline
        Foursquare  &639,015  &639,014  &0 &0  &1   \\ \hline
        Friendster &5,689,609  &5,689,498  &0  &0 &3 \\ \hline
        Lastfm  &1,192,094   &1,191,812 &0   &0  &2\\ \hline
    \end{tabular}
    \end{center}
    \vspace*{-0.4cm}
  \caption{The Structure of AutoTrees of real graphs}
\vspace*{-0.2cm}
\label{tbl:autotree}
}
\end{table}

\begin{table}[h]
{\scriptsize 
    \begin{center}
   \begin{tabular}{|l@{}|r|r|r|r|r|} \hline
       {\bf Graph}  & { $|V({\mathcal {AT}})| $}  & { singleton}  & { non-singleton } & {avg size} & {depth }  \\       \hline\hline
         ag2-49 & 1 &0 &1 &4,851 &0   \\ \hline
        cfi-200 &1 &0 &1 &2,000 &0  \\ \hline
        difp-21-0-wal-rcr & 16,928 &16,927 &0 &0 & 1
  \\ \hline
        fpga11-20-uns-rcr  &2,441 &2,418  &22 &121.91  &1
  \\ \hline
        grid-w-3-20 &1  &0  &1 &8,000   &0  \\ \hline
        had-256  &1  &0  &1 &1,024   &0  \\ \hline
        mz-aug-50 &1   &0  &1  &1,000  &0
   \\ \hline
        pg2-49  &1  &0 &1 &4,902   &0   \\ \hline
        s3-3-3-10    &12,999  &12,974  &0  &0  &2 \\ \hline
    \end{tabular}
    \end{center}
    \vspace*{-0.4cm}
  \caption{The Structure of AutoTrees of benchmark graphs}

\label{tbl:autotree-bm}
}
\end{table}

\begin{table*}[t]
{\scriptsize 
    \begin{center}
    \begin{tabular}{|l@{}||@{}r|r||@{}r|r||@{}r|r||@{}r|r||@{}r|r||@{}r|r|}
      \hline
      \multirow{2}{*} {\bf Graph} &\multicolumn{2}{|c|} {\bf \nauty} &\multicolumn{2}{|c|} {\bf {\CL}+n}  &\multicolumn{2}{|c|} {\bf \traces}  &\multicolumn{2}{|c|} {\bf {\CL}+t} &\multicolumn{2}{|c|} {\bf \bliss} &\multicolumn{2}{|c|} {\bf {\CL}+b}   \\
        \cline{2-13}
       & time &memory & time &memory & time &memory & time &memory & time &memory & time &memory  \\
        \hline
        Amazon  &- &- & 1.19 &280.18 &- &-  &{\bf 1.18} &280.21 &10.88 &158.22 &{\bf 1.18}   &280.23\\ \hline
        BerkStan  &- &- &{\bf 2.59} &575.16 &- &- & 2.6 &575.17 &3,510.85 &694.52 &2.62  &575.26\\ \hline
        Epinions  &- &- &{\bf 0.14} &56.11  &0.43 &84.66  &{\bf 0.14} &55.12 &7.75 &35.86 &{\bf 0.14}  &56.12\\ \hline
        Gnutella  &- &- &0.09 &31.26  &{\bf 0.01} &34.73  &0.09 &31.26 &4.78 &28.19 &0.08  &31.26\\ \hline
        Google  &- &- &2.7 &553.42   &-  & &{\bf 2.69} &553.49 &- &- &{\bf 2.69}  &553.55\\ \hline
        LiveJournal &- &- &15.95 &3,051.48 &- & &{\bf 15.92} &3,051.77 &- &- &15.94 &3,051.96\\ \hline
        NotreDame  &- &- &9.28 &221.98  &-  & &{\bf 0.84} &221.98 &509.32 &202.65 &0.96  &222\\ \hline
        Pokec  &- &- & 6.9 &1,549.93  &-  & &{\bf6.8} &1,549.94 &293.5 &767.44 &6.85  &1,549.95\\ \hline
        Slashdot0811 &- &- &0.15 &57.62  &{\bf 0.13} &65.87 &0.15 &57.62 &5.34 &33.94 &0.15 &57.62\\ \hline
        Slashdot0902 &- &- &0.16 &59.04 &{\bf 0.05}  &58.86 &0.15 &59.04 &5.93 &38.36 &0.16 &59.04\\ \hline
        Stanford   &- &- &{\bf1.34} &193.79  &-  & & 1.36 &193.79 &311.04 &203.99 &1.35  &193.89\\ \hline
        WikiTalk  &- &- &4.04 &1,548.52  &- & &{\bf 3.93} &1,548.55 &- &- &3.96  &1,548.58\\ \hline
        wikivote &9.5 &40.27 &0.02 &6.43  & {\bf 0.01} &6.52 &0.02 &6.42 &0.07 &3.8 &0.02 &6.42\\ \hline
        Youtube   &- &- & 2.33 &769.7 &- & &{\bf 2.3} &769.76 &4,623.3 &517.62 &2.33  &773.65\\ \hline
        Orkut   &- &- &  25.76  &5,182.92 &166.07 &10,856 &{\bf 25.47} &5,182.93 &340.45 &2,224.17 &25.86 &5,182.94\\ \hline
        BuzzNet  &- &- &0.48 &145.05 &{\bf 0.07} &135.22 &0.45 &145.05 &268.16 &69.69 &0.45 &145.05\\ \hline
        Delicious  &- &- &0.94 &384  &2,546.96  &11,535.7 &{\bf 0.89} &384 &1,302.8 &293.54 &0.95 &384 \\ \hline
        Digg   &- &- &  1.83  &565.94 &5.61 &1,012 &{\bf1.82} &565.94 &2,088.65 &418.79 &{\bf1.82} &565.94\\ \hline
        Flixster &- &- &4.26 &1,626.57  &{\bf 1.05} &1,518 &4.2 &1,626.57 &- &-  &4.17 &1,626.57\\ \hline
        Foursquare &- &- &1.32 &449  &{\bf 0.28} &407.55 &1.37 &449 &1,457.7 &340.51  &1.36 &449\\ \hline
        Friendster &- &- &  9.12  &3,337.77 &24.71 &5,607.55 &{\bf8.87} &3,337.77 &- &-  &8.95 &3,337.77\\ \hline
        Lastfm   &- &- &2.2 &847.63  &{\bf 0.6} &713.47 &2.24 &847.63&5,264.58 &553.64  &2.23 &847.63 \\ \hline
    \end{tabular}
    \end{center}
\vspace*{-0.4cm}
\caption{Performance of \nauty, {\CL}+n,  \traces, {\CL}+t,
  \bliss,  and {\CL}+b on real-world networks}
\label{tbl:performance}
\vspace*{-0.4cm}
}
\end{table*}

\begin{table}[t]
{ \small
    \begin{center}
    \begin{tabular}{|l||@{}r|r||@{}r|r|}
      \hline
      \multirow{2}{*} {\bf Graph} &\multicolumn{2}{|c|} {\bf $|S|=10$} &\multicolumn{2}{|c|} {\bf $|S|=100$}    \\
        \cline{2-5}
       & number &time & number &time    \\
        \hline
        Amazon  &1 &0.12 &1 &0.1  \\ \hline
        BerkStan &16 &0.12 &1.12E23 &0.12  \\ \hline
        Epinions &2 &0.01 &840 &0.01 \\ \hline
        Gnutella  &1 &0.01 &1 &0.01 \\ \hline
        Google &40 &0.19 &1.43E25 &0.18  \\ \hline
        LiveJournal &30 &1.39 &1.19E37 &1.53 \\ \hline
        NotreDame  &88 &0.04 &63,360 &0.04 \\ \hline
        Pokec  &1 &0.5 &302,400 &0.52 \\ \hline
        Slashdot0811 &1 &0.01 &192 &0.01 \\ \hline
        Slashdot0902 &2 &0.01 &4,608 &0.01 \\ \hline
        Stanford  &6 &0.04 &1.23E15 &0.04  \\ \hline
        WikiTalk  &1 &0.49 &1 &0.48 \\ \hline
        wikivote  &8.82E15 &0 &2.94E15 &0\\ \hline
        Youtube   &1 &0.27 &1 &0.28 \\ \hline
        Orkut  &4 &1.01 &2.91E10 &1.01  \\ \hline
        BuzzNet  &80 &0.02 &7.36E88 &0.02 \\ \hline
        Delicious  &19 &0.09 &787,968 &0.09  \\ \hline
        Digg   &1 &0.15 &1 &0.16 \\ \hline
        Flixster &1 &0.13 & 1 &0.14 \\ \hline
        Foursquare &6.64E6 &0.13  &4.44E71 &0.13 \\ \hline
        Friendster  &1 &1.64 &1 &1.62 \\ \hline
        Lastfm  &1 &0.29 &1 &0.28   \\ \hline
    \end{tabular}
    \end{center}
\vspace*{-0.4cm}
\caption{SSM on seed set $S$ by IM}
\label{tbl:im}

}
\end{table}

\begin{table}[t]
{ \scriptsize
    \begin{center}
    \begin{tabular}{|l@{}||@{}r|r|r||@{}r|r|r|}
      \hline
      \multirow{2}{*} {\bf Graph} &\multicolumn{3}{|c|} {\bf maximum clique} &\multicolumn{3}{|c|} {\bf triangle}    \\
        \cline{2-7}
       &number &cluster &max  &number &cluster &max  \\
        \hline
        Amazon  &610 &584 &3 &3,986,507 &3,837,711 &120  \\ \hline
        BerkStan &4 &4 &1 &64,690,980 &10,487,015 &735,000  \\ \hline
        Epinions &18 &18 &1 &1,624,481 &1,622,749 &35 \\ \hline
        Gnutella  &16 &16 &1 &2,024 &2,017 &2\\ \hline
        Google &8 &2 &4 &13,391,903 &6,325,254 &4,200 \\ \hline
        LiveJournal &589,824 &36,864 &16  &177,820,130 &158,645,941 &198,485 \\ \hline
        NotreDame  &1 &1 &1 &8,910,005 &2,629,782 &2,268,014 \\ \hline
        Pokec  &6 &6 &1 &32,557,458 &32,545,137 &84 \\ \hline
        Slashdot0811 &52 &52 &1 &551,724 &550,747 &46 \\ \hline
        Slashdot0902 &104 &104 &1 &602,588 &600,239 &242 \\ \hline
        Stanford  &10 &6 &2 &11,329,473 &4,041,344 &42,504 \\ \hline
        WikiTalk  &141 &141 &1 &9,203,518 &9,165,115 &780 \\ \hline
        wikivote  &23 &23 &1 &608,389 &608,366 &6\\ \hline
        Youtube   &2 &2 &1 &3,056,537 &3,036,649 &445 \\ \hline
        Orkut  &20 &20 &1 &- &- &-\\ \hline
        BuzzNet  &12 &12 &1 &30,919,848 &30,914,434 &71 \\ \hline
        Delicious  &9 &9 &1 &487,972 &478,909 &132  \\ \hline
        Digg   &192 &192 &1 &62,710,797 &62,685,651 &407 \\ \hline
        Flixster &752 &752 & 1 &7,897,122 &7,114,518 &192\\ \hline
        Foursquare &8 &8 &1 &21,651,003 &21,646,991 &13 \\ \hline
        Friendster  &120 &120 &1 &8,722,131 &8,604,990 &563 \\ \hline
        Lastfm  &330 &330 &1 &3,946,212 &3,930,145 &100   \\ \hline
    \end{tabular}
    \end{center}
\vspace*{-0.4cm}
\caption{Subgraph clustering by SSM}
\label{tbl:cluster}

}
\end{table}

\section{Experimental studies}
\label{sec:exp}

We conducted extensive experimental studies using 22 large real graphs
and 9 benchmark graphs to test how \CL improves \nauty
\cite{mckay1981practical}, \bliss \cite{junttila2007engineering}, and
\traces \cite{piperno2008search}, using their latest distributed
versions, i.e., \nauty-2.6r10, \traces-2.6r10 ({\small
  \url{http://pallini.di.uniroma1.it/}}) and \bliss-0.73 ({\small
  \url{http://www.tcs.hut.fi/Software/bliss/index.html}}).
Below, we use {\CL}+{\sl X} to indicate that {\sl X} is used to
compute canonical labeling for non-singleton leaf nodes in ${\mathcal
  AT}$. We have {\CL}+n, {\CL}+b and {\CL}+t, where n, b, and t are
for \nauty, \bliss, and \traces.
%
All   algorithms are implemented in C++ and complied by gcc 4.8.2,
and tested on machine with 3.40GHz Intel Core i7-4770 CPU, 32GB RAM
and running Linux. Time unit used is second and we set time limit
as 2 hours.

\stitle{Datasets}: The 22 large real-world graphs include social
networks (Epinions, LiveJournal, Pokec, Slashdot0811, Slashdot0902,
wikivote, Youtube, Orkut, BuzzNet, Delicious, Digg, Flixster,
Foursquare and Friendster), web graphs (BerkStan, Google, NotreDame,
Stanford), a peer-to-peer network (Gnutella), a product co-purchasing
network (Amazon), a communication network (WikiTalk), and a music
website (lastfm).
All these datasets are available online.
\comment{
Here, Amazon, BerkStan, Epinions, Google, Gnutella, LiveJournal,
NotreDame, Pokec, Stanford, Slashdot0811, Slashdot0902, WikiTalk and
wikivote are downloaded from Stanford large network dataset collection
({\small \url{http://snap.stanford.edu/data}}); Both Youtube and Orkut are
from {\small \url{socialnetworks.mpi-sws.org}};
and BuzzNet,
Delicious, Digg, Flixster, Foursquare, Friendster and Lastfm are
from {\small \url{socialcomputing.asu.edu}}.
}
The detailed information of the real-world datasets are summarized in
Table~\ref{tbl:summarization}, where, for each graph, the 2nd and 3rd
columns show the numbers of vertices and edges\footnote{for each
  dataset, we remove directions if included and delete all self-loops
  and multi-edges if exist.}, the 4th and 5th columns show the sizes
of max degree and average degree of each graph, and the 6th and 7th
columns show the numbers of cells and singleton cells in the orbit
coloring of each graph.  As shown in Table~\ref{tbl:summarization},
the majority of the cells in the orbit coloring are singleton cells. This property makes \DivideP and \DivideS effective since the partition (Theorem~\ref{the:simplify}) are more likely to happen when subgraphs get smaller.

For the 9 benchmark graphs, we select the largest one in each family
of graphs used in \bliss collection
\cite{junttila2007engineering}. Detail descriptions of each benchmark
graph can be found in \cite{junttila2007engineering}. Similarly,
summarization   are given in Table~\ref{tbl:sum_bm}.

Below, we first demonstrate the structure of AutoTrees constructed, and
use the observations made to explain the efficiency and performance of our
approaches {\CL}+{\sl X}, which will be confirmed when we illustrate
the performance of {\CL}+{\sl X} and {\sl X}.

\stitle{The Structure of AutoTree}: Table~\ref{tbl:autotree}
demonstrates the structure of  AutoTrees constructed for real graphs
by {\CL}+{\sl X}. Note that for the same graph, three {\CL}+{\sl X} algorithms construct
the same AutoTree.
The 2nd, 3rd, 4th columns show the numbers of total nodes, singleton leaf nodes, non-singleton leaf nodes in AutoTree,
respectively. The 5th column shows the average size (number of vertices) of each non-singleton
leaf node and the 6th column shows the depth of AutoTree. Several
interesting observations can be made.  First, in 15 out of 22
datasets, AutoTree contains only singleton leaf nodes. In these datasets,
there is no need to exploit existing approaches to discover automorphism group and canonical
labeling, i.e., the three {\CL}+{\sl X} algorithms on these graphs can be done in polynomial time and the performances are almost the same.
The AutoTree ${\mathcal {AT}(G,\pi)}$, the automorphism group $Aut(G,\pi)$ and the canonical labeling $C(G,\pi)$ can be achieved
with only an equitable coloring at the root in AutoTree.
Second, in the remaining 7
datasets that contain non-singleton leaf nodes, there are only a small
number of non-singleton leaf nodes and these non-singleton leaf nodes
are  of small sizes. Transferring the problem of discovering
the canonical labeling for a massive graph to finding the canonical
labeling for a few small subgraphs improves the efficiency and
robustness significantly. This observation also explains the phenomenon that all {\CL}+{\sl X} consume  almost the same amount of memory in each datasets:
AutoTree is the most space-consuming structure when there are
only a few small non-singleton leaf nodes.  Third, AutoTrees are usually with low depths. Since both \DivideP and \DivideS cost $O(m_s)$ for a graph with $m_s$ edges and all subgraphs in the same depth in AutoTree  are vertex disjoint, constructing ${\mathcal AT} (G,\pi)$ costs $O(m)$. Similarly, with the canonical labeling of all non-singleton leaf nodes, achieving the canonical labeling for all tree nodes in ${\mathcal AT} (G,\pi)$ only costs $O(m\cdot ln m)$. Forth, comparing the 3rd column in Table~\ref{tbl:autotree} and the 7th column in Table~\ref{tbl:summarization}, \DivideP and \DivideS can further partition some automorphic vertices into singleton leaf nodes in AutoTree, which can further improve the efficiency.

Table~\ref{tbl:autotree-bm} demonstrates the structure of AutoTrees constructed for benchmark graphs by {\CL}+{\sl X}. Different from those constructed for real graphs, AutoTrees of most benchmark graphs contain only the root node. Revisit Table~\ref{tbl:sum_bm}, most benchmark graphs are highly regular and contain none singleton cells, which makes \CL and AutoTree useless in improving the performance.

\comment{
For synthetic graphs, we use SNAP library \cite{snap} to generate
networks by PowerLaw (PL) \cite{molloy1995critical} with $n$ vertices and
exponent $\alpha$, by Watts-Strogatz (WS) \cite{watts1998collective}
with $n$ vertices where each vertex is connected to $k$ nearest
neighbors in the ring topology and each edge is rewired with
probability $p$, by Erd{\H{o}}s-R{\'e}nyi
We do not consider Erd{\H{o}}s-R{\'e}nyi random graph model
\cite{erdds1959random} since the graphs are significantly different
from real-world massive graphs, and we neglect Barab{\'a}si-Albert
preferential attachment model \cite{barabasi1999emergence} since the
maximum clique is known to be a $(k+1)$-clique, where $k$ is the
number of edges each new vertex link to existing vertices.
}

\begin{table}[t]
{ \scriptsize 
    \begin{center}
   \begin{tabular}{|l|r@{}|r@{}|r@{}|r@{}|r@{}|r@{}|} \hline
       {\bf Graph}  & \nauty & {\CL}+n & \traces & {\CL}+t & \bliss & {\CL}+b \\       \hline\hline
        ag2-49  &{\bf 0.04} &0.07 &{\bf 0.04} &{\bf 0.04} &0.21 &0.19  \\ \hline
        cfi-200 &27.62 &1.79 &{\bf 0.02} &{\bf 0.02} &0.66 &0.67 \\ \hline
        difp-21-0-wal-rcr  &8.52 &0.02 &{\bf 0.01} &0.02 &0.05 &0.02  \\ \hline
        fpga11-20-uns-rcr  &0.09 &0.01 &0.23 &0.01 &  ${\bf <0.01}$ &0.01  \\ \hline
        grid-w-3-20 & ${\bf<0.01}$ &0.02 & ${\bf<0.01}$ &0.01 &0.03 &0.02   \\ \hline
        had-256 &0.08 &0.12 & ${\bf<0.01}$ &0.05 &0.51 &0.51    \\ \hline
        mz-aug-50 &4.35 &4.35 &${\bf <0.01}$ &{\bf 0.001} &0.01 &0.01   \\ \hline
        pg2-49  &{\bf 0.04} &0.08 &{\bf 0.04} &{\bf 0.04} &0.25 &0.21    \\ \hline
        s3-3-3-10  &0.01 &0.02 & ${\bf< 0.01}$ &0.02 &0.02 &0.02  \\ \hline

    \end{tabular}
    \end{center}
  \vspace*{-0.4cm}
  \caption{Performance
    on benchmark graphs}
\label{tbl:performance_bm}

}
\end{table}

\comment{

\begin{table}[h]
{\scriptsize 
    \begin{center}
   \begin{tabular}{|l|r|r|r|r|r|} \hline
       {\bf Graph}  & { $|V({\mathcal {AT}})| $}  & { singleton}  & { non-singleton } & {avg size} & {depth }  \\       \hline\hline
        Amazon & 396,506 & 392,931 &3,572 &2.92 &2   \\ \hline
        BerkStan &591,589 & 572,549 &10,302  &10.94 &2   \\ \hline
        Epinions & 76,040 &75,008 &412 &2.11 &2   \\ \hline
        Gnutella &62,576 &62,564 &11  &2 &1  \\ \hline
        Google  &788,287 &756,107 &24,562 &4.87 &2  \\ \hline
        LiveJournal &3,990,848  &3,963,325 &26,588 &2.75  &2\\ \hline
        NotreDame &309,214 &302,277 &2,837  &8.27 &2   \\ \hline
        Pokec &1,631,976 &1,631,181 &755  &2.15  & 2 \\ \hline
        Slashdot0811 &77,353  &76,908 &219  &2.06   &2   \\ \hline
        Slashdot0902 &82,088 &81,605 &270 &2.08  &2\\ \hline
        Stanford  &234,538 &227,007 &3,654  &15.02   &2\\ \hline
        WikiTalk  &2,393,484  &2,389,034 &2,670 &2 &2   \\ \hline
        wikivote  &7,099  &7,075  &20  &2 & 2  \\ \hline
        Youtube   &1,151,179  &1,128,567 &4,639   &0 & 2 \\ \hline
        Orkut    &3,071,705  &3,070,936 &764  &0  & 2\\ \hline
        BuzzNet  &101,153  &101,137 &13  &0  & 2\\ \hline
        Delicious &536,559  &532,639 &482   &8.56  &2\\ \hline
        Digg  &770,865 &770,259 &450 &0 &2  \\ \hline
        Flixster &2,524,554  &2,523,301 &26  &0  &2\\ \hline
        Foursquare  &639,015  &639,014  &0 &0  &1   \\ \hline
        Friendster &5,689,541 &5,689,434  &28  &0 &2 \\ \hline
        Lastfm  &1,192,047  &1,191,765 &23  &0  &2\\ \hline
    \end{tabular}
    \end{center}
  \caption{The Structure of AutoTree of real graphs}
\label{tbl:autotree}
\vspace{-0.2cm}
}
\end{table}

}

\stitle{The efficiency on real datasets}: Table~\ref{tbl:performance}
shows the efficiency of {\CL}+{\sl X} and {\sl X} on real graphs. The
2nd, 4th, 6th, 8th, 10th and 12th columns show the running time of
\nauty, {\CL}+n, \traces, {\CL}+t, \bliss and {\CL}+b,
  respectively.
In Table~\ref{tbl:performance}, the symbol of ``-'' indicates that the
algorithm cannot get the result in 2 hours, and the champion on each
dataset is  in bold. Several points can be made. First,
among  22 datasets, {\CL}+{\sl X}  outperform {\sl X} in 14 datasets significantly. Specifically, in 3 datasets (Google, LiveJournal
and WikiTalk), none of previous approaches can achieve the results,
and in 10 datasets, none of previous approaches can obtain the results
in 100 seconds. For the remaining 8 datasets, \traces performances the
best, however, its advantage over {\CL}+t  is marginal. Second,
if we take \CL as a preprocessing procedure, \CL improves the efficiency
and robustness of \nauty, \bliss and \traces
significantly. Third, among the 6 algorithms, only {\CL}+{\sl X} algorithms
can achieve results in all datasets, and furthermore, {\CL}+{\sl X} can
get the results in all datasets in less than 26 seconds.
We explain the efficiency and robustness of \CL. By constructing an AutoTree, \CL is able to
discover canonical labeling for only few small subgraphs instead of directly finding the canonical labeling for a massive graph for two reasons. (1) The AutoTree construction, including graph partition and canonical labeling generation, is of low cost. (2) Finding canonical labeling for small subgraphs is always efficient. Forth, for the graphs whose AutoTrees contain no non-singleton leaf nodes, all the three {\CL}+{\sl X} algorithms perform similarly.

The 3rd, 5th, 7th, 9th, 11th and 13th columns illustrate the max
memory consumptions of \nauty, {\CL}+n, \traces, {\CL}+t,
\bliss and {\CL}+b, respectively. First, it is interesting to
find that Algorithms {\CL}+n, {\CL}+t
and {\CL}+b  consume almost the same amount of memory in each
dataset, confirming our analysis when demonstrating the structure of AutoTrees.
Second, \bliss consumes the least amount of
memory in most datasets.

\stitle{The efficiency on benchmark datasets}:
Table~\ref{tbl:performance_bm} shows the efficiency of \CL and its
comparisons on benchmark graphs. The 2nd, 3rd, 4th, 5th, 6th and 7th
columns show the running time of \nauty, {\CL}+n, \traces,
{\CL}+t, \bliss and {\CL}+b, respectively. Worth noting that
due to the accuracy of the timers provided in \nauty, \traces and
\bliss,
we equate $<0.01$ with any value in
$[0,0.01)$.
From Table~\ref{tbl:performance_bm}, almost all algorithms perform
well in all datasets. Among these 6 algorithms,  \traces and
{\CL}+t are the best two approaches. Although \traces performs
the best in more datasets then {\CL}+t, {\CL}+t is more
robust than \traces. Specifically, {\CL}+t can achieve the
result in at most 0.04s in any benchmark dataset tested, while
\traces spends   0.23s in dealing with
fpga11-20-uns-rcr.

In conclusion, since AutoTrees constructed for real graphs are of low depths and  non-singleton leaf nodes in AutoTrees are few and small, our   \CL reduces substantial redundant computations and significantly improves the performance for massive real graphs by introducing small extra cost for constructing AutoTrees. Due to the small sizes and regularity of benchmark graphs, the improvements are not remarkable.

\comment{
\begin{figure}[t]
\begin{center}
  \includegraphics[width=0.7\columnwidth,height=2.8cm]{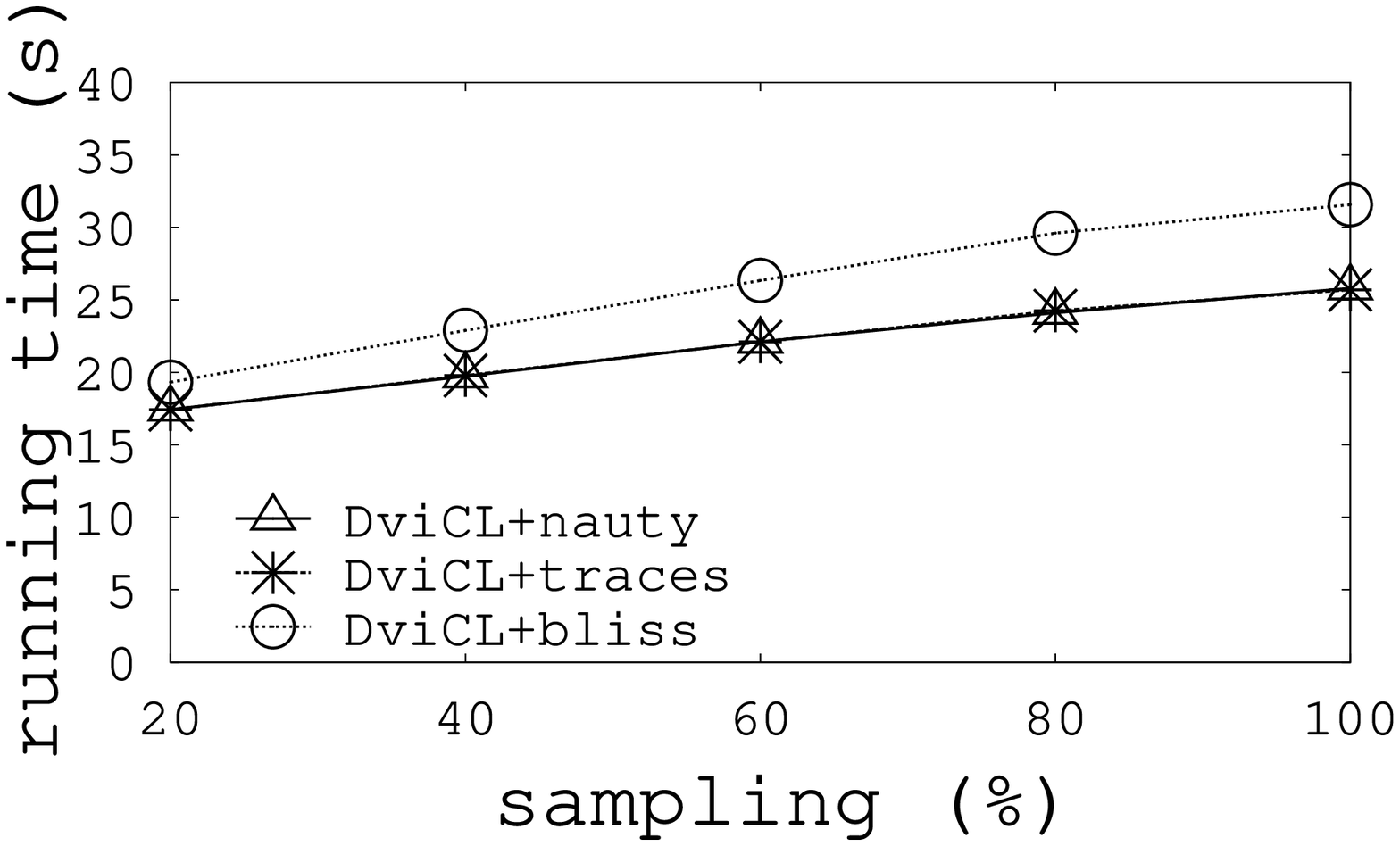}
\end{center}
\vspace*{-0.2cm}
\caption{The scalability of \CL }
\label{fig:scalability}
\vspace*{-0.6cm}
\end{figure}

\stitle{The scalability of \CL}: We study the scalability of \CL using LiveJournal. To test the scalability, we sample 5 subgraphs with 20\%, 40\%, 60\%, 80\% and 100\% edges, respectively. Fig.~\ref{fig:scalability} shows that \CL algorithms scales well.
}

\stitle{Applications of SSM}: We study the applications of SSM. First, given a seed set $S$ by influence maximization, we estimate the number of sets that have the same max influence as $S$. Here, $S$ is obtained by PMC\cite{ohsaka2014fast}, one of the best performing algorithms for IM, and seed number $k$, i.e., $|S|$, is set as $10$ and $100$, respectively.
Table~\ref{tbl:im} demonstrates the results. The 2nd and 4th columns show the number of candidate seed sets when $|S|=10$ and $|S|=100$, respectively. The 3rd and 5th columns show the running time for estimation. Several observations can be made. First, for a large number of graphs tested, numerous candidate sets can be found. Second, it is efficient to estimate the number of candidate sets. The reasons are as follows, 1) the most time consuming part in Algorithm \SSM is invoking SM on non-singleton leaf nodes (Line~3 in Algorithm~\ref{alg:SSM}); 2)in AutoTrees, non-singleton leaf nodes are few and are of small sizes.  

Second, we study subgraph clustering by SSM. Given a set of subgraphs in a graph $G$, all these subgraphs can be clustered s.t., each cluster contains subgraphs that are mutually symmetric. We consider the set of all maximum cliques and all triangles, and estimate the number of clusters and the size of the maximum cluster. Table~\ref{tbl:cluster} illustrates the results. The 2nd, 3rd and 4th columns shows the total number, the number of clusters and the size of the maximum cluster for maximum cliques, respectively. The 5th, 6th and 7th columns shows the statistics for triangles, respectively. 
It is interesting to find that, 1) both the maximum cliques and triangles are diverse; 2) given a single maximum clique or a triangle, it is possible to find several symmetric ones by SSM.

\section{Conclusion}
\label{sec:conclusion}

In this paper, we study graph isomorphism and automorphism
detection for massive
graphs. Different from the state-of-the-art algorithms that adopt an
individualization-refinement schema for canonical labeling, we propose
a novel efficient canonical labeling algorithm \CL following the
divide-and-conquer paradigm.  With \CL, a tree-shape index, called
AutoTree, is constructed for the given colored graph
$(G,\pi)$. AutoTree ${\mathcal AT}(G,\pi)$ provides insights into the symmetric structure of
$(G,\pi)$ in addition to the automorphism group and canonical
labeling.
We show that ${\mathcal AT}(G,\pi)$ can be used (1)  to
find all possible seed sets for influence maximization and
(2) to find all subgraphs in a graph $G$ that are symmetric to a given subgraph that exist in $G$.
%
%
We conducted comprehensive experimental studies to demonstrate the
efficiency and robustness of our approach \CL.
First, non-singleton leaf nodes in AutoTrees constructed are few and small, and the AutoTrees are of low depths. Thus, the extra cost for AutoTree construction is low and worthy.
Second, {\CL}+{\sl X}
outperforms {\sl X}, where {\sl X} is for \nauty, \traces and
\bliss, in 14 out of 22 datasets significantly. For the remaining 8
datasets, only \traces can beat {\CL}+{\sl X}, whereas the advantages
are marginal. Third, among these 6 algorithms tested, all of {\CL}+{\sl X} can achieve the results in all datasets in less than 26 seconds, while {\sl X} is
inefficient in most datasets.

\stitle{ACKNOWLEDGEMENTS}: This work was supported by the grants from RGC 14203618, RGC 14202919, 
RGC 12201518, RGC 12232716, RGC 12258116, RGC 14205617 and  NSFC 61602395.

{\small
\bibliographystyle{abbrv}
\bibliography{ref}
}

\end{document}